\documentclass[prb,aps,nofootinbib,amssymb,twocolumn,superscriptaddress,10pt]{revtex4-2}
\usepackage[T1]{fontenc}
\usepackage{graphicx}
\usepackage{xcolor}
\usepackage{dcolumn}
\usepackage{comment}
\usepackage{amsmath,amssymb,bbold,bm}
\usepackage{hyperref}
\usepackage{amsfonts}
\usepackage{float,latexsym}
\usepackage{multirow}
\usepackage{color}
\usepackage[normalem]{ulem}
\usepackage{cleveref}
\usepackage{physics}
\usepackage{tabularx}
\usepackage{array}
\usepackage[caption=false]{subfig}
\usepackage{siunitx}
\usepackage{gensymb}
\usepackage{pifont}
\usepackage{placeins}
\usepackage{mathtools}
\usepackage{longtable}
\usepackage{multirow}
\usepackage{nicematrix}
\usepackage{tikz}
\usepackage{xstring}
\usepackage{makecell}
\usepackage{xr}
\usepackage{ifthen}
\usetikzlibrary{decorations.markings}
\usetikzlibrary{positioning}
\usetikzlibrary{arrows.meta}
\usepackage{tikz-feynman}
\usetikzlibrary{calc}
\tikzfeynmanset{warn luatex=false}
\let\vec\mathbf

\usepackage{chemformula}
\usepackage{amsthm}
\usepackage[most]{tcolorbox}

\makeatletter
\def\maketag@@@#1{\hbox{\m@th\normalfont\normalsize#1}}
\makeatother

\crefname{appendix}{Appendix}{Appendices}
\crefname{equation}{Eq.}{Eqs.}
\crefname{figure}{Fig.}{Figs.}
\crefname{table}{Table}{Tables}
\crefname{section}{Section}{Sections}
\crefname{enumi}{Point}{Points}
\creflabelformat{appendix}{[#2#1#3]}
\newcounter{mybox}
\crefname{mybox}{Box}{Boxes}
\crefmultiformat{figure}{Figs.~#2#1\xdef\mycreffirstarg{#1}#3}%
 { and~#2#1#3}{, #2#1#3}{ and~#2#1#3}

\newcommand{\citeSI}[1]{(\cref{#1})}
\newcommand{\citeSINoPara}[1]{\cref{#1}}

\newcommand{\cre}[2]{\hat{#1}^\dagger_{#2}}
\newcommand{\des}[2]{\hat{#1}_{#2}}

\newcommand{\vk}{\vec{k}}

\newcommand{\vp}{\vec{p}}

\newcommand{\vQ}{\vec{Q}}
\newcommand{\vq}{\vec{q}}
\newcommand{\vG}{\vec{G}}
\newcommand{\vR}{\vec{R}}

\def\kk{\mathbf{k}}
\def\qq{\mathbf{q}}

\def\QQ{\mathbf{Q}}
\def\RR{\mathbf{R}}
\def\rr{\mathbf{r}}
\def\xx{\mathbf{x}}

\def\ba#1\ea{\begin{align}#1\end{align}}  

\definecolor{colorhhy}{rgb}{0.9, 0.17, 0.31}

\definecolor{colorlmr}{rgb}{0.1, 0.2, 0.7}

\newcommand{\titlePaper}{
Emergent Interacting Phases in the Strong Coupling Limit of Twisted $M$-Valley Moir\'e Systems: Application to SnSe${}_2$
}

\crefname{appendix}{Appendix}{Appendices}
\crefname{equation}{Eq.}{Eqs.}
\crefname{figure}{Fig.}{Figs.}
\crefname{table}{Table}{Tables}
\crefname{section}{Section}{Sections}
\creflabelformat{appendix}{[#2#1#3]}

\makeatletter
\AtBeginDocument{\let\LS@rot\@undefined}
\makeatother

\makeatletter
\renewcommand\onecolumngrid{
\do@columngrid{one}{\@ne}%
\def\set@footnotewidth{\onecolumngrid}
\def\footnoterule{\kern-6pt\hrule width 1.5in\kern6pt}%
}

\allowdisplaybreaks 

\makeatletter     
\renewcommand\onecolumngrid{
\do@columngrid{one}{\@ne}%
\def\set@footnotewidth{\onecolumngrid}
\def\footnoterule{\kern-6pt\hrule width 1.5in\kern6pt}%
}

\makeatletter
\AtBeginDocument{\let\LS@rot\@undefined}
\makeatother 
\renewcommand{\arraystretch}{1.2}

\begin{document}
\title{\titlePaper}
    \author{Ming-Rui Li}
	\affiliation{Institute for Advanced Study, Tsinghua University, Beijing 100084, China}
    \affiliation{Department of Physics, Princeton University, Princeton, NJ 08544, USA}
    \author{Dumitru Calugaru}
   \affiliation{Rudolf Peierls Centre for Theoretical Physics, University of Oxford, Oxford OX1 3PU, United Kingdom}
   \author{Yi Jiang}
   \affiliation{Donostia International Physics Center (DIPC), Paseo Manuel de Lardizábal. 20018, San Sebastián, Spain}
   \author{Hanqi Pi}
   \affiliation{Donostia International Physics Center (DIPC), Paseo Manuel de Lardizábal. 20018, San Sebastián, Spain}
\author{Ammon Fischer}
        \affiliation{Max Planck Institute for the Structure and Dynamics of Matter, Luruper Chaussee 149, 22761
Hamburg, Germany} 
\author{Henning Schl\"omer}
\affiliation{Department of Physics and Arnold Sommerfeld Center for Theoretical Physics (ASC), Ludwig-Maximilians-Universit\"at M\"unchen, Theresienstr. 37, M\"unchen D-80333, Germany}
\affiliation{Munich Center for Quantum Science and Technology (MCQST), Schellingstr. 4, M\"unchen D-80799, Germany}
\author{Lennart Klebl}
\affiliation{Institute for Theoretical Physics and Astrophysics
and Würzburg-Dresden Cluster of Excellence ct.qmat,
University of Würzburg, 97074 Würzburg, Germany}
\author{Maia G. Vergniory}
\affiliation{Département de Physique et Institut Quantique, Université de Sherbrooke, Sherbrooke, J1K 2R1 Québec, Canada}
 \affiliation{Donostia International Physics Center (DIPC), Paseo Manuel de Lardizábal. 20018, San Sebastián, Spain}

    \author{Dante M. Kennes}
        \affiliation{Max Planck Institute for the Structure and Dynamics of Matter, Luruper Chaussee 149, 22761
Hamburg, Germany} 
\affiliation{
Center for Computational Quantum Physics (CCQ), The Flatiron Institute, New York, New York 10010, USA}
\affiliation{Institut f¨ur Theorie der Statistischen Physik, RWTH Aachen, 52056 Aachen, Germany and JARA - Fundamentals of Future Information Technology}
    \author{Siddharth A. Parameswaran}
    \affiliation{Rudolf Peierls Centre for Theoretical Physics, University of Oxford, Oxford OX1 3PU, United Kingdom}
    \author{Hong Yao}
    \affiliation{Institute for Advanced Study, Tsinghua University, Beijing 100084, China}
\author{B.~Andrei Bernevig}
\email{bernevig@princeton.edu}
\affiliation{Department of Physics, Princeton University, Princeton, NJ 08544, USA}
\affiliation{Donostia International Physics Center (DIPC), Paseo Manuel de Lardizábal. 20018, San Sebastián, Spain}
\affiliation{IKERBASQUE, Basque Foundation for Science, 48013 Bilbao, Spain}
\author{Haoyu Hu}
\email{hh5463@princeton.edu}
\affiliation{Department of Physics, Princeton University, Princeton, NJ 08544, USA}
\let\oldaddcontentsline\addcontentsline



\begin{abstract}
We construct an interacting Wannier model for both AA-stacked and AB-stacked twisted \ch{SnSe2}, revealing a rich landscape of correlated quantum phases. For the AA-stacked case, the system is effectively described by a three-orbital triangular lattice model, where each orbital corresponds to a valley and exhibits an approximate one-dimensional hopping structure due to a new momentum-space non-symmorphic symmetry. 
By exploring the interacting phase diagram using a combination of theoretical methods, including Hartree-Fock mean-field theory and exact solutions of the spin model in certain limits, we identify several exotic quantum phases. These include a dimerized phase with finite residual entropy, valence bond solids, and quantum paramagnetism. In the AB-stacked case, the system realizes an interacting kagome lattice model, where the Wannier orbitals associated with the three valleys form three sublattices. 
In the strong coupling regime, we use cluster mean-field methods to demonstrate the emergence of a classical spin liquid phase due to the frustrated lattice structure.
The high tunability of the moir\'e system, which allows control over both the filling and interaction strength (via twist angle), renders twisted \ch{SnSe2} a versatile platform for realizing a wide range of exotic correlated quantum phases.
\end{abstract}
\maketitle

{\it Introduction.}
The discovery of moiré superlattices has revolutionized the study of strongly correlated systems by providing an unprecedented level of control for engineering electronic phases through the variation of twist angles and stacking configurations ~\cite{AND21,KEN21,GOM22,MAK22,jiang20242dtheoreticallytwistablematerial}.
The moiré modulation quenches the band structure and drives the system into the strong interaction regime. The high tunability further enables the realization of various exotic many-body phases. In particular, twisted bilayer graphene ~\cite{BIS11} has been shown to host unconventional superconductivity~\cite{CAO18a,YAN19,LU19,STE20,SAI20,DE21a,OH21,TIA23,DI22a,CAL22d}, correlated insulators~\cite{CAO18,KER19,XIE19,SHA19,JIA19,CHO19,POL19,YAN19,LU19,STE20,SAI20,SER20,CHE20b,WON20,CHO20,NUC20,CHO21,SAI21,LIU21c,PAR21c,WU21a,CAO21,DAS21,TSC21,PIE21,STE21,CHO21a,XIE21d,DAS22,NUC23,YU23c}, 
and other exotic phases \cite{TOM19,CAO20,ZON20,LIS21,BEN21,LIA21c,ROZ21,SAI21a,LU21,HES21,DIE23,HUB22,GHA22,JAO22,PAU22,GRO22,ZHO23a}. 
It has also emerged as a simulator for topological heavy-fermion physics~\cite{SON22,HU23i,HU23,RAI23a,CAL24,HER24,MER24,YU23a,BAT24,PhysRevB.106.245129,CHO23,PhysRevB.109.125404,herzogarbeitman2024topologicalheavyfermionprinciple,Singh2024}. More recently, multilayer rhombohedral graphene was also shown to exhibit highly nontrivial correlated and topological physics~\cite{Han2024,LU2024,Choi2025,Han2025,doi:10.1073/pnas.2017366118,PhysRevB.109.205122,kwan2023moirefractionalcherninsulators,yu2024moirefractionalcherninsulators,PhysRevB.110.115146,PhysRevLett.133.206503,PhysRevLett.133.206502,PhysRevLett.133.206504,VinasBostrom2024,kwan2024fractionalchernmosaicsupermoire,Xia2025,park2025ferromagnetism,bernevig2025berrytrashcanmodelinteracting,li2025multibandexactdiagonalizationiteration,xie2025unconventionalorbitalmagnetismgraphenebased}. Beyond graphene-based systems, transition metal dichalcogenide bilayers offer another avenue to realize a wide variety of exotic phases, including correlated insulators~\cite{WU18c,TAN20,Wang2020,Wei2025}, unconventional superconductors~\cite{PhysRevLett.130.126001,GUO24,XIA24a,PhysRevResearch.5.L012034,PhysRevB.111.L060501,Kim2025,guerci2024topologicalsuperconductivityrepulsiveinteractions,qin2024kohnluttingermechanismsuperconductivitytwisted,tuo2024theorytopologicalsuperconductivityantiferromagnetic,PhysRevB.111.014507,fischer2024theoryintervalleycoherentafmorder,xu2025signaturesunconventionalsuperconductivitynear,xu2025chiralsuperconductivityspinpolarized}, integer and fractional Chern insulators~\cite{li2021quantum,devakul2022quantum,xie2022valley,tao2024valley,regnault2011fractional,wu2012zoology,LI21d,XU23,PAR23,CAI23,ZEN23,crepel2023anomalous,RED23a,RED23,WAN24a,JIA24,YU24a,JU24,xu2024maximally,morales2024magic,zhang2024universalmoiremodelbuildingmethodfitting,wang2024phase,PhysRevLett.134.076503,park2025ferromagnetism,park2025observation,chang2025evidencecompetinggroundstates}, and other intriguing emerging phases~\cite{WAN22,doi:10.1073/pnas.2321665121,Zong2025}. These discoveries highlight the growing potential of moir\'e materials as a versatile platform for exploring correlated and topological quantum physics.

However, to date, most moir\'e superlattices have been constructed by twisting monolayers with low-energy electronic states located near the 
$K$~\cite{BIS11,WU19b,WU18c,DEV21,ZHA24a} or $\Gamma$~\cite{ANG21,CLA22a} points of the Brillouin zone (BZ). Recently, a new class of moir\'e systems based on monolayers with low-energy states at the $M$ points of the BZ has garnered significant interest and has been actively explored~\cite{cualuguaru2024new,PhysRevB.107.085127,eugenio2025tunablettuhubbardmodels,MAH24,JIA24b,bao2024anisotropicbandflatteningtwisted,luo2025singlebandsquarelatticehubbard,sun2025singlebandtriangularlatticehubbard}.
These $M$-valley systems feature three time-reversal-preserving valleys related by three-fold rotational symmetry. Specifically, twisted bilayers of 1T-\ch{SnSe2}, among other materials, have been proposed as an experimentally realizable candidate \cite{cualuguaru2024new,lei2024moirebandtheorymvalley} 
In this work, we construct interacting Wannier models based on {\it ab initio} calculations of both AA-stacked and AB-stacked twisted \ch{SnSe2} and explore the corresponding correlated physics. 
In the AA-stacked case, emergent momentum-space non-symmorphic symmetries~\cite{cualuguaru2024new,DE21,CHE22,ZHA23b,XIA24} lead to quasi-one-dimensional intra-valley hopping, 
while the system acquires a two-dimensional character through inter-valley interactions. 
Combining various theoretical methods, in some physical limits, we identify several exotic phases, including a dimerized phase with finite residual entropy, valence bond solids, and quantum paramagnetism.
For the AB-stacked \ch{SnSe2}, the Wannier centers of the three valleys form a kagome lattice. 
The specific lattice geometry
 enables us to map the system in the strong coupling limit to a kagome Ising model.
 The frustration inherent in the kagome lattice gives rise to strong fluctuations of Ising variables, and stabilizes a classical spin liquid phase.

\begin{figure}
    \centering
    \includegraphics[width=1.0\linewidth]{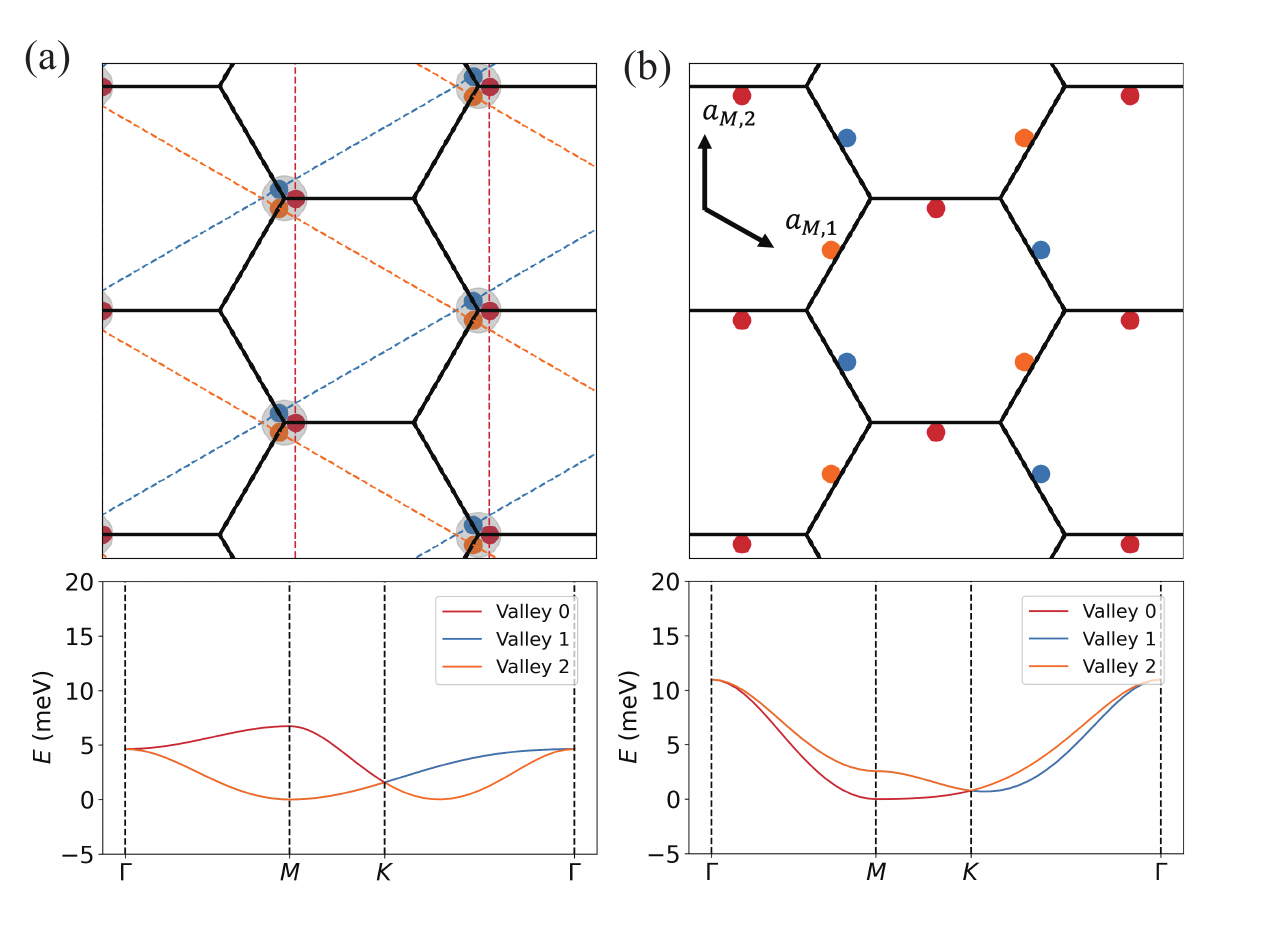}
\caption{Wannier centers (upper panel) and lowest‐band dispersions at $\theta=3.89^\circ$ (lower panel) of the three valleys for AA-stacked (a) and AB-stacked (b) \ch{SnSe2}. In the upper panel, solid black lines denote the moiré unit cell boundary. 
The Wannier centers of valleys $0$, $1$, and $2$ are marked by red, blue, and orange dots, respectively. In panel (a), the dominant intra-valley hoppings are indicated by dashed lines. In panel (b), the three sublattices form an approximate kagome lattice. 
}

    \label{fig:main:wannier}
\end{figure}

{\it Interacting Wannier models.~~}
For both AA-stacked and AB-stacked twisted \ch{SnSe2}, the lowest conduction band at each valley and for each spin is topologically trivial~\cite{cualuguaru2024new}, allowing the construction of exponentially localized Wannier orbitals~\cite{PhysRevB.85.115415,MAR12} \citeSI{sec:wann_orb}. The corresponding interactions between Wannier orbitals are obtained by projecting the screened Coulomb repulsion~\cite{PhysRevB.103.205414} to the Wannier basis \citeSI{sec:wann_model}.

The Wannier centers differ between the two stacking configurations. 
In \cref{fig:main:wannier}, we use dots of three different colors to indicate the Wannier centers corresponding to three distinct valleys for AA-stacked (panel (a)) and AB-stacked (panel (b)) \ch{SnSe2}. In the AA-stacked case, the Wannier centers of the three valleys are nearly coincident, enabling an effective description in terms of a three-orbital triangular lattice Hubbard model (\cref{app:sec:wannier_aa}). 
In contrast, for AB-stacked \ch{SnSe2}, the Wannier centers form an approximate kagome lattice, giving rise to an effective kagome lattice model where the three valleys correspond to three sublattices.
These models exhibit novel features that distinguish them from conventional triangular and kagome lattice models, as discussed in the following. 

For the AA-stacked \ch{SnSe2}, the kinetic term of the system can be written as (\cref{app:sec:wannier_aa})
\begin{align}
\label{eq:main:hopping}
      H_{t}= \sum_{\vR ,\Delta \vR, \eta, s} t^{\eta}_{\Delta \vR} \cre{d}{\vR,\eta,s}\des{d}{\vR+\Delta \vR,\eta,s},
\end{align}
where $\hat{d}^{(\dagger)}_{\vR,\eta,s}$ annihilates (creates) an electron at moir\'e unit cell labeled by $\vR$ of valley $\eta$ and spin $s$, and $t^\eta_{\Delta\vR}$ is the hopping amplitude for valley $\eta$ along $\Delta \vR$. 
Due to the approximate non-symmorphic symmetry~\cite{cualuguaru2024new}, the hopping exhibits a quasi-1D character, with the dominant hopping terms given by $t_{\Delta\vR}^\eta= t \delta_{\Delta\vR, \pm C_{3z}^\eta \bm{a}_{M,2}}$. In \cref{fig:main:wannier} (a), we have indicated the dominant hopping directions for the three valleys using dashed lines of different colors. 
The dominant interaction terms are the local intra-valley (inter-valley) repulsions that act between electrons occupying the same (different) Wannier orbitals within the same unit cell as marked by gray-shaded circles in~\cref{fig:main:wannier}~(a).
The corresponding Hamiltonian takes the form 
\ba 
\label{eq:main:HU}
    H_U = \sum_{\vR, \eta, \eta'}\frac{U_{\eta\eta'}}{2}n_{\vR,\eta}n_{\vR,\eta'} ,
\ea 
where the density operator is defined as  
$
    n_{\vR,\eta}=\sum_{s}\cre{d}{\vR,\eta,s}\des{d}{\vR,\eta,s}.
$
Since the Wannier centers of the three valleys within the same unit cell are very close to each other, the 
{interaction} approximately preserves a $U(6)$ symmetry, with 
$U_{\eta \eta'}\approx U$. 
The subleading interactions are long-range density-density couplings, which do not exceed $10\%$ of the on-site Hubbard interaction (\cref{app:sec:wannier_aa}). 
With only the nearest-neighbor 1D hoppings and the density-density interactions, the model also has a particle-hole symmetry. 
(\cref{app:sec:interacting_model_symmetry}). 

We now discuss the Wannier model of AB-stacked \ch{SnSe2}, where the kinetic term takes the same form of \cref{eq:main:hopping} with only intra-valley hopping. Unlike the conventional kagome model, nearest-neighbor inter-valley hopping is forbidden. 
The dominant interactions can be written as  
\begin{align}
\label{eq:main:HV}
    H_V =& \frac{1}{2} \sum_{\vR,\Delta\vR,\eta\eta'} V_{\eta\eta'}(\Delta\vR) n_{\vR, \eta}(\vR) n_{\vR+\Delta\vR,\eta'},
\end{align}
which consist of an on-site intra-valley Hubbard interaction $V_{\eta,\eta}(\mathbf{0})=U$ and a nearest-neighbor inter-valley density-density interaction of strength $V_1$,
where the nearest neighbor is defined with respect to the kagome lattice geometry. 
%
Next, we investigate the correlation physics and quantum phases captured by our interacting Wannier models. 
Due to the uncertainty of the overall interaction scale (\cref{sec:wann_model}), we take the dielectric constant $\epsilon$ as a tuning parameter. 

\begin{figure*}[ht]
    \centering
\includegraphics[width=0.9\linewidth]{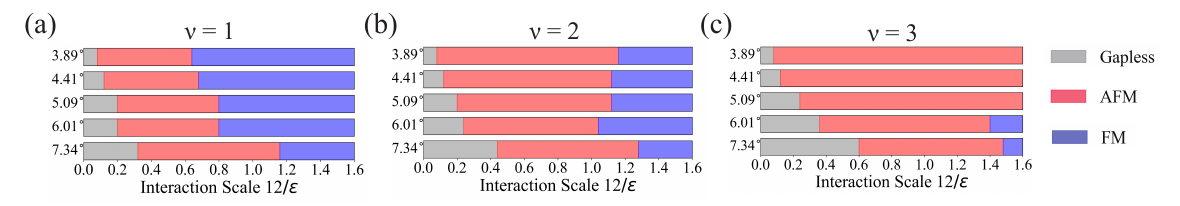}
    \caption{
    (a) (b) (c) Hartree-Fock phase diagram for $ \nu = 1, 2, 3$. In the phase diagram, the gray, red, and blue regions correspond to the weak-coupling (gapless symmetric), strong-coupling (gapped AFM), and flat-band regimes (gapped FM), respectively.}
    \label{fig:main:phase_diagram}
\end{figure*}

{\it Hartree-Fock phase diagram of AA-stacked \ch{SnSe2}.}
We begin with the phase diagram of AA-stacked \ch{SnSe2}, obtained via momentum-space, self-consistent Hartree-Fock. Due to particle-hole symmetry that is present in the model with only nearest-neighbor hoppings and the density-density interactions, we focus on integer fillings $\nu = 1, 2, 3$, where $\nu$ represents the number of electrons per moir\'e unit cell out of a maximum of six.
%
The phase diagrams for different interaction strengths (characterized by different dielectric constants $\epsilon$) are presented in \cref{fig:main:phase_diagram} (a,b,c), and can be categorized into three distinct regions: the weak coupling region, the strong coupling region, and the flat-band region.

In the weak-coupling region, our Hartree-Fock simulations
reveal only gapless phases \citeSI{sec:HFphase_diagram} \cite{tba2025a}. As the interaction strength increases, the system enters the strong-coupling regime, where charge degrees of freedom become gapped due to the strong interaction. However, virtual hopping processes give rise to a Heisenberg interaction of order $t^2/U$. 
We observe the emergence of robust antiferromagnetic spin order in this regime, suggesting that the system could be effectively described by a Heisenberg model.
Further increasing the interaction strength drives the system into the ultra-strong coupling regime. 
In this regime, the Heisenberg exchange $t^2/U$ is strongly suppressed due to the large interaction, distinguishing it from the previous strong-coupling regime.
The finite quantum geometry of the bands leads to a ferromagnetic, valley-polarized state~\cite{PhysRevB.103.205414,KAN19}. We also note that the ultra-strong coupling regime for $\nu=3$ emerges at $12/\epsilon > 1.6$ in the small-angle regime (as detailed in \citeSINoPara{sec:HFphase_diagram}).

\begin{figure*}[ht]
    \centering
\includegraphics[width=1.0\linewidth]{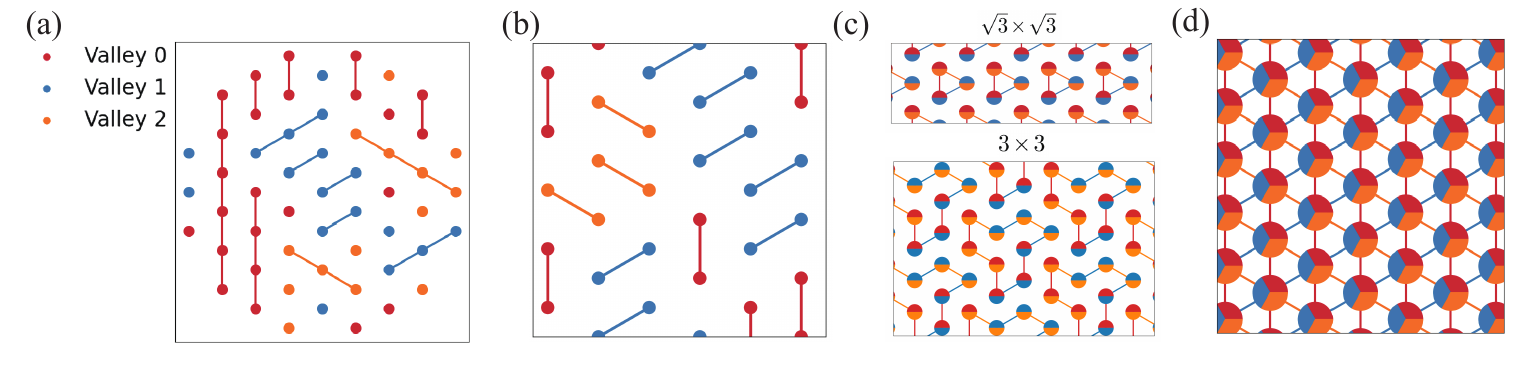}
    \caption{(a) A random $\{n_{\vR,\eta}\}$ configuration for $\nu=1$. 
    Each dot represents a unit cell of the system where red, orange and blue denote the unit cell has one electron in valley $0,1,2$ respectively. The dots connected by different colors denote the formation of spin chains with various lengths in different valleys. 
    (b) (c) (d) characterize the ground state configurations at filling $\nu=1,2,3$ in the strong coupling limit, respectively. The dots marked with one, two, or three colors represent unit cells occupied by one, two, or three electrons from different valleys, respectively. }
    \label{fig:main:spin_config}
\end{figure*}

{\it Quantum spin model in the strong coupling limit of $AA$-stacked \ch{SnSe2}.} 
We now focus on the strong coupling region. In this regime, the effective quantum spin model can be solved exactly in certain limits. 
By employing second-order perturbation theory, we derive the following quantum spin model (\cref{sec:spin_strong_coupling}) applicable to integer fillings $\nu = 1,2,3$:
\begin{align}\label{eq:effective_spin_ham}
  &H_{\nu}=\sum_{\vR,\Delta\vR,\eta,\eta'}
  P_{L}^\nu \frac{\delta V_{\eta\eta'}(\Delta\vR)}{2} n_{\vR,\eta}n_{\vR+\Delta\vR,\eta'}
  P_L^\nu  \nonumber\\ 
  &
+J\sum_{\vR,\Delta\vR,\eta}
        P_L^{\nu}
       \bigg[\bm{S}_{\vR,\eta}\cdot \bm{S}_{\vR+\Delta\vR,\eta} +\frac{n_{\vR,\eta}n_{\vR+\Delta\vR,\eta}}{4}\bigg] 
    P_L^{\nu},
\end{align}
Here, $P_L^\nu$ denotes the projection operator that restricts the system to the low-energy subspace with $\nu$ electrons per moiré unit cell, and $S^\mu_{\vR,\eta} = \frac{1}{2} \sum_{s,s'} \cre{d}{\vR,\eta,s} \sigma^\mu_{s,s'} \des{d}{\vR,\eta,s'}$ represents the spin operator for valley $\eta$.  
The model includes two types of interactions (\cref{sec:spin_strong_coupling}): The first, $\delta V_{\eta,\eta'}(\Delta\vR)$, is a density-density term from the projected Coulomb repulsion that favors valley polarization. The second is an intra-valley Heisenberg interaction, derived via second-order perturbation theory, with coupling $J = 2t^2 / (U - V)$, where $V$ is the nearest-neighbor interaction strength. 
Since the hopping is 1D-like, the corresponding spin coupling is also 1D-like, thus we only keep the dominant nearest-neighbor 1D coupling with $\Delta \vR= \pm C_{3z}^{\eta}\bm{a}_{M,2}$ for each valley $\eta$. The Heisenberg coupling term in general favors AFM correlations, while the $\delta V$ term favors valley polarization.
The density operator 
$n_{\vR,\eta} = \sum_{s} \cre{d}{\vR, \eta,s} \des{d}{\vR,\eta,s}$ is a good quantum number for all $\RR,\eta$ of the spin model allowing us to exactly solve the model in two limiting cases (see \cref{sec:spin_strong_coupling}). At relatively small $12/\epsilon$, $J$-term dominates over $\delta V_{\eta\eta'}(\Delta\vR)$, allowing us to approximate $\delta V_{\eta\eta'}(\Delta\vR) \approx 0$. In contrast, at relatively large $12/\epsilon$, the $\delta V_{\eta\eta'}(\Delta\vR)|$ becomes dominant over $J$, and we can treat $J$ as perturbation.

We first analyze the exact solution in the limit where $\delta V_{\eta\eta'}(\Delta\vR) = 0$.
For a given density distribution $\{n_{\vR,\eta}\}$, the system can be effectively described as a collection of open-boundary spin chains.  
Taking $\nu = 1$ as an example, a random configuration of $\{n_{\vR,\eta}\}$ is illustrated in \cref{fig:main:spin_config}~(a), where each dot represents a moir\'e unit cell, and the color indicates the valley index of the electron in the corresponding cell. 
Since the spin-spin coupling is intra-valley and 1D-like, we connect moir\'e unit cells that interact with each other by a straight line. 
As a result, the system can be viewed as a collection of open-boundary spin chains of varying lengths, each consisting of a sequence of sites occupied by electrons from the same valley and aligned along a specific direction.
Since the energy of open-boundary nearest-neighbor Heisenberg spin chains can be computed exactly using either the Bethe Ansatz or the Density Matrix Renormalization Group (DMRG)
\cite{PhysRev.112.309,des1966anisotropic,alcaraz1987surface,SCHOLLWOCK201196}, we can determine the ground state that minimizes the energy for any given configuration $\{n_{\vR,\eta}\}$. 
For open-boundary spin chains~\cref{eq:effective_spin_ham} in the $\delta V_{\eta\eta'}(\Delta \RR)=0$ limit, we observe that the energy per electron is minimized when the chain length is 2, indicating the formation of a spin-singlet dimer (\citeSINoPara{sec:spin_strong_coupling}). 
Based on this insight we construct the exact ground states explicitly.

For $\nu=1$, each moir\'e unit cell hosts a single electron. In the ground state configurations, all electrons must pair into spin-singlet dimers with electrons from one of their nearest neighbors~\cite{PhysRevLett.86.1881}. However, to satisfy the ground state condition, two dimers of the same valley must not be nearest neighbors along the $\pm C_{3z}^{\eta} \bm{a}_{M,2}$ direction; otherwise, they would merge to form a spin chain of length 4, increasing the energy~\citeSI{sec:spin_strong_coupling}. 
One of the ground state configurations is shown in~\cref{fig:main:spin_config}~(b).
However, we find that the ground states are highly degenerate, corresponding to a finite zero-temperature entropy.
The entropy can be computed exactly for small system sizes. For larger systems, we estimate the entropy by mapping the spin model to an interacting Majorana fermion model and applying a cumulant/perturbation expansion as described in \citeSINoPara{sec:spin_strong_coupling}. 
By combining both numerical and analytical approaches, we estimate the zero-temperature entropy to be approximately $S \approx 0.307k_B$. We expect that the high ground-state degeneracy will be lifted by incorporating interaction terms that were omitted in the current zeroth order spin model.

At $\nu = 2$, the same dimer constraints must be satisfied. The only difference is that each unit cell now contains two electrons. 
The ground states are characterized by two symmetry-inequivalent valence bond solid (VBS) patterns, with $\sqrt{3} \times \sqrt{3}$ and $3 \times 3$ supercells, respectively.
In \cref{fig:main:spin_config} (c), we illustrate the patterns of the two VBS states, where each dot is marked with two colors indicating that each moir\'e unit cell contains two electrons of two different valleys. 
All other 4 ground states can be generated by applying symmetry transformations to the two given VBS states (\cref{app:sec:vbsnu=2}). 

For $\nu = 3$, each site is occupied by three electrons, one from each valley. Unlike the cases of $\nu = 1$ and $\nu = 2$, it is not possible for all electrons to form length-2 spin-singlet dimers at $\nu=3$. 
As we prove in \citeSINoPara{sec:spin_strong_coupling}, the electrons tend to form infinitely long spin chains in the ground state, as illustrated in Fig.~\ref{fig:main:spin_config}(d). 
Accordingly, the system can be effectively described as a set of one-dimensional spin chains, each exhibiting a quantum disordered ground state driven by strong quantum fluctuations.

In the opposite limit, where $\delta V_{\eta\eta'}(\Delta\vR)$ dominates over $J$.  the prevalence of $\delta V_{\eta\eta'}$ tends to stabilize valley polarization state (see \cref{sec:gnd_state_spin_nu_1}) 
such that each moir\'e unit cell is occupied by electrons of the same valley flavor.
Dimer formation is thus forbidden, and the system is described by a set of one-dimensional spin chains after including the finite $J$, each favoring a quantum disordered ground state~\citeSI{sec:spin_strong_coupling}.

\begin{figure}
    \centering
    \includegraphics[width=1.0\linewidth]{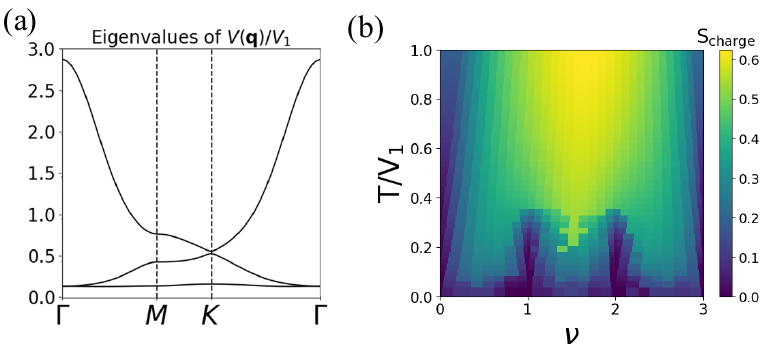}
    \caption{
    (a) Eigenvalues of the density-density interaction matrix, $V_{\eta\eta'}(\qq)$ (obtained from the realistic models), for AB-stacked SnSe$_2$ (see \cref{sec:ab_strong_coupling_charge_model}).
    (b) Entropy from charge fluctuations $S_{charge}$ at various fillings $\nu$ (per moir\'e unit cell) and temperatures $T$ for the smallest angle, $\theta = 3.89^\circ$, for AB stacking. $V_1$ denotes the strength of nearest-neighbor density-density interactions (cf. \cref{tab:parameter_val_AB}). 
    }
    \label{fig:main:Ab}
\end{figure}

{\it Flat-band limit of AB-stacked \ch{SnSe2} and the classical ``spin'' liquid.}
We now explore interaction effects in AB-stacked \ch{SnSe2}. 
Due to the offset between the Wannier centers of the three valleys, nontrivial behaviors associated with the charge degrees of freedom emerge even in the strong-coupling limit, where interactions dominate over the kinetic terms. We therefore focus on the effect of the interaction term and set the kinetic term to zero. 
In this regime, the system is effectively described by a classical charge model, with the corresponding Hamiltonian given in \cref{eq:main:HV}, along with an additional chemical potential term $-\mu \sum_{\vR,\eta} n_\eta(\vR)$ that controls the filling of the system. 
The density-density interaction model can be mapped to a kagome Ising model. 
We treat $n_{\eta}(\vR)$ as an Ising variable, with fillings $0$ and $1$ mapped to Ising values $-1$ and $+1$, respectively.
Double-occupied states are neglected due to strong on-site Hubbard repulsion. 
Then, $V_{\eta\eta'}(\Delta\vR)$ represents the effective Ising coupling, while the chemical potential acts as an effective magnetic field \citeSI{sec:ab_strong_coupling}. 
We also perform a Fourier transformation of the interaction matrix $V_{\eta\eta'}(\Delta\vR)$ and plot the corresponding eigenvalues at each momentum in \cref{fig:main:Ab}~(a). 
We observe a flat band arising from the kagome geometry. 
The flatness also suggests that at $\nu = 3/2$, corresponding to the zero-magnetic-field limit of the effective Ising model, the system hosts quasi-degenerate ground states, reflecting strong charge fluctuations \citeSI{sec:ab_strong_coupling}. 
At zero magnetic field, the kagome Ising model with nearest-neighbor coupling is known to exhibit a classical spin liquid phase~\cite{kano1953antiferromagnetism,PhysRev.79.357}, where spin in this context refers to the Ising variable associated with the density operator. Such a classical spin liquid phase is characterized by a non-zero entropy appearing at zero temperature, indicating strong fluctuations of the classical Ising variables.

We perform cluster mean-field simulations of the strong-coupling model without the kinetic term across different fillings and temperature regimes \citeSI{sec:ab_strong_coupling}. Since the strong coupling limit is more likely to be realized at a small angle, we focus on the smallest angle, i.e., $\theta=3.89^\circ$. 
In \cref{fig:main:Ab}~(b), we illustrate the entropy arising from charge fluctuations. We observe an enhancement of entropy over an extended temperature range near $\nu = 3/2$ out of the 6 flavors (valley and spin) of electrons per unit cell, which corresponds to the zero magnetic field limit of the kagome Ising model. 
At zero temperature, however, the entropy is quenched even at $\nu = 3/2$ due to realistic long-range interactions beyond nearest neighbors (\cref{sec:app:wannier_ab_interaction}), which stabilize long-range order. 
At integer fillings $\nu = 1, 2, 3$, the entropy is quenched at relatively higher temperatures due to the formation of a valley-polarized state. 
Since the Wannier centers of the three valleys are spatially offset from one another, electrons tend to occupy the same valley to minimize inter-valley density repulsion.
\citeSI{sec:gnd_classical_charge}.
Our results suggest classical spin liquid behavior---arising from the geometric frustration of the kagome lattice---to emerge within the strong coupling phase diagram of AB-stacked twisted \ch{SnSe2}.


{\it Summary and discussion.~~}
In this work, we construct an interacting Wannier model for both twisted AA-stacked and AB-stacked \ch{SnSe2}. In the AA-stacked configuration, the system is described by an interacting three-orbital triangular lattice model, where each orbital corresponds to a valley and develops an approximate 1D hopping structure. We explore the strong-coupling interacting phase diagram at zeroth order (meaning with only the most important terms in the Hamiltonian) and identify various nontrivial quantum phases, including a dimerized phase with non-zero entropy, valence bond solids, and quantum paramagnetism. 
Further results will be presented in Refs.~\cite{tba2025a,tba2025b}. 
In the AB-stacked configuration, the system forms a kagome lattice model, where three sublattices originate from three valleys. In the strong coupling regime, we demonstrate the existence of a ``classical spin liquid''{} phase. 

The high tunability of the moir\'e system allows us to control both the filling and the interaction strength (by tuning the twist angles), enabling a systematic exploration of the phase diagram of correlated systems. Our study provides the first theoretical framework for twisted \ch{SnSe2} and predicts a range of exotic quantum phases that could be realized in experiments.

\vspace{1em}  

\begin{acknowledgments}
B.A.B. and H.H. were supported by the Gordon and Betty Moore Foundation through Grant No. GBMF8685 towards the Princeton theory program, the Gordon and Betty Moore Foundation’s EPiQS Initiative (Grant No. GBMF11070), the Office of Naval Research (ONR Grant No. N00014-20-1-2303), the Global Collaborative Network Grant at Princeton University, the Simons Investigator Grant No. 404513, the NSF-MERSEC (Grant No. MERSEC DMR 2011750), the Simons Collaboration on New Frontiers in Superconductivity (Grant No. SFI-MPS-NFS-00006741-01), the Princeton collaborative network, and the Schmidt Foundation at the Princeton University.
M.L. and H.Y. were supported in part by the NSFC under Grant Nos. 12347107 and 12334003
(WS, CT, and HY), and by MOSTC under Grant No.
2021YFA1400100. H.Y. acknowledges the support in part by the New Cornerstone Science Foundation through the Xplorer Prize. 
Y.J. was supported by the European Research Council (ERC) under the European Union’s Horizon 2020 research and innovation program (Grant Agreement No. 101020833), as well as by the IKUR Strategy under the collaboration agreement between Ikerbasque Foundation and DIPC on behalf of the Department of Education of the Basque Government.
H.P. was supported by the Ministry for Digital Transformation and of Civil Service of the Spanish Government through the QUANTUM ENIA project call - Quantum Spain project, and by the European Union through the Recovery, Transformation and Resilience Plan - NextGenerationEU within the framework of the Digital Spain 2026 Agenda. 
AF and DMK acknowledge funding by the DFG  within the Priority Program SPP 2244 ``2DMP'' -- 443274199. 
D.C. and S.A.P.  acknowledge support from the UKRI Horizon Europe Guarantee Grant No. EP/Z002419/1 (for an ERC Consolidator Grant to S.A.P.). D.C. also gratefully acknowledges the support provided by the Leverhulme Trust. 
M.G.V received financial support from the Canada Excellence Research Chairs Program for Topological Quantum Matter, ID2022-142008NB-I00 projects funded by
MICIU/AEI/10.13039/501100011033 and FEDER, UE and Diputacion Foral de Gipuzkoa Programa Mujeres y Ciencia.
\end{acknowledgments}
\bibliography{MWannierRef.bib,tbg.bib,Twistronics.bib}

\renewcommand{\addcontentsline}[3]{}
\bibliographystyle{apsrev4-2}
\let\addcontentsline\oldaddcontentsline

\renewcommand{\thetable}{S\arabic{table}}
\renewcommand{\thefigure}{S\arabic{figure}}
\renewcommand{\theequation}{S\arabic{section}.\arabic{equation}}
\onecolumngrid
\pagebreak
\thispagestyle{empty}
\newpage
\begin{center}
	\textbf{\large Supplementary Information for ''\titlePaper{}``}\\[.2cm]
\end{center}

\appendix
\renewcommand{\thesection}{\Roman{section}}
\tableofcontents
\let\oldaddcontentsline\addcontentsline
\newpage

\section{Continuous model}

\begin{figure}
    \centering
    \includegraphics[width=1.0\linewidth]{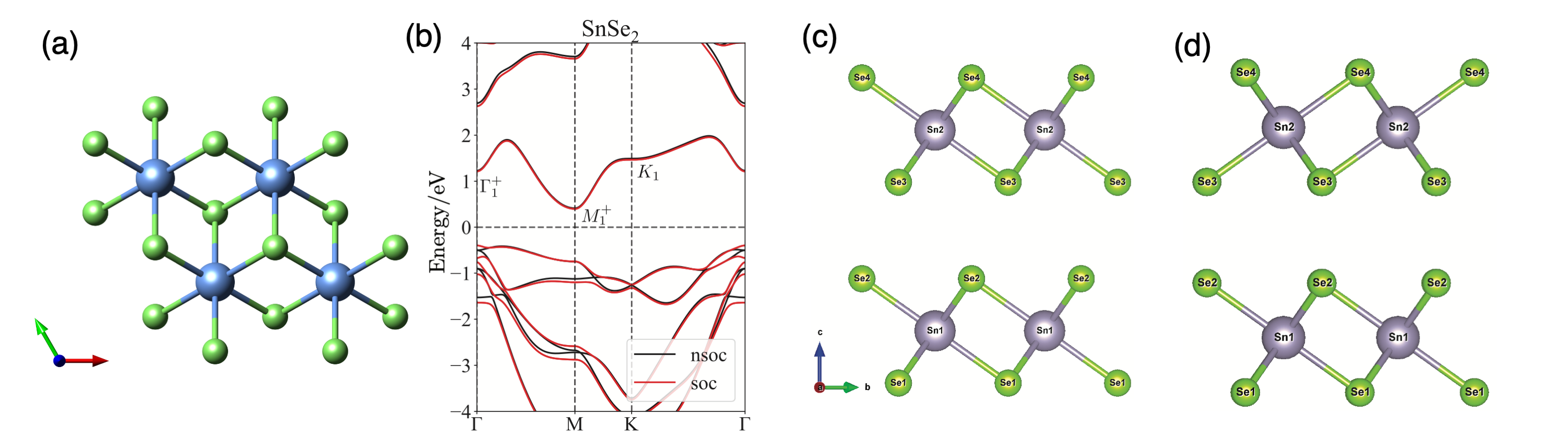}
    \caption{Lattice structure (a) and band structures (b) of monolayer SnSe2. AA (c) and AB (d) stacking patterns. }
    \label{fig:lattice_structure}
\end{figure}
\begin{figure}
    \centering
    \includegraphics[width=1.0\linewidth]{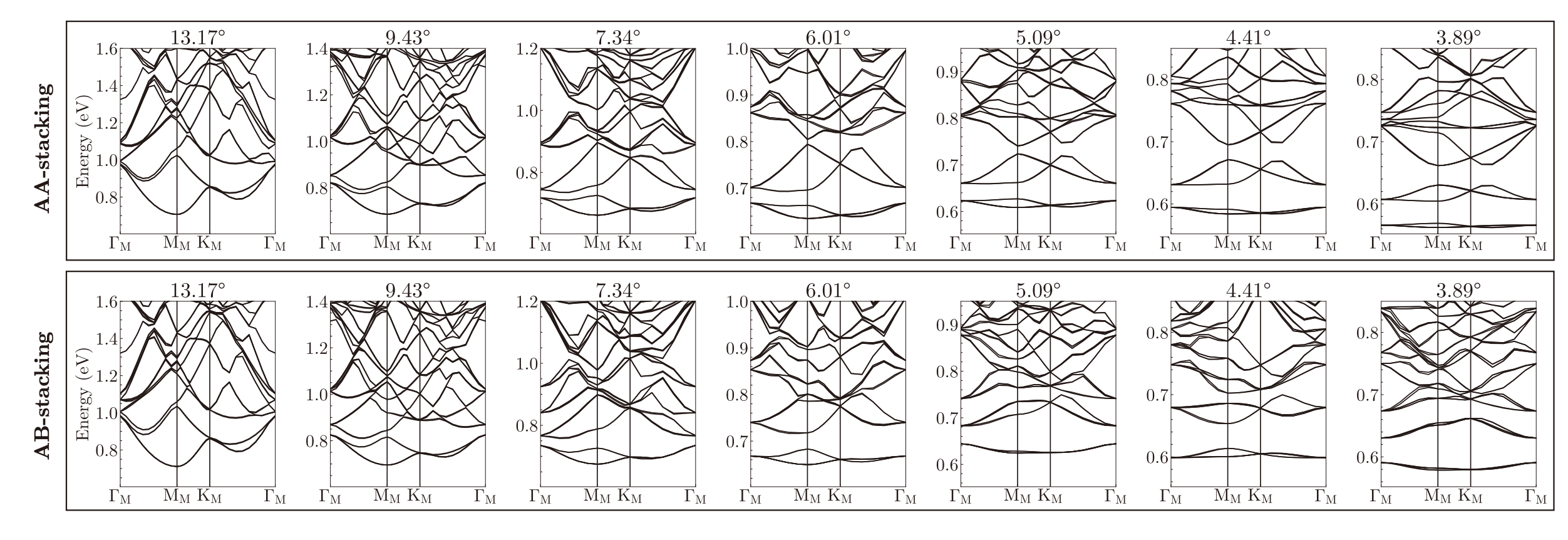}
    \caption{The {\it ab initio}s band structure of twisted AA-stacking and AB-stacking \ch{SnSe2} at different angles. Figures are adapted from Ref.~\cite{calugaru2024mtwist}.}
    \label{fig:twist_band_dft}
\end{figure}

We consider the moir\'e system formed by twisting two layers of SnSe$_2$.
The crystal structure and band structure of monolayer \ch{SnSe2} are presented  in~\cref{fig:lattice_structure} (a) (b). The monolayer \ch{SnSe2} is characterized by  the trigonal space group $P\bar{3}m1$ (No.164) in its non-magnetic form, the corresponding Shubnikov space group is $P\bar{3}m11'$ [Shubnikov Space Group (SSG) 164.86]. The lattice vectors of the monolayer \ch{SnSe2} are
\begin{equation}
	\vec{a}_1=a \left(1, 0 \right), \quad 
	\vec{a}_2=a \left( -\frac{1}{2}, \frac{\sqrt{3}}{2} \right),
\label{eq:hexagonal_cell_basis}
\end{equation}
with lattice constant $a=\SI{3.811}{\angstrom}$. The atoms are located at
\begin{equation}
	\label{app:eqn:atom_positions}
	\mathrm{Sn}:\ \left(0,0,0\right),\quad \mathrm{Se}:\ \left(\frac{1}{3}, \frac{2}{3}, z\right),\,\left(\frac{2}{3}, \frac{1}{3}, -z\right).  
\end{equation}
The bottom conduction band of the monolayer band structure has band minimuma at the three $M$ points, which form three valleys of the moir\'e systems. 
After twisting, the systems realizes a relatively flat isolated conduction band as shown in~\cref{fig:twist_band_dft}. The moir\'e lattice vectors and reciprocal lattice vectors are 
\begin{align}
    &\bm{a}_{M,1} =a_M(\frac{\sqrt{3}}{2},-\frac{1}{2}) ,\quad \bm{a}_{M,2} =a_M(0,1) \nonumber\\ 
    &\bm{b}_{M,1} = \frac{4\pi}{\sqrt{3}a_M}(1,0)
    ,\quad \bm{b}_{M,2}=\frac{4\pi}{\sqrt{3}a_M}(\frac{1}{2},\frac{\sqrt{3}}{2})
\end{align}
where $a_M = a/(2\sin(\theta/2))$ with $\theta$ the twist angle. 
The moir\'e system can be described by the monolayer electron operator near the three $M$ valleys
\begin{align}
    \cre{a}{C_{3z}^\eta \bm{K}_{M}^l +\vp, s,l}
\end{align}
where $l= \pm 1 $ denotes top and bottom layers respectively, $s$ is the spin index and $\bm{K}_M^l$ denotes the momentum of $M$ valley of layer $l$. For future convenience, we also introduce the following new electron operators 
\begin{align}
    \cre{c}{\vk,\vQ,\eta,s} = \cre{a}{C_{3z}^\eta \bm{K}_M^l+\vk -\vQ,s,l},\quad \vk \in \text{MBZ},\quad  \vQ \in \mathcal{Q}_{(\eta+l)\text{mod}3}
\end{align}
where MBZ denotes the moir\'e Brillouin zone, and we have defined the $\mathcal{Q}$ lattice as
\begin{align}
    &\mathcal{Q}_n = \mathbb{Z} \bm{b}_{M,1} + \mathbb{Z} \bm{b}_{M,2}  +\vq_n\nonumber\\ 
    &\vq_0 = \frac{\bm{b}_{M,1}}{2},\quad 
    \vq_1 =C_{3z}\vq_0,\quad \vq_2 = C_{3z}\vq_1 
\end{align}

\begin{figure}
    \centering
    \includegraphics[width=0.9\linewidth]{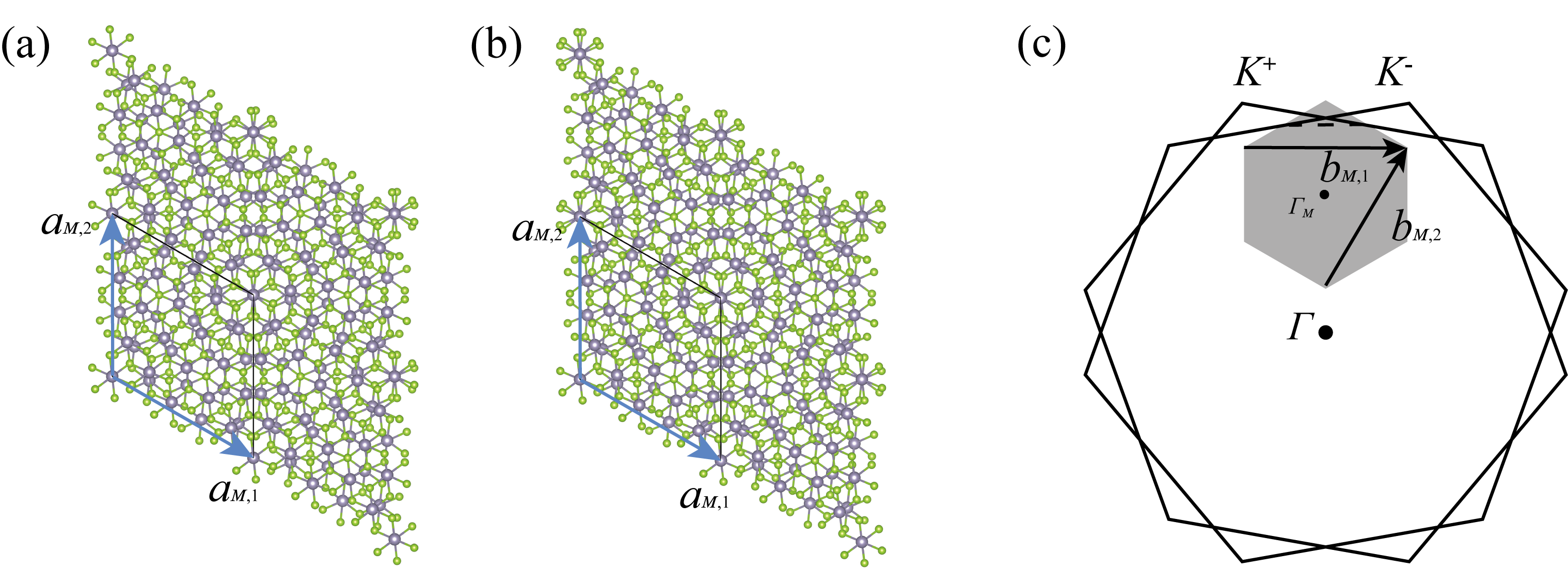}
    \caption{The lattice structure of the twisted AA (a) and AB (b)-stacking bilayer SnSe${}_2$. The gray atoms represent the Sn atoms while the green atoms represent the Se atoms. (c) The monolayer BZ and MBZ. }
    \label{fig:moire_lattice_structure_mbz}
\end{figure}

The continuous model of the moir\'e system is then described by the following Hamiltonian
\begin{equation}
	\mathcal{H} = \sum_{\vk \in \mathrm{MBZ}} \sum_{s_1,s_2,\eta} \sum_{\vQ,\vQ'\in \mathcal{Q}_{(\eta \pm 1)\text{mod}3}} h^\eta_{\vQ,\vQ'} \left( \vk \right)  \cre{c}{\vk,\vQ,\eta,s} \des{c}{\vk,\vQ',\eta, s},
\end{equation} 
The isolated bottom conduction band has approximate spin $SU(2)$ symmetry since it is mainly contributed by the $s$ orbital of the Sn and the $p$ orbital of Se which both have negligible spin-orbit coupling. Thus, our moir\'e Hamiltonian has spin $SU(2)$ symmetry. The full Hamiltonian and the corresponding value have been derived in Ref.~\cite{calugaru2024mtwist}.

For the moir\'e systems, there are two stacking patterns, which we denote by AA-stacking (by small $\theta$)  and AB-stacking (at $\pi+ \theta$, with small $\theta$) as shown in~\cref{fig:lattice_structure} (c) (d). 
\begin{enumerate}
    \item For AA-stacking, the system has $C_{3z}$ and $C_{2x}$ symmetries. The underlying non-magnetic space group is $\mathrm{P}312 \,(\text{No.\,149})$ whose grey Shubnikov extension is $\mathrm{P}3121' \,(\text{SSG }149.22)$.
    \item For AB-stacking, the system features $C_{3z}$ and $C_{2y}$ symmetries.  The underlying non-magnetic  space group is $\mathrm{P}321 \,(\text{No.\,150})$ whose grey Shubnikov extension is $\mathrm{P}3211' \,(\text{SSG }150.26)$.
\end{enumerate}
In addition, for the AA-stacking, at the small angle, the system also features an approximate zero-twist symmetry $\tilde{M}_z$ as obtained in \cite{calugaru2024mtwist}
\begin{align}
    \tilde{M}_z \cre{c}{\vk,\vQ,\eta,s} \tilde{M}_z^{-1} = \cre{c}{\vk + \vq_{\eta}, \vQ + \vq_{\eta}, \eta,s},
\end{align}
where the Hamiltonian obeys
\begin{align}
    h_{\vQ,\vQ'}(\vk)&=h_{\vQ+\vq_{\eta},\vQ'+\vq_{\eta}}(\vk+\vq_\eta).
\end{align}
We show the twisted band structures obtained from DFT calculations in \cref{fig:twist_band_dft}.

\section{Wannier orbitals}\label{sec:wann_orb}
We focus on the interacting physics of the lowest six conduction bands ($6 = 3 \times 2$, with three corresponding to the valleys and two to the spin indices). Since the lowest band for each valley and spin possesses trivial topology~\cite{calugaru2024mtwist}, Wannier orbitals can be constructed by smoothening the gauge~\cite{PhysRevB.85.115415}.

\subsection{Smooth gauge}
We adapt the method introduced in~\cite{PhysRevB.85.115415} to smooth the gauge and construct Wannier orbitals. In this section, we provide a brief review of the procedure.
We begin by discussing gauge smoothening for a one-dimensional (1D) system, achieved by constructing a parallel-transport gauge. Starting with the Bloch function:
\begin{align}
   \hat{d}_{\vk}^\dag =  \sum_a u_a(\vk) \hat{c}_{\vk,a}^\dag ,
\end{align}
where $\hat{d}_{\vk}^\dag$ is the electron operator in the band basis with wavefunction $u_a(\vk)$, and $\hat{c}_{\vk,a}^\dag$ represents the electron operator with momentum $\vk$ and orbital $a$. The embedding matrix \cite{wang2016hourglass} is defined via 
\begin{align}
\hat{c}_{\vk+\bm{G},a} = \sum_b V^{\bm{G}}_{ab} \hat{c}_{\vk,b}.
\end{align} 
Our goal is to perform a $U(1)$ transformation on the wavefunction $u_a(\vk)$ such that $u_a(\vk)$ becomes a continuous function throughout the entire Brillouin zone. As the bands are topologically trivial, this is possible.
We first smooth the gauge for $\vk \in \{k_j, {k_j=j\frac{2\pi}{N_1}} \text{ for } {j=0,...,N_1}\}$ by requiring the overlap $\langle u(k_{j})|u(k_{j+1})\rangle$ to be real and positive, where $|u(k_j)\rangle=(u_{1}(k_j), u_2(k_j),\dots u_{M}(j_j))^T$ and $M$ is the number of orbitals. This is achieved by applying a $U(1)$ phase rotation, $|u'(k_{j+1})\rangle = e^{-i\beta_{j+1}}|u(k_{j+1})\rangle$, where $\beta_{j+1} = \text{Arg}(\langle u'(k_j)|u(k_{j+1})\rangle)$, for each $|u'(k_{j+1})\rangle$ sequentially. Such procedure corresponds to unwinding the Berry phase. It starts with the initial condition $|u'(0)\rangle = |u(0)\rangle$. By this method, all discontinuities are transformed to the boundary of the Brillouin zone
\begin{align}
    \langle u'(2\pi) | V^{2\pi}|u'(0)\rangle =\sum_{a,b} u'^*_a(2\pi)V_{ab}^{2\pi} u'_{b}(0) = e^{i\phi_c  }
\end{align}
where $\phi_c$ will be the Wannier center. 
We can then remove this discontinuity by letting 
\begin{align}
    |\tilde{u}(k_j)\rangle  = |u'(k_j)\rangle e^{i \phi_c k_j/(2\pi)},
\end{align} 
which leads to a continuous Bloch wavefunction $\tilde{u}(\vk)$. We can show this by noticing that the overlapping matrix satisfies
\begin{align}
    M(k_j) = \langle \tilde{u}(k_{j+1}) | \tilde{u}(k_j)\rangle = e^{-i\phi_c/N_1} \approx 1 + O(\frac{1}{N_1})
\end{align}
\begin{align}
    M(k=2\pi) = \langle \tilde{u}(0)| V^{2\pi}|\tilde{u}(2\pi)\rangle = e^{-i\phi_c} e^{i\phi_c} = 1.
\end{align}
However, we note that after smoothing the gauge, the Wilson loop, which can be evaluated as
\begin{align}
W = \sum_j (-1)\, \arg\left[ \langle \tilde{u}(k_{j+1}) | \tilde{u}(k_j) \rangle \right] = \sum_j \frac{\phi_c}{N_1} = \phi_c,
\end{align}
can still be nonzero as long as $\phi_c \ne 0$. Each discrete $k$-point contributes $\phi_c / N_1$ to the Wilson loop, and summing over all contributions yields the total Wilson loop, which corresponds to the Wannier center $\phi_c$. 


\begin{figure}
    \centering
    \includegraphics[width=0.4\linewidth]{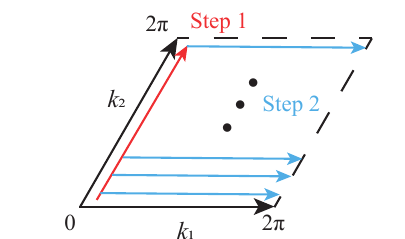}
    \caption{Illustration for the gauge smoothing method in 2D. We first smooth the gauge along the line $\vk = (0,k_2)$, then perform the same process for each $k_2$ to completely fix the gauge for the 2D BZ.}
    \label{fig:gauge_smoothing}
\end{figure}

We next consider the 2D system with $\vk \in \{(k_{1,i},k_{2,j}), {k_{1,i}=i\frac{2\pi}{N_1}, \text{for }i=0,\dots N_1,\quad k_j=j\frac{2\pi}{N_2}} \text{ for } {j=0,...,N_2}\}$. A simple illustration is also shown in \cref{fig:gauge_smoothing}. The gauge is first smoothened along the line $\vk = (0,k_2)$, following the same procedure as described for the 1D case. Specifically, we require the overlap $\langle u(k_1=0, k_{2,j}) | u(k_1=0, k_{2,j+1}) \rangle$ to be real and positive. As before, this can be achieved by applying a $U(1)$ phase rotation. 
\begin{align}
    |u'(k_{1}=0,k_{2,j+1})\rangle=e^{-i\beta_{j+1}}|u(k_{1}=0,k_{2,j+1})\rangle,
\end{align}
 where $\beta_{j+1}=\text{Arg}\left(\langle {u}(k_{1}=0,k_{2,j})|{u}(k_{1}=0,k_{2,j+1})\rangle\right)$, and we set the initial point as $|u'(k_1=0,k_2=0)\rangle=|u(k_1=0,k_2=0)\rangle$. A $U(1)$ transformation is then applied to remove the discontinuity 
\begin{align}
    |\tilde{u}(k_1=0,k_2)\rangle  = |u'(k_1=0,k_2)\rangle e^{i \phi_{c0} k_2/(2\pi)}
\end{align}
where $\langle u'(k_1=0,k_2=2\pi) | V^{\textbf{b}_2}|u'(k_1=0,k_2=0)\rangle = e^{i\phi_{c0}}$. We then let $|\tilde{u}(k_1,k_2)\rangle = |u(k_1,k_2)\rangle$ for $k_1 \ne 0$. 

Then within the gauge on $k_1=0$, we perform the same process along the $\mathbf{b}_1$ direction for each fixed $k_2$ separately to completely fix the gauge. 
 Specifically, we apply the $U(1)$ phase rotation
\begin{align}
    |\tilde{u}'(k_{1,j},k_{2})\rangle=e^{-i\beta_{k_2,j+1}}|\tilde{u}(k_{1,j+1},k_{2})\rangle,
\end{align}
where $\beta_{k_2,j+1}=\text{Arg}\left(\langle \tilde{u}(k_{1,j},k_{2})|\tilde{u}_{}(k_{1,j+1},k_{2})\rangle\right)$. The initial point for $k_1$ is set to be $|\tilde{u}'_{}(k_1=0,k_2)\rangle=|\tilde{u}(k_1=0,k_2)\rangle$. We restore the periodicity with 
\begin{align}
    |\tilde{\tilde{u}}(k_1,k_2)\rangle  = |\tilde{u}'(k_1,k_2)\rangle e^{i \phi_{c,k_2} k_1/(2\pi)},\label{eq:gauge_fixing}
\end{align}
where $\langle \tilde{u}'(k_1=2\pi,k_2) | V^{\textbf{b}_1}|\tilde{u}'(k_1=0,k_2)\rangle = e^{i\phi_{c,k_2}}$. 
Due to the trivial topology of the band, the above procedure leads to a Bloch function  $|\tilde{\tilde{u}}(k_1,k_2)\rangle$ with periodicity and smooth gauge.

\subsection{Constructing Wannier orbitals}\label{app:sec:construct_wannier}

Within the smoothing-gauge method, we are now able to obtain the Wannier function for the bottom band in the valley $\eta = 0$ and spin $\uparrow$
\begin{align}\label{app:eq:wannier}
    \cre{d}{ \vk,\eta=0,s=\uparrow} = \sum_{\vQ} u^{\eta=0}_{\vQ}(\vk)\cre{c}{\vk,\vQ,\eta=0,s=\uparrow}.
\end{align}
where $u^{\eta=0}_{\vQ}(\vk)$ denotes the wave function we obtained by smoothing the gauge, notice that $u^{\eta=0}_{\vQ}(\vk)$ is $\tilde{\tilde{u}}$ in \cref{eq:gauge_fixing} and we remove the tilde for convenience. 
Since the system has $SU(2)$ spin symmetry, it is sufficient to only consider spin $\uparrow$ while the Wannier orbital of spin $\downarrow$ shares the same wavefunction with spin $\uparrow$ 
\begin{align}
    \cre{d}{ \vk,\eta=0,s=\downarrow} = \sum_{\vQ} u^{\eta=0}_{\vQ}(\vk)\cre{c}{\vk,\vQ,\eta=0,s=\downarrow}.
\end{align}
    
In addition, three valleys are connected by the following spinless $C_{3z}$ rotation
\begin{align}
C_{3z}\hat{c}^\dagger_{\vk,\vQ,\eta,s}C_{3z}^{-1}=
\hat{c}^\dagger_{C_{3z}\vk,C_{3z}\vQ,(\eta+1)\text{mod}3,s}.
\end{align}
So we obtain the wavefunction of the other two valleys $\eta=1,2$ via $C_{3z}$ rotation. By acting with a $C_{3z}$ rotation on the Wannier orbital of valley $\eta=0$, we find that
\begin{align}
     C_{3z}^{\eta} \cre{d}{\vk, \eta=0,s} C_{3z}^{-\eta} = 
     \sum_{ \vQ}u_{\vQ}^{\eta=0 }(\vk) 
     \cre{c}{C_{3z}^\eta\vk, C_{3z}^\eta\vQ,\eta,s}.
\end{align} 
Then one can define the electron operators of the other two valleys as
\begin{align}
    \cre{d}{\vk,\eta,s}  \equiv
    \sum_{\vQ}u_{C_{3z}^{-\eta}\vQ}^{\eta=0 }(C_{3z}^{-\eta}\vk) 
     \cre{c}{\vk, \vQ,\eta,s} e^{-i\vk\cdot \vR_\eta}
      = 
    \sum_{\vQ}u_{\vQ}^{\eta }(\vk) 
     \cre{c}{\vk, \vQ,\eta,s} 
\end{align}
with
\begin{align}
    u_{\vQ}^{\eta}(\vk) =
    u_{C_{3z}^{-\eta}\vQ}^{\eta=0 }(C_{3z}^{-\eta}\vk)e^{-i\vk\cdot \vR_\eta},
    \label{eq:wannier_valley_1_2}
\end{align} 
We have introduced an additional displacements $\vR_\eta \in \mathbb{Z}\bm{b}_{M,1} + \mathbb{Z}\bm{b}_{M,2}$, 
which are introduced for convenience and will be discussed later in this section, near \cref{eq:shift_AA}. 

Finally, we define the real-space Wannier orbital of valley $\eta$ as
\begin{align}
  \cre{d}{\vR,\eta,s} = \frac{1}{\sqrt{N}}\sum_{\vk\in \text{MBZ}} \cre{d}{\vk, \eta,s}e^{-i\vk\cdot \vR }
\end{align}
where $N$ is the number of moir\'e unit cells of the system. 

\subsection{Wannier orbital for AA-stacked SnSe$_2$}
\begin{figure}
    \centering
\includegraphics[width=0.8\linewidth]{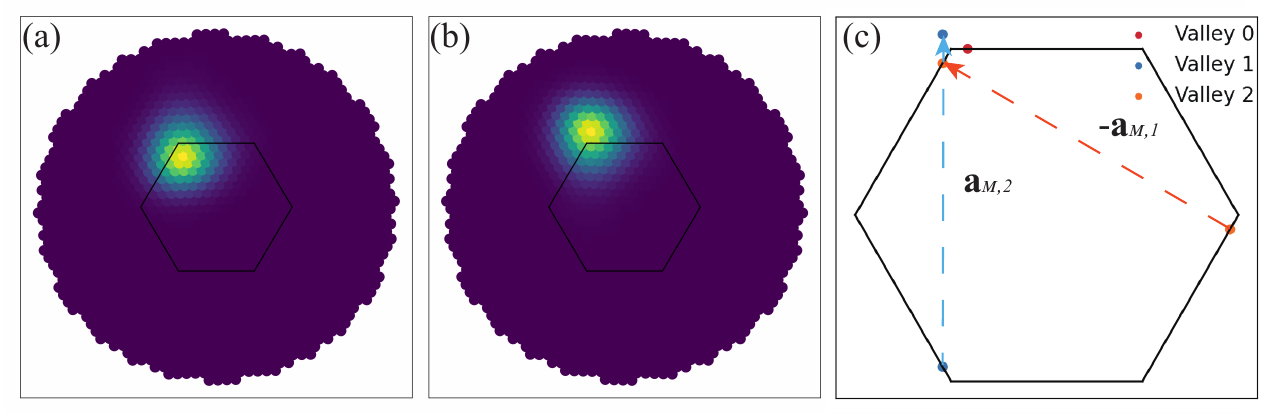}
    \caption{The amplitude distribution of the Wannier wave-function ($|\psi(\bm{r})|^2$ with $\psi(\bm{r})$ the real-space Wannier function) of valley 0 for AA-stacked SnSe${}_2$ with twisting angle $3.89^{\circ}$ at top layer (a) and bottom layer (b). (c) The Wannier centers of three valleys (three blue dots) where the black lines denote the unit-cell boundary. We apply displacements \cref{eq:shift_AA} (labeled by dashed lines) to make the three Wannier centers align closest to each other.}.
    \label{fig:wannier_AA}
\end{figure}

For AA-stacked SnSe$_2$, the displacements introduced in~\cref{eq:wannier_valley_1_2} are chosen as
\begin{align}
\label{eq:shift_AA}
 \vR_0=\bm{0},\quad   \vR_1 = \bm{a}_{M,2},\quad \vR_2 = -\bm{a}_{M,1},
\end{align}
ensuring that the three Wannier centers are closely aligned and nearly coincide at the same position, as shown in \cref{fig:wannier_AA}. The distributions of the Wannier functions and the Wannier centers of the three valleys are plotted in \cref{fig:wannier_AA} and listed in \cref{app:tab:wcaa}. 

We next discuss the symmetry properties of the Wannier orbitals. For a given symmetry $g \in \{C_{3z},C_{2x}\}$, we define the representation matrix $D[g]$ as 
\begin{align}
    g\cre{d}{\vR,\eta,s} g^{-1} \equiv\sum_{\eta'} \cre{d}{g(\vR+\mathbf{r}_{\eta})-\mathbf{r}_{\eta'},\eta',s}D[g]_{\eta'\eta}
\end{align}
where $\rr_\eta$ denotes the Wannier center
(\cref{app:tab:wcaa}). Due to the $C_{3z}$ symmetry, we also have $\rr_\eta = C_{3z}^{\eta} \rr_{0} +\RR_\eta$. 
We observe that 
\begin{align}
    C_{3z} 
    \cre{d}{\vR,\eta,s} C_{3z}^{-1} 
    = &\frac{1}{\sqrt{N}}\sum_{\kk\in \text{MBZ},l} 
    \sum_{\QQ\in\mathcal{Q}_{(\eta+l)\text{mod}3}}
    u_\QQ^{\eta}(\kk)\hat{c}_{C_{3z}\kk, C_{3z}\QQ, (\eta+1)\text{ mod }3, s}^\dag  e^{-i\kk\cdot\RR}\nonumber\\
    =& \frac{1}{\sqrt{N}}
    \sum_{\kk \in \text{MBZ},l} \sum_{\QQ\in \mathcal{Q}_{(\eta+l+1)\text{mod}3}}u_{C_{3z}^{-1}\QQ}^{\eta}(C_{3z}^{-1}\kk) \cre{c}{\kk,\QQ,(\eta+1)\text{ mod }3, s} e^{-i\kk\cdot C_{3z}\RR} 
\end{align} 
We use \cref{eq:wannier_valley_1_2} and observe that 
\begin{align}
    u_{C_{3z}^{-1}\QQ}^{\eta}(C_{3z}^{-1}\kk) 
    = u_{C_{3z}^{-\eta-1}\QQ}^{\eta=0}(C_{3z}^{-\eta-1}\kk) e^{-iC_{3z}^{-1}\kk\cdot\RR_\eta} 
    = u _{\QQ}^{\eta+1}(\kk) e^{i\kk\cdot(\RR_{\eta+1}-C_{3z}\RR_\eta)} =
     u _{\QQ}^{\eta+1}(\kk) e^{-i\kk\cdot(C_{3z}\rr_\eta - \rr_{(\eta+1) \text{ mod } 3})}
\end{align}
We also note that an additional prefactor $\RR_\eta$ has been introduced in \cref{eq:wannier_valley_1_2} to ensure that the Wannier centers of the three valleys of the same moir\'e unit cell are positioned close to one another (as we discussed near \cref{eq:shift_AA}). 
Then 
\begin{align}
     C_{3z} 
    \cre{d}{\vR,\eta,s} C_{3z}^{-1}  = &
    \frac{1}{\sqrt{N}}
    \sum_{\kk \in \text{MBZ},l}
     \sum_{\QQ\in\mathcal{Q}_{(\eta+1+l)\text{mod}3}}u_{\QQ}^{\eta+1}(\kk) \cre{c}{\kk,\QQ,(\eta+1)\text{ mod }3, s} e^{-i\kk\cdot [C_{3z}( \RR +\rr_\eta) - \rr_{(\eta+1)\text{ mod }3}]} \nonumber\\ 
     =& 
    \cre{d}{C_{3z}(\RR + \rr_\eta )- \rr_{(\eta+1)\text{ mod} 3}, (\eta+1)\text{ mod }3,s}
\end{align}
Then we have 
\begin{align}
    D[C_{3z}] = \begin{bmatrix}
        & & 1 \\
        1  \\ 
        & 1 
    \end{bmatrix}
\end{align}
We use the $C_{3z}$ transformation to generate the Wannier orbitals of valley $1,2$ from the Wannier orbital of valley $0$. After generating the Wannier orbitals, we also calculate their representation matrix of $C_{2x}$ symmetry, which gives 
\ba 
 D[C_{2x}] = \begin{bmatrix}
        1 \\ & & 1 \\ 
        & 1 
    \end{bmatrix}
\ea

\begin{figure}
    \centering
    \includegraphics[width=0.5\linewidth]{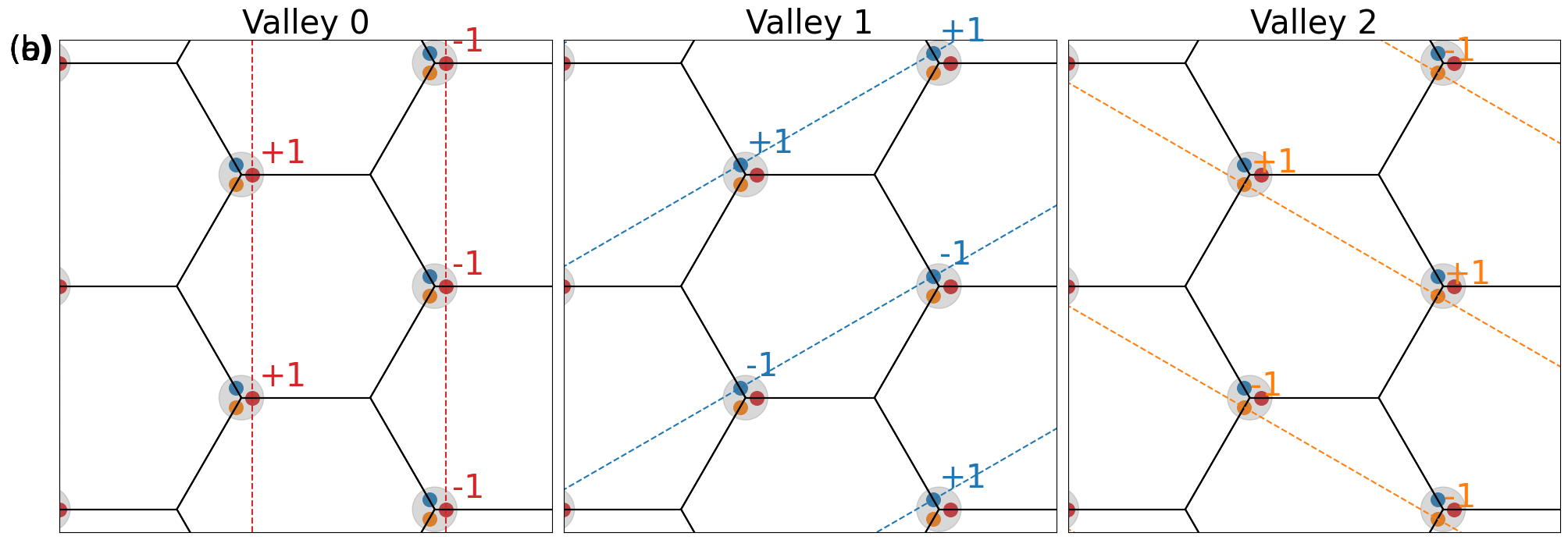}
    \caption{Eigenvalues under zero-twist symmetry (\cref{eq:zero_twist_symmetry}) of Wannier orbital of each valley and each moir\'e unit cell.}
    \label{fig:zero_twist_symmetry}
\end{figure}

In addition, as we discussed in Ref.~\cite{calugaru2024mtwist}, the AA-stacked SnSe$_2$ model also has an approximate zero-twist symmetry $\tilde{M}_z$. 
In the real-space, it behaves as (see also \cref{fig:zero_twist_symmetry})
\begin{align}
    &\tilde{M}_z \cre{d}{\vR= n\bm{a}_{M,1} +m\bm{a}_{M,2}, \eta=0, s} \tilde{M}_z ^{-1} = (-1)^{n}  \cre{d}{\vR= n\bm{a}_{M,1} +m\bm{a}_{M,2}, \eta=0, s} \nonumber\\ 
    &\tilde{M}_z \cre{d}{\vR= n\bm{a}_{M,2} +m(-\bm{a}_{M,1}-\bm{a}_{M,2}), \eta=1, s} \tilde{M}_z ^{-1} = (-1)^{n}  \cre{d}{\vR= n\bm{a}_{M,2} +m(-\bm{a}_{M,1}-\bm{a}_{M,2}), \eta=1, s} \nonumber\\ 
    &\tilde{M}_z \cre{d}{\vR= n(-\bm{a}_{M,1} -\bm{a}_{M,2}) +m\bm{a}_{M,1}, \eta=2, s} \tilde{M}_z ^{-1} = (-1)^{n}  \cre{d}{\vR= n(-\bm{a}_{M,1} -\bm{a}_{M,2}) +m\bm{a}_{M,1}, \eta=2, s} 
    \label{eq:zero_twist_symmetry}
\end{align}
with $n,m\in\mathbb{Z}$. 
For the nearest-neighbor hoppings in valley $\eta$, only the hopping along the $C_{3z}^\eta \bm{e}_y$ direction is permitted by the effective zero-twist symmetry. This constraint results in an effective one-dimensional band structure for SnSe$_2$.

\begin{table}[h!]\label{app:tab:wcAA}
\centering
\renewcommand{\arraystretch}{1.2}
\begin{tabular}{|c|c|c|c|c|}
\hline
$\theta$ & $\text{WC}_{\eta=0}$ & $\text{WC}_{\eta=1}$ & $\text{WC}_{\eta=2}$ & Spread $/a_M$ \\
\hline
3.89° & [-0.275  0.362] & [-0.362  0.362] & [-0.362  0.275] & 0.268 \\
\hline
4.41° & [-0.272  0.364] & [-0.364  0.364] & [-0.364  0.272] & 0.294 \\
\hline
5.09° & [-0.272  0.364] & [-0.364  0.364] & [-0.364  0.272] & 0.343 \\
\hline
6.01° & [-0.268  0.366] & [-0.366  0.366] & [-0.366  0.268] & 0.394 \\
\hline
7.34° & [-0.264  0.368] & [-0.368  0.368] & [-0.368  0.264] & 0.494 \\
\hline
\end{tabular}
\caption{Wannier center position and the spread of the Wannier function of the three valleys for AA-stacked SnSe$_2$ with angles $3.89^\circ, 4.41^\circ, 5.09^\circ, 6.01^\circ, 7.34^\circ$. The coordinates and the spread are listed in the unit of $(\bm{a}_{M,1}, \bm{a}_{M,2})$ and $a_M$ respectively. }\label{app:tab:wcaa}
\end{table}

\subsection{Wannier orbital for AB-stacked SnSe$_2$}
\begin{figure}
    \centering
\includegraphics[width=0.8\linewidth]{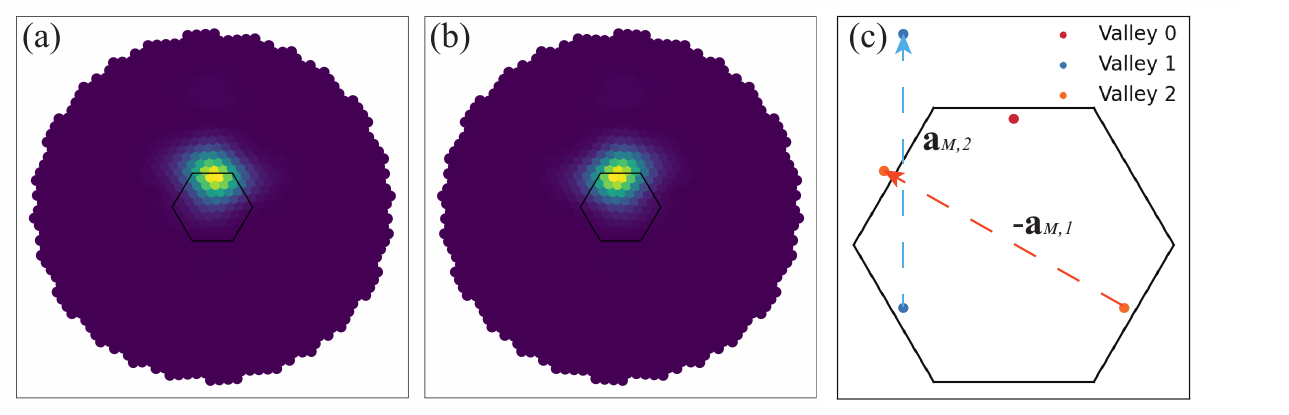}
    \caption{The configuration of the Wannier wave-function of valley 0 for AB-stacked SnSe${}_2$ with twisting angle $3.89^{\circ}$ at top layer (a) and bottom layer (b). (c) The Wannier center positions of the three valleys (three blue dots) where the black lines denote the unit-cell boundary.  We apply displacements \cref{eq:shift_AB} (labeled by dashed lines) to make the three Wannier centers align closest to each other.}.
    \label{fig:wannier_AB}
\end{figure}
For AB stacked SnSe$_2$, the displacements introduced in~\cref{eq:wannier_valley_1_2} are taken to be
\begin{align}\label{eq:shift_AB}
    \vR_0 = 0,\quad\vR_1 = \bm{a}_{M,2},\quad \vR_2 = -\bm{a}_{M,1},
\end{align}
such that three Wannier centers are closest to each other. The distributions of the Wannier function and the Wannier centers of the three valleys are plotted in \cref{fig:wannier_AB} and listed in \cref{app:tab:wcab}.

The symmetry representation matrix of the Wannier orbital is defined as 
\begin{align}
    g\cre{d}{\vR,\eta,s} g^{-1} =\sum_{\eta'} \cre{d}{g(\vR+\mathbf{r}_{\eta})-\mathbf{r}_{\eta'},\eta',s}D[g]_{\eta'\eta}
\end{align}
We have
\begin{align}
    D[C_{3z}] = \begin{bmatrix}
        & & 1 \\
        1  \\ 
        & 1 
    \end{bmatrix}
    ,\quad 
    D[C_{2y}] =
    \begin{bmatrix}
        1 \\ 
        & & 1 \\
        & 1 
    \end{bmatrix}
\end{align}
In addition, we comment that the zero-twist symmetry is no longer a good symmetry for the AB-stacked case \cite{calugaru2024mtwist}. 

\begin{table}[h!]
\centering
\renewcommand{\arraystretch}{1.2}
\begin{tabular}{|c|c|c|c|c|}
\hline
$\theta$ & $WC_{1,\eta=0}$ & $WC_{1,\eta=1}$ & $WC_{1,\eta=2}$ &Spread $/a_M$ \\
\hline
3.89° & [0.00     0.458] & [-0.458  0.542] & [-0.542  0.00   ] & 0.272\\
\hline
4.41° & [0.00     0.464] & [-0.464  0.536] & [-0.536  0.00   ] & 0.289 \\
\hline
5.09° & [0.00    0.470] & [-0.470  0.530] & [-0.530  0.00  ] & 0.321 \\
\hline
6.01° & [0.00     0.474] & [-0.474  0.526] & [-0.526  0.00   ] & 0.365 \\
\hline
7.34° & [0.00     0.483] & [-0.483  0.517] & [-0.517  0.00   ] & 0.442 \\
\hline
\end{tabular}
\caption{Wannier center position and the spread of the Wannier function of the three valleys for AB-stacked SnSe$_2$ with angles $3.89^\circ, 4.41^\circ, 5.09^\circ, 6.01^\circ, 7.34^\circ$. The coordinates and the spread are listed in the unit of $(\bm{a}_{M,1}, \bm{a}_{M,2})$ and $|a_M|$ respectively. }\label{app:tab:wcab}
\end{table}

\section{Interacting Wannier Models}
\label{sec:wann_model}
In this appendix, we discuss the construction of an interacting model from the Wannier orbitals obtained in ~\cref{sec:wann_orb}. 

\subsection{Tight-binding model} 
Since the bands under consideration are topologically trivial, we construct the Wannier orbitals by smoothing the gauge as discussed above. Consequently, the band basis and the Wannier basis differ by a $U(1)$ phase. Then the tight-binding model can be expressed as
\begin{align}
\label{eq:def_H_t}
    H_t = \sum_{\vk,\eta,s} \epsilon_{\vk,\eta} \cre{d}{\vk,\eta, s}\des{d}{\vk,\eta, s}
\end{align}
with $\epsilon_{\vk,\eta}$ the dispersion of the lowest band obtained from the continuous model. The real-space hopping can be obtained by the following Fourier transformation
\begin{align}
    t_{\Delta\vR}^\eta = \frac{1}{N}\sum_\vk \epsilon_{\vk,\eta} e^{-i\vk \cdot \Delta\vR}
\end{align}


\subsection{Interactions}
We consider the screened Coulomb repulsion with~\cite{PhysRevB.103.205414}
\begin{align}
    V(\vq) =  2\pi \xi^2 U_\xi \frac{\tanh(\xi |\vq|/2)}{\xi|\vq|},
\end{align}
where $\xi$ the screening length and $\epsilon$ the dielectric constant, and $U_\xi = e^2/(\epsilon \xi)$. We take the following setup
\begin{align}
    \xi = 10 \text{nm} \,, \quad \quad \epsilon=12 \,,
\end{align}
which leads to $U_\xi = e^2/(\epsilon \xi) = 12$meV. 
The screened Coulomb repulsion reads
\begin{align}
\label{eq:screened_coulomb_repulsion}
    H_{I} = \frac{1}{2\Omega_{tot}} \sum_{\vG,\vq}V(\vq+\vG) \delta \rho_{-\vq-\vG} \delta \rho_{\vq +\vG} 
\end{align}
where $\Omega_{tot}$ denotes the size of the sample and 
\begin{align}
    \delta \rho_{\vq+\vG} = \rho_{\vq+\vG} - \langle 0| \rho_{\vq+\vG}|0\rangle 
\end{align}
We define the vacuum state $|0\rangle$ as the charge neutrality where all Wannier orbitals are empty. To obtain the interactions for the Wannier model, we introduce the density operator and project it onto the Wannier basis
\begin{align}
    \rho(\vq, \vG) = \sum_{\vk, \vQ, \eta,s,l }\cre{c}{\vk,\vQ, s,l}\des{c}{\vk+\vq, \vQ - \vG, s,l} \approx \sum_{\vk, \vQ, \eta,s}u_{\vQ}^{\eta,*}(\vk) u_{\vQ-\vG}^{\eta}(\vk+\vq) \cre{d}{\vk,\eta,s} \des{d}{\vk+\vq,\eta,s} 
\end{align}
where $    \vq \in MBZ, \vG \in \mathbb{Z}\bm{b}_{M,1}+\mathbb{Z}\bm{b}_{M,2}$,
and the form factor
\begin{align}
     M^{\eta}_{\vq+\vG}(\vk) = \sum_{\vQ}u_{\vQ}^{\eta,*}(\vk) u_{\vQ-\vG}^{\eta}(\vk+\vq)\label{app:eq:formfactor}
\end{align}
where the summation of the $\vQ$ has been taken with respect to the $\vQ$-lattice of the corresponding valley. 


The density operator can then be written as
\begin{align}
\label{eq:density_operator_from_form_factor}
    \rho(\vq,\vG) =\sum_{\vQ,\vk,\eta,s} M^{\eta}_{\vq+\vG}(\vk) 
 \cre{d}{\vk, \eta,s} \des{d}{\vk+\vq,\eta,s} 
\end{align} 
We notice that, 
\begin{align}
\rho(-\vq,-\vG) = \rho(\vq,\vG)^\dag = \sum_{\vQ,\vk,ij,\eta,s}M^{\eta, *}_{\vq+\vG}(\vk) \cre{d}{\vk+\vq,\eta,s} \des{d}{\vk,\eta,s} 
\end{align}
which indicates the hermiticity relation of $M^\eta$
\begin{align}
    M_{-\vq-\vG}^{\eta}(\vk) = M^{\eta, *}_{\vq+\vG}(\vk-\vq) 
\end{align} 

Then we are able to project the interaction onto the $\cre{d}{}$ electron basis. Combining~\cref{eq:density_operator_from_form_factor,eq:screened_coulomb_repulsion}, we obtain
\begin{align}
    H_I \approx & \frac{1}{2\Omega_{tot}} \sum_{\vG,\vq,\eta ,\eta',s,s',\vk,\vk'} V(\vq+\vG) M^{\eta}_{\vq+\vG}(\vk) M^{\eta'}_{-\vq-\vG}(\vk') \cre{d}{\vk,\eta,s}\des{d}{\vk+\vq,\eta,s} 
     \cre{d}{\vk',\eta',s'}\des{d}{\vk'-\vq,\eta',s'} \nonumber\\ 
= & 
     \frac{1}{2N} \sum_{\vq,\eta ,\eta',s,s',\vk,\vk'} V_{\eta\eta'}(\vq,\vk,\vk') \cre{d}{\vk,\eta,s}\des{d}{\vk+\vq,\eta,s} 
     \cre{d}{\vk',\eta',s'}\des{d}{\vk‘-\vq,\eta',s'} 
\end{align}
where we have introduced the momentum-space interaction in the $d$ electron basis 
\begin{align}
    V_{\eta\eta'}(\vq,\vk,\vk') = \frac{1}{\Omega_{0}} \sum_{\vG} 
V(\vq+\vG)  M^{\eta}_{\vq+\vG}(\vk) M^{\eta'}_{-\vq-\vG}(\vk')
= \frac{1}{\Omega_{0}}\sum_{\vG}V(\vq+\vG) M_{\vq+\vG}^{\eta}(\vk) 
M_{\vq+\vG}^{\eta',*}(\vk'-\vq) 
\end{align}

It is more convenient to define 
\begin{align}
{V}_{\eta\eta'}(\vq,\vk_1,\vk_2) = \tilde{V}_{\eta\eta'}(\vq, \vk_1,\vk_2-\vq)  
\end{align}
where 
\begin{align}
    \tilde{V}_{\eta\eta'}(\vq, \vk_1,\vk_2) = \frac{1}{\Omega_{0}}\sum_{\vG}V(\vq+\vG) M_{\vq+\vG}^{\eta}(\vk_1) 
M_{\vq+\vG}^{\eta',*}(\vk_2)
\end{align}
Then we have 
\begin{align}
    H_I=\frac{1}{2N} \sum_{\vq,\eta ,\eta',s,s',\vk,\vk'} \tilde{V}_{\eta\eta'}(\vq,\vk,\vk') \cre{d}{\vk,\eta,s}\des{d}{\vk+\vq,\eta,s} 
     \cre{d}{\vk'+\vq,\eta',s'}\des{d}{\vk',\eta',s'} \label{eq:def_H_I_mom}
\end{align}

Finally, we perform Fourier transformation which gives the following interaction in the real space 
\begin{align}
\label{eq:def_H_I}
    H_I = \frac{1}{2}\sum_{\eta,\eta',s,s',\Delta\vR, \vR,\mathbf{r}_1,\mathbf{r}_2}
    V_{\eta\eta'}(\Delta\vR,\Delta\mathbf{r}_1,\Delta \mathbf{r}_2) \cre{d}{\vR+\Delta\mathbf{r}_1,\eta,s}\des{d}{\vR,\eta,s} 
    \cre{d}{\vR+\Delta\vR,\eta',s'}\des{d}{\vR+\Delta \vR+\Delta\mathbf{r}_2 ,\eta',s'} 
\end{align}
The real-space interaction tensor is defined as 
\begin{align}
 V_{\eta\eta'}(\Delta\vR,\Delta\mathbf{r}_1,\Delta \mathbf{r}_2) =&\frac{1}{N^3}
 \sum_{\vq,\vk,\vk'} \tilde{V}_{\eta\eta'}(\vq,\vk,\vk') e^{i\vk\Delta\mathbf{r}_1 -i\vk'\Delta\mathbf{r}_2  +i\vq\Delta \vR}\label{app:eq:real_2_mom}
\end{align}

\subsection{Wannier model of AA-stacked SnSe$_2$}\label{app:sec:wannier_aa}
In this section, we discuss the Wannier model for AA-stacked SnSe$_2$. 
\subsubsection{Tight-binding model}
\begin{figure}
    \centering
    \includegraphics[width=1.0\linewidth]{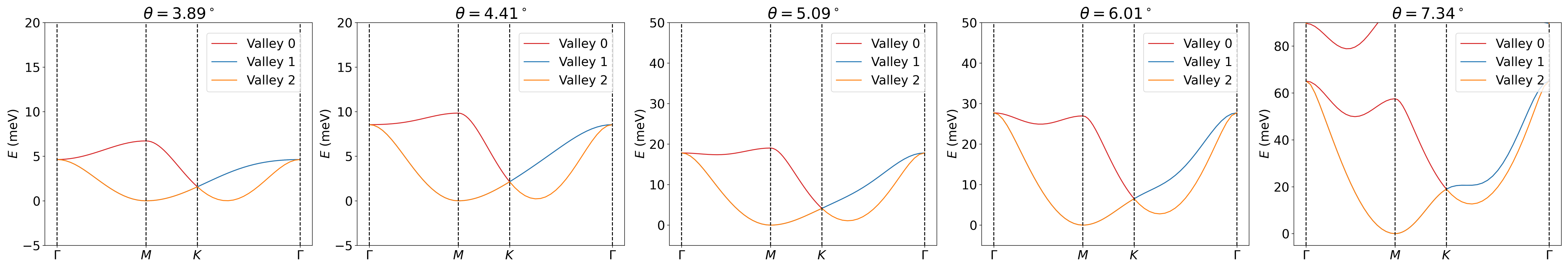}
    \caption{Lowest dispersion of the AA-stacked SnSe$_2$. Red, blue and orange denote valley $\eta=0,1,2$ respectively. }
    \label{fig:disp_AA}
\end{figure}

The tight-binding model can be expressed as
\begin{align}
   H_{t}= \sum_{\vR ,\Delta \vR} t^{\eta}_{\Delta \vR} \cre{d}{\vR,\eta,s}\des{d}{\vR+\Delta \vR,\eta,s} \label{app:eq:real_tb_aa}
\end{align}
where the dominant hopping is the nearest-neighbor hopping along the $C_{3z}^{\eta} \bm{a}_{M,2}$ direction (see the discussion near \cref{eq:zero_twist_symmetry}). 
For the $3.89^\circ$ twisted bilayer $\text{SnSe}_2$ system, the dominant hopping strength is
\begin{align}
\label{eq:nn_t}
    t^{\eta}_{ \pm  C_{3z}^\eta\bm{a}_{M,2}} =t= 1.399\text{meV}, .
\end{align}
For valley $\eta$, in addition to the dominant nearest-neighbor hopping along the $C_{3z}^\eta \bm{a}_{M,2}$ direction, there also exist weaker nearest-neighbor hoppings along the $C_{3z}^{(\eta+1),\text{mod}, 3} \bm{a}_{M,2}$ and $C_{3z}^{(\eta+2),\text{mod}, 3} \bm{a}_{M,2}$ directions. These weaker terms, which also break the zero-twist symmetry, have a strength of approximately $-0.23$ meV. Furthermore, the next-nearest-neighbor hopping has strength $\sim -0.03$meV. 
The dominant hoppings for all angles are given in Table.~\ref{tab:parameter_val_AA}.  And the dispersions of the lowest band for all angles are shown in \cref{fig:disp_AA}.

\subsubsection{Interactions} 
We first discuss the interactions for the smallest angle considered, $\theta = 3.89^\circ$. The interactions for larger angles are qualitatively similar. 
The dominant interactions take the form of density-density interactions and can be written as 
\begin{align}
    H_{V} = \sum_{\eta\eta',\vR,\Delta\vR,ss'}\frac{V_{\eta\eta'}(\Delta\vR)}{2}
n_{\vR,\eta,s}n_{\vR+\Delta\vR,\eta',s'},
\label{eq:density_density_int}
\end{align}
where $n_{\vR,\eta,s} =  \cre{d}{\vR,\eta,s}\des{d}{\vR,\eta,s}$. 
Since 
\begin{align}
  & \frac{1}{2}\sum_{\vR,ss'} V_{\eta\eta'}(\Delta\vR) n_{\vR,\eta,s}n_{\vR+\Delta\vR,\eta',s'} +   \frac{1}{2}\sum_{\vR,ss'} V_{\eta'\eta}(-\Delta\vR) n_{\vR,\eta',s}n_{\vR-\Delta\vR,\eta,s'} \nonumber\\ 
   =&  \frac{1}{2}
   \sum_{\vR,ss'}\bigg[
   V_{\eta\eta'}(\Delta\vR)
   +V_{\eta'\eta}(-\Delta\vR)
   \bigg]n_{\vR,\eta,s}n_{\vR+\Delta\vR,\eta',s'} \, , \label{app:eq:ddinteraction}
\end{align}
we could therefore take $V_{\eta\eta'}(\Delta\vR) = V_{\eta'\eta}(-\Delta\vR)$. 

The density-density interactions are approximately $U(6)$ symmetric and can be approximately expressed as
\begin{align}
\label{eq:U6_int_def}
    H_{V,U(6)} = \sum_{\eta\eta',\vR,\Delta\vR,ss'} \frac{V(\Delta\vR)}{2}
n_{\vR,\eta,s}n_{\vR+\Delta\vR,\eta',s'},\quad V(\Delta\vR) = \frac{1}{9}\sum_{\eta\eta'}V_{\eta\eta'}(\Delta\vR)
\end{align}
where we also have $V(\Delta\vR)= V(-\Delta\vR)$. 
The approximate $U(6)$ symmetry originates from the close proximity of the Wannier orbitals corresponding to the three valleys within the same unit cell (see \cref{fig:wannier_AA}), resulting in similar interaction strengths between different valleys. 

The on-site Hubbard interactions $U_{\eta\eta'} = V_{\eta\eta'}(\Delta\vR = \bm{0})$ takes values of
\begin{align}
    &U_{\eta\eta'} = \delta_{\eta,\eta'}86.4\text{meV} +(1-\delta_{\eta,\eta'})83.6\text{meV} 
\end{align}
We can observe the violation of the $U(6)$ symmetry in the on-site term is around 
\begin{align}
\frac{1}{V(\Delta\vR=0)}\sqrt{\sum_{\eta\eta'}[V_{\eta\eta'}(\Delta\vR=0)-V(\Delta\vR=0)]^2/9} \approx 2\%
\end{align}
The average value of nearest-neighbor Coulomb repulsion 
\begin{align}
    V_1 = \frac{1}{9}\sum_{\eta\eta'}V_{\eta\eta'}(\Delta\vR),\quad \Delta\vR = \text{nearest neighbor}
\end{align}
is around $\sim 10$meV. 
We list the dominant interaction strength in Table.~\ref{tab:parameter_val_AA}.

The other interactions include the spin-spin couplings
\begin{align}
\label{eq:spin_spin_coupling}
    H_{Spin} = \sum_{\eta,\vR,\Delta\vR}J_{\eta}(\Delta\vR)\bm{S}_{\vR,\eta}\cdot\bm{S}_{\vR+\Delta\vR,\eta},\quad \bm{S}^\mu_{\vR,\eta} 
    = \sum_{ss'} \cre{d}{\vR,\eta,s}\frac{\sigma^\mu_{ss'}}{2} \des{d}{\vR,\eta,s'}
\end{align}
with 
\begin{align}
    J_\eta(\Delta\vR) = -V_{\eta\eta}(\bm{0},\Delta\vR,\Delta\vR)
\end{align}
and density-dependent hopping
\begin{align}
     H_{nhop} = \sum_{\eta\eta',\vR,\Delta\vR,\Delta\mathbf{r}_1,ss'}K_{\eta\eta'}(\Delta\vR,\Delta\mathbf{r}_1)\cre{d}{\vR,\eta,s}\des{d}{\vR+\Delta\mathbf{r}_1,\eta,s}n_{\vR+\Delta\vR,\eta,s},\quad K_{\eta\eta'}(\Delta\vR,\Delta\mathbf{r}_1) = V_{\eta'\eta}(-\Delta\vR,\bm{0},\Delta\mathbf{r}_1)
\end{align}
The spin-spin couplings and density-dependent hopping are at the order of $1$ meV. All other interactions are smaller than $1$ meV. 
In practice, for the real-space Hartree-Fock mean-field calculation of the AA-stacked model, we take the following truncation
\begin{align}
    &|\Delta \mathbf{r}_1|,|\Delta \mathbf{r}_2| \le |\bm{a}_{M,1}| \nonumber\\ 
    &|\Delta \vR| \le \sqrt{3}|\bm{a}_{M,1}|,\label{app:eq:interacting_cutoff_aa}
\end{align}
where only the nearest neighbors are considered for 
$\Delta \mathbf{r}_{1,2}$ and both nearest neighbor and next-nearest neighbors are considered for $\Delta \vR$.


\begin{table}[h!]
\centering
\renewcommand{\arraystretch}{1.2}
\begin{tabular}{|c|c|c|c|c|c|}
\hline
  & $t^1_{ \pm \bm{a}_{M,2}}$[meV] & $U_{\eta\eta}$[meV] & $U_{\eta\ne\eta'}$[meV] & $U_{\eta\eta}/w_{b}$ & $V_1$[meV] \\
\hline
3.89° & 1.40 & 86.38 & 83.62 & 12.00 & 8.65 \\ 
\hline
4.41° & 2.22 & 92.34 & 89.82 & 8.54 & 11.39\\ 
\hline
5.09° & 4.22 & 95.34 & 93.14 & 4.50 & 15.36\\ 
\hline
6.01° & 5.89 & 103.03 & 100.98 & 3.38 & 20.83\\
\hline
7.34° & 11.35 & 108.39 & 106.64 & 1.64 & 29.08\\
\hline
\end{tabular}
\caption{Parameter values of the Wannier model for AA-stacked SnSe2 with $\xi=10$nm and $\epsilon=12$. Here $w_b$ represents the bandwidth of the single-particle band structure.
}
\label{tab:parameter_val_AA}
\end{table}

Finally, we discuss the uncertainty in the interaction strength. The current interactions are calculated for a screening length of $\xi = 10 \text{nm}$ and a dielectric constant of $\epsilon = 12$. In practice, the screening length can be experimentally tuned. To account for this variability, we provide the onsite Hubbard $U$ values for different screening lengths in Table~\ref{tab:parameter_val_AA} and Table~\ref{tab:hubbard_var_xi}.
\begin{table}[h!]
\centering
\renewcommand{\arraystretch}{1.2}
\begin{tabular}{|c|c|c|c|c|c|}
\hline
 $U_{\eta\eta}$ & 3.89° & 4.41° & 5.09° & 6.01° & 7.34° \\
\hline
$\xi=5$nm & 72.34 & 78.12 & 81.10 & 88.62 & 93.96 \\
\hline
$\xi=10$nm & 86.38 & 92.34 & 95.34 & 103.03 & 108.39\\ 
\hline
$\xi=15$nm & 91.61 & 97.60 & 100.59 & 108.31 & 113.71 \\
\hline
\end{tabular}
\caption{Hubbard interaction for different screening lengths $\xi=5,10,15$nm.
}
\label{tab:hubbard_var_xi}
\end{table}
Secondly, the dielectric constant in experiments can deviate from $12$, potentially introducing additional scaling factors to the interaction strength.
Finally, the screening effect from the high-energy bands (that we have dropped in our model) could also reduce the interaction strength. To test this,  we consider the monolayer SnSe$_2$. We found that the bare on-site Hubbard repulsion, obtained from DFT calculations, for the Sn $s$-orbital (which forms the lowest moir\'e conduction bands) is $9.4 \text{eV}$. 
However, when including the screening effects from all other orbitals via constrained random phase approximation within DFT calculations (see Ref.\cite{PhysRevB.111.125163}), the on-site Hubbard repulsion for the Sn $s$-orbital is significantly reduced to $3.2 \text{eV}$.

Given the uncertainty in the overall interaction strength, we use the interactions derived for $\xi = 10 \text{nm}$ and $\epsilon = 12$ as a baseline. We then explore the phase diagram by introducing a global scaling factor to the interaction terms, which is equivalent to tuning the dielectric constant $\epsilon$. 

\subsubsection{Symmetry}\label{app:sec:interacting_model_symmetry}

The AA-stacking SnSe$_2$ has 
\begin{align}
    C_{3z},\quad C_{2x},\quad \mathcal{T}
\end{align}
symmetries. The system also has an approximate zero-twist symmetry
$\tilde{M}_z$~\cite{calugaru2024mtwist}. 
In addition, for each valley, the system also develops $U(1)\times SU(2)=U(2)$ symmetry, which leads to a
\begin{align}
    U(2)\times U(2) \times U(2) 
\end{align}
symmetry. Moreover, the interaction term of the system has an approximate $U(6)$ symmetry with a small symmetry breaking of around $3\%$ of the on-site Hubbard interactions. 

Finally, if we only keep the nearest-neighbor 1D hopping (\cref{eq:nn_t}) and the density-density interaction (\cref{eq:density_density_int}), which are the two dominant terms in the model, the system exhibits additional effective symmetries. First, such a system possesses particle-hole symmetry, characterized by the transformation
\begin{align}
    \mathcal{P} \cre{d}{C_{3z}^\eta(n\bm{a}_{M,1}+m\bm{a}_{M,2}),\eta, s}\mathcal{P} \rightarrow  (-1)^m \des{d}{C_{3z}^\eta(n\bm{a}_{M,1}+m\bm{a}_{M,2}),\eta, s},\quad n,m\in\mathbb{Z}
\end{align}
The additional $(-1)^m$ is introduced to keep the sign of the hopping term during the particle-hole transformation. This can be easily seen with the hopping term (with only 1D nearest-neighbor hopping) given as
\begin{align}
      H_{t}= t\sum_{\vR ,\eta,s}\sum_{n=\pm 1 } \cre{d}{\vR,\eta,s}\des{d}{\vR+nC_{3z}^\eta(\bm{a}_{M,2}),\eta,s} 
\end{align}
Under particle-hole transformation, we find 
\begin{align}
\label{eq:particle_hole_sym}
    \mathcal{P}H_{t}\mathcal{P}  = 
    t\sum_{\vR ,\eta,s}\sum_{n=\pm 1 }(-1) \des{d}{\vR,\eta,s}\cre{d}{\vR+nC_{3z}^\eta(\bm{a}_{M,2}),\eta,s} 
    =  t\sum_{\vR ,\eta,s}\sum_{n=\pm 1 }\cre{d}{\vR+nC_{3z}^\eta(\bm{a}_{M,2}),\eta,s}  \des{d}{\vR,\eta,s} = H_t
\end{align}
The density-density interaction term under particle-hole symmetry becomes
\begin{align}
\label{eq:particle_hole_density_density}
  & \mathcal{P}  H_{V} \mathcal{P}= \sum_{\eta\eta',\vR,\Delta\vR,ss'}\frac{V_{\eta\eta'}(\Delta\vR)}{2}
(1-n_{\vR,\eta,s})(1-n_{\vR+\Delta\vR,\eta',s'})\nonumber\\ 
=&H_{V} - \sum_{\eta,s,\vR}
\bigg[ \sum_{\Delta\vR, \eta',s'}V_{\eta\eta'}(\Delta\vR)
\bigg] n_{\vR,\eta,s} + \sum_{\eta\eta',\vR,\Delta\vR,ss'}\frac{V_{\eta\eta'}(\Delta\vR)}{2}
\end{align}
We observe that $\mathcal{P} H_V \mathcal{P}$ and $H_V$ differ only by a chemical potential term and a constant term. 
Thus, the model has a particle-hole symmetry. 
Under the particle-hole transformation, the filling of the system becomes
\begin{align}
    \frac{1}{N}\sum_{\vR,\eta,s}n_{\vR,\eta,s} = \frac{1}{N}\sum_{\vR,\eta,s}(1-n_{\vR,\eta, s})  = 6- \frac{1}{N}\sum_{\vR,\eta,s}n_{\vR,\eta, s}
\end{align}
Due to the particle-hole symmetry, it is sufficient to consider the ground states at fillings $\nu = 1, 2, 3$. The ground states at fillings $\nu = 4, 5, 6$ can be obtained by applying a particle-hole transformation to the ground states of $\nu = 1, 2, 3$, respectively. 

Besides particle-hole symmetry, the system with nearest-neighbor 1D hopping and density-density interactions \cref{app:eq:ddinteraction} exhibits a
\ba 
\label{eq:additional_U2_symmetry}
U(2)^{L_0} \times U(2)^{L_1} \times U(2)^{L_2}
\ea 
symmetry, where $L_\eta$ denotes the size of the system along the $C_{3z}^\eta \bm{a}_{M,1}$ direction. For valley $\eta$, hopping occurs exclusively along the $C_{3z}^\eta \bm{a}_{M,2}$ direction. Consequently, for each valley, the system can be represented as multiple 1D chains, where electrons are restricted to hopping within a chain and cannot hop between chains. 
This restriction gives rise to a $U(2)$ symmetry for each chain of electrons within a given valley. As a result, the system exhibits a total symmetry of $U(2)^{L_0} \times U(2)^{L_1} \times U(2)^{L_2}$, where $L_\eta$ denotes the number of one-dimensional chains of valley $\eta$.

\subsection{Wannier model of AB-stacked SnSe$_2$} \label{app:sec:wannier_AB}
\subsubsection{Tight-binding model}
The tight-binding model for AB stacking can be written as 
\begin{align}
    H_t = \sum_{\vk,\eta,s} \epsilon_{\vk,\eta}\cre{d}{\vk,\eta,s}\des{d}{\vk,\eta,s} 
    = \sum_{\vR,\Delta\vR,\eta,s} t_{\Delta\vR}^\eta \cre{d}{\vR,\eta,s}\des{d}{\vR+\Delta\vR,\eta,s}
\end{align} 
Unlike AA stacking where only the nearest-neighbor intra-valley hopping along one direction is dominant, all six nearest-neighbor (intra-valley) hoppings are relevant in AB stacking.

\begin{figure}
    \centering
    \includegraphics[width=0.4\linewidth]{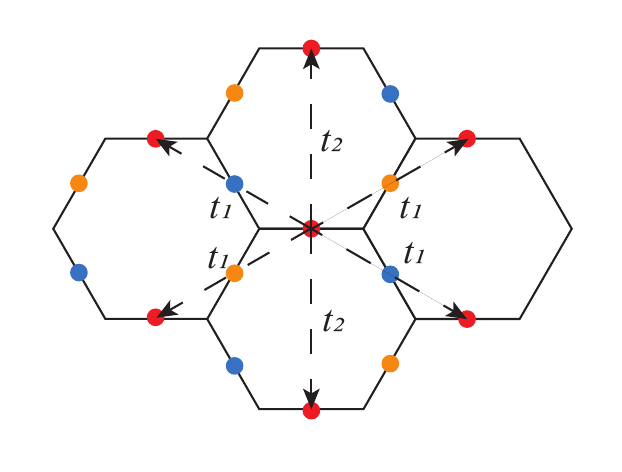}
    \caption{An illustration of the hoppings on the effective kagome lattice of the AB stacking.}
    \label{fig:ab_wannier}
\end{figure}

\begin{figure}
    \centering
    \includegraphics[width=1.0\linewidth]{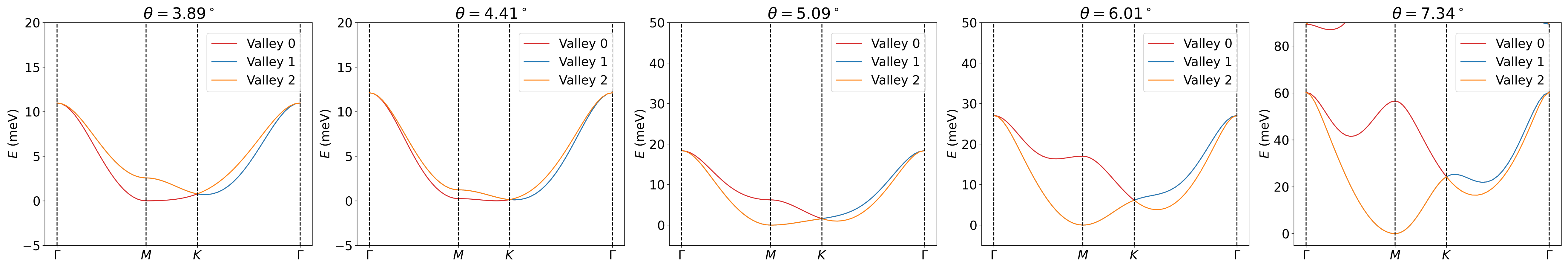}
    \caption{Dispersion of the AB-stacked SnSe$_2$. Red, blue and orange denote valley $0,1,2$ respectively. }
    \label{fig:disp_AB}
\end{figure}

At $\theta=3.89^\circ$ and valley $\eta=0$, we find 
\begin{align}
\label{eq:def_t1_t2_AB}
t^{\eta=0}_{\pm \bm{a}_{M,1}}=t^{\eta=0}_{\pm(\bm{a}_{M,1}+\bm{a}_{M,2})}= t_1=1.30\text{meV},\quad 
    t^{\eta=0}_{\pm \bm{a}_{M,2}} = t_2=0.65\text{meV}
\end{align}
where all other hoppings are smaller than $0.07$meV. An illustration of the hoppings is shown in \cref{fig:ab_wannier}.
The parameters of other angles are given in~\cref{tab:parameter_val_AB}. 
It is worth mentioning that, for $\theta = 3.89, 4.41$, we have $|t_1|> |t_2|$. However, for $\theta = 5.09,6.01, 7.34$, we have $|t_1|<|t_2|$. We show the dispersions of the model at all angles in~\cref{fig:disp_AB}, and we can also observe the differences of dispersion for small and large angles.

The tight-binding model also exhibits an additional inversion symmetry. Due to time-reversal symmetry, the energy dispersion satisfies 
\begin{align}
\label{eq:ab_inv_sym_disp}
    \epsilon_{\vk,\eta} = \epsilon_{-\vk,\eta}
\end{align}
We can define the following inversion transformation
\begin{align}
    \mathcal{I}\cre{d}{\vk,\eta,s}\mathcal{I}^{-1} = \cre{d}{-\vk,\eta,s}
\end{align}
With this symmetry, the tight-binding Hamiltonian exhibits the symmetry of the kagome lattice, corresponding to the space group SG 191 ($P6/mmm$) but the hoppings are different from the nearest-neighbor kagome lattice model. This is because the three valleys correspond to the three sublattices of the kagome lattice. In the kagome model, nearest-neighbor hopping occurs between different sublattices, which corresponds to inter-valley hopping in our system. However, such hopping is forbidden by the $U(1)$ valley symmetry present in our model.

\subsubsection{Interactions}\label{sec:app:wannier_ab_interaction}
We begin by discussing the interactions for the smallest angle considered, $\theta = 3.89^\circ$. The interactions for larger angles are qualitatively similar. 

We first comment that, unlike the AA stacking model, the Wannier centers in the AB stacking model are approximately arranged in a kagome lattice configuration (see \cref{fig:wannier_AB}). 

The dominant interactions take the form of density-density interactions and can be written as 
\begin{align}
    H_{V} = \sum_{\eta\eta',\vR,\Delta\vR,ss'}\frac{V_{\eta\eta'}(\Delta\vR)}{2}
n_{\vR,\eta,s}n_{\vR+\Delta\vR,\eta',s'}
\label{eq:density_density_int1}
\end{align}
where $n_{\vR,\eta,s} =  \cre{d}{\vR,\eta,s}\des{d}{\vR,\eta,s}$. 
The dominant interactions are intra-valley Hubbard interactions 
\begin{align}
\label{eq:AB_HU}
    H_{U} = \sum_{\vR,\eta,ss'}\frac{U}{2} n_{\vR,\eta,s}n_{\vR,\eta,s'}
\end{align}
with 
\begin{align}
    U=87.2\text{meV with }\xi=10\text{nm and }\epsilon=12.
\end{align} 
The sub-dominant interactions are nearest-neighbor (with respect to the approximate kagome lattice) inter-valley density-density interactions
\begin{align}
\label{eq:AB_HV}
    H_V = &\sum_{\vR,ss'}V_1 (n_{\vR,0,s}n_{\vR,1,s'} + n_{\vR,0,s}n_{\vR+\bm{a}_{M,1},1,s'}+
n_{\vR,0,s}n_{\vR+\bm{a}_{M,1}+\bm{a}_{M,2},2,s'}
+n_{\vR,0,s}n_{\vR,2,s'} \nonumber\\
&+n_{\vR,1,s}n_{\vR,2,s'}
+n_{\vR,1,s}n_{\vR+\bm{a}_{M,2},2,s'}
    )
\end{align}
with $V_1 = 35.5$meV. This takes the same form as the nearest-neighbor density-density interactions on a kagome lattice. The next-nearest-neighbor density-density interactions are around $13$meV. 
If we only consider on-site and nearest-neighbor interactions, the interaction term $H_U +H_V$ has an enlarged symmetry group (SG 191 $P6/mmm$), which is the symmetry group of the kagome lattice.

The spin-spin couplings, density-dependent hopping, and all other interactions are around $2$meV and are much weaker than the density-density interactions.

 The dominant interaction parameters of all five angles are listed in Table.~\ref{tab:parameter_val_AB}. 
Uncertainty in the interaction scales also exists in the AB stacking model. As with the AA stacking model, we use the interactions derived for $\xi = 10  \text{nm}$ and $\epsilon = 12$ as a baseline and explore the phase diagram by tuning the $\epsilon$.

\begin{table}[h!]
\centering
\renewcommand{\arraystretch}{1.2}
\begin{tabular}{|c|c|c|c|c|c|}
\hline
 & $t_1$ & $t_2$ & $U$ & $V_1$ & $U_{\eta\eta}/w_b$ \\
\hline
3.89° & 1.30 & 0.64 & 87.20 & 35.50 & 7.95  \\
\hline
4.41° & 1.36 & 1.05 & 94.39 & 42.21 & 7.78  \\
\hline
5.09° & 1.25 & 2.61 & 99.12 & 49.76 & 5.39  \\
\hline
6.01° & 0.84 & 4.36 & 105.50 & 59.63 & 3.90  \\
\hline
7.34° & 0.26 & 9.88 & 110.99 & 70.75 & 1.84  \\
\hline
\end{tabular}
\caption{Parameter values of the Wannier model for AB-stacked SnSe${}_2$ with $\xi=10$nm and $\epsilon=12$. $w_b$ represent the band width.}\label{tab:parameter_val_AB}
\end{table}

\subsubsection{Symmetry}\label{sec:AB_symmetry}
The AB-stacking SnSe$_2$ has 
\begin{align}
    C_{3z},\quad C_{2y},\quad \mathcal{T}
\end{align}
symmetries. In addition, for each valley, the system also develops $U(1)\times SU(2)=U(2)$ symmetry, which leads to a
\begin{align}
    U(2)\times U(2) \times U(2) 
\end{align}
symmetry.  


In addition, with only nearest-neighbor and on-site density-density interactions, the model has an enlarged symmetry characterized by SG 191, as we discussed near~\cref{eq:ab_inv_sym_disp}

\section{Hartree-Fock phase diagram of AA-stacked SnSe$_2$} \label{sec:HFphase_diagram}
In this appendix, we study the interacting Wannier model of AA-stacked SnSe$_2$ within the Hartree-Fock approximation, focusing on filling factors $\nu = 1, 2, 3$ (electrons per moiré unit cell) at various interaction strengths. 
Next, we demonstrate that, within the strong coupling limit, the system can be mapped onto an effective spin model, allowing for the exact determination of the ground states.

\begin{figure}
    \centering
\includegraphics[width=1.0\linewidth]{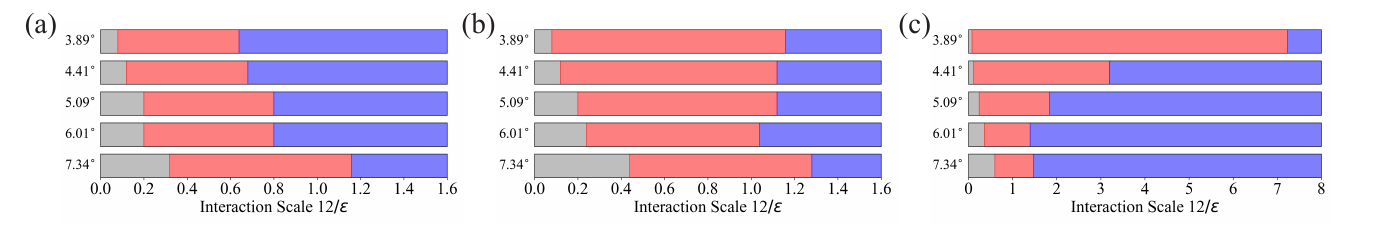}
    \caption{Momentum space Hartree-Fock phase diagram for AA stacked bilayer SnSe${}_2$ with different twisting angles and filling number. (a) $\nu=1$ (b) $\nu=2$ (c) $\nu=3$. The gray/red/blue phase represents the symmetric gapless/AFM/FM VP phase.  
    }
    \label{fig:phase_diagram_AA_mom}
\end{figure}

\begin{figure}
    \centering
\includegraphics[width=0.96\linewidth]{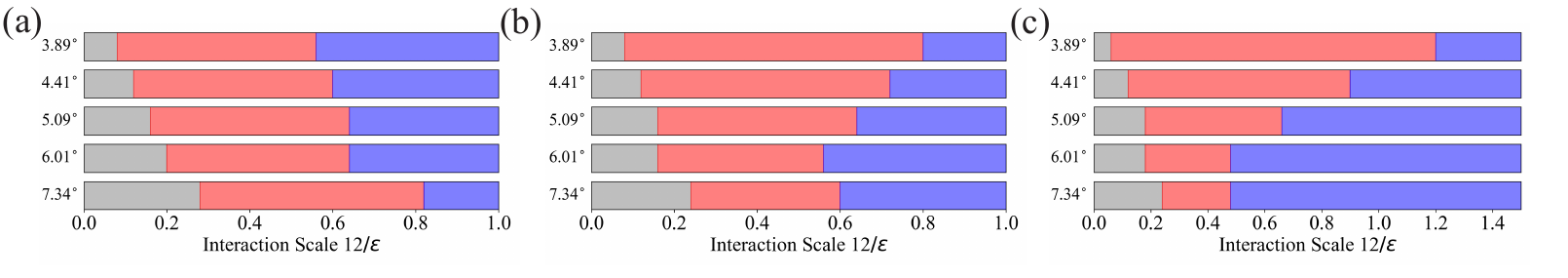}
    \caption{   Real-space Hartree-Fock phase diagram for AA stacked bilayer SnSe${}_2$ with different twisting angles and filling number. (a) $\nu=1$ (b) $\nu=2$ (c) $\nu=3$. The gray/red/blue phase represents the symmetric gapless/AFM/FM VP phase.  
    }
    \label{fig:phase_diagram_AA}
\end{figure}


Compared with the real-space formulation, the momentum-space Hamiltonian retains all long-range interactions, whereas the real-space version applies an explicit cutoff to the interaction range, to make the calculation physically tractable. To map out the phase behavior at filling factors $\nu=1,\,2,\,3$, we therefore rely primarily on momentum-space Hartree–Fock (HF) calculations and use the real-space results only to verify our findings. In both representations, the full Hamiltonian is
\begin{align}
    H=H_t+H_I,
\end{align}
with $H_t$ defined in \cref{app:eq:real_tb_aa} (real space) or \cref{eq:def_H_t} (momentum space), and $H_I$ specified by \cref{eq:def_H_I} together with the interaction-range cutoff in \cref{app:eq:interacting_cutoff_aa} (real space) or by  \cref{eq:def_H_I_mom} (momentum space). Based on the HF solutions, the phase diagram naturally separates into three regimes—weak coupling, strong coupling, and the flat-band limit—as illustrated in Fig.~\ref{fig:phase_diagram_AA_mom} and Fig.~\ref{fig:phase_diagram_AA}.

For $\nu = 1, 2, 3$, we conducted HF simulations using $2 \times 3$, $\sqrt{3} \times \sqrt{3}$, and $2 \times 2$ supercells, respectively. These choices of supercells are motivated by the exact solutions of the effective spin model, discussed later in \cref{sec:spin_strong_coupling}. These supercells are sufficiently large to capture the ground state predicted by the spin model in the strong coupling limit.

When the interaction strength is small (large $\epsilon$ such that $U/w_b\lesssim1$ with $w_b$ being the bandwidth, for all angles), the system is in the weak-coupling limit. Our Hartree-Fock simulations in this regime suggest a symmetric gapless phase for all three fillings, represented by the gray area in Fig.~\ref{fig:phase_diagram_AA_mom}. However, we cannot rule out the possibility of a gap opening if the supercell size is further increased. A more detailed study of this region will be discussed in the following paper\cite{tba2025c}.
 As the interaction strength increases, the system transitions to a regime with gapped ground states exhibiting antiferromagnetic spin order as observed in our mean-field simulations. We refer to this regime as the strong coupling limit.

\begin{figure}
    \centering
\includegraphics[width=0.9\linewidth]{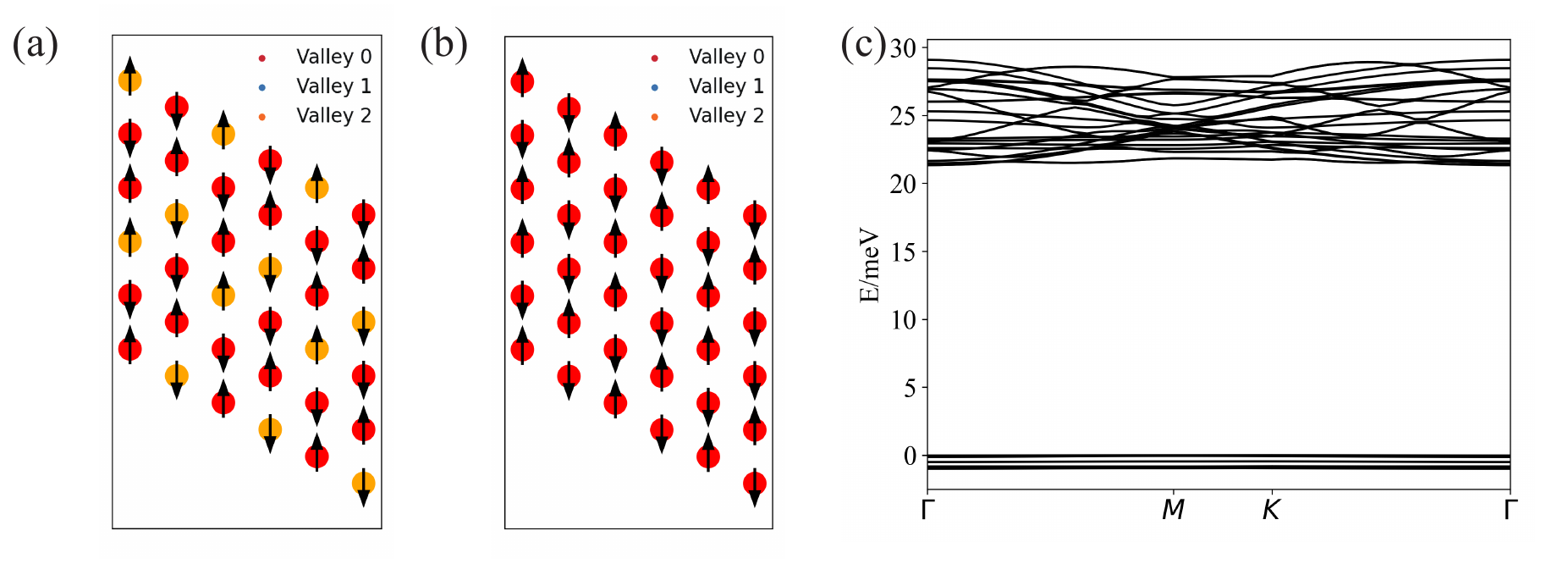}
    \caption{(a)/(b) The valley and spin distribution of the ground state for $\theta=3.89^\circ$ AA stacked bilayer SnSe${}_2$ with $\nu=1$,
 $12/\epsilon=0.2/0.6$. The color of the sites is determined by the valley components, and valley $\eta=0/1/2$ are colored as red/blue/orange separately, and the arrow represents the spin direction. (a) and (b) both show AFM order but have different valley patterns which suggests a large ground state quasi-degeneracy.  (c) The Hartree-Fock spectrum for $\theta=3.89^\circ$ AA stacked bilayer SnSe${}_2$ with $12/\epsilon=0.3$ with real-space mean-field solution. The large charge gap indicates the Hubbard physics. Below the gap, there are 6 quasi-degenerate approximate flat bands with bandwidth $0.1$meV, which is consistent with the electron filling.}
    \label{fig:dimer_hf}
\end{figure}

\begin{figure}
    \centering
    \includegraphics[width=0.9\linewidth]{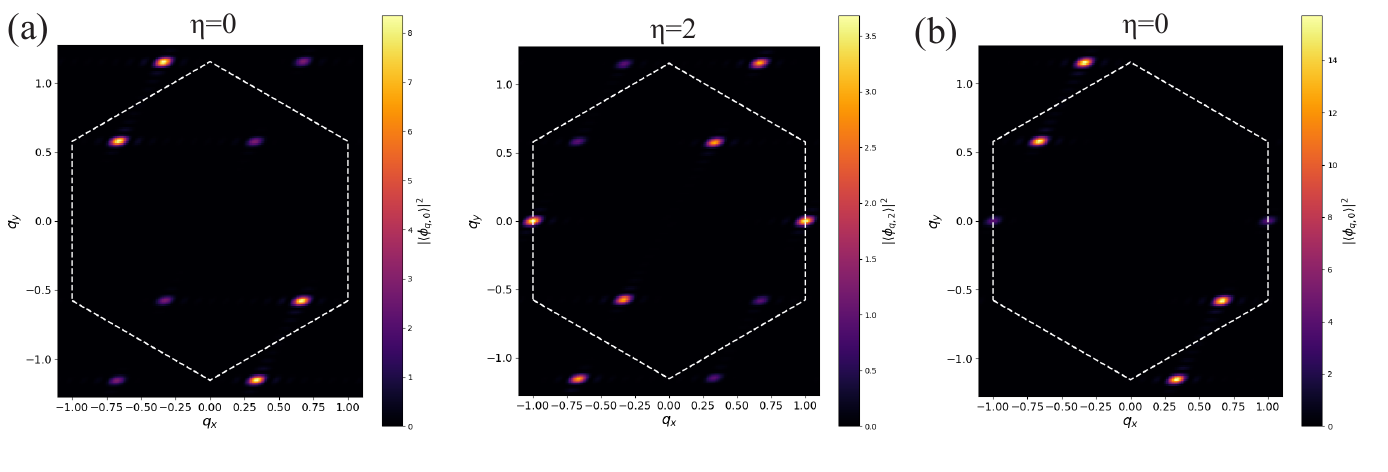}
    \caption{Expectation value $|\phi_{\qq,\eta}^z|^2$ (\cref{eq:def_S_q_eta}). Panel (a) corresponds to the Hartree-Fock state in \cref{fig:dimer_hf} (a). Panel (b) corresponds to the Hartree-Fock state in \cref{fig:dimer_hf} (b).
    In addition, since the magnetic orders in \cref{fig:dimer_hf} (a) and (b) develop along $z$ direction, $\phi_{\qq,\eta}^{x},\phi_{\qq,\eta}^{y}$ are zero.  
    }
    \label{fig:afm_form_fac}
\end{figure}

At filling $\nu = 1$, in the strong coupling region, we observe quasi-degenerate ground states in our simulations.
These quasi-degenerate ground states all exhibit antiferromagnetic (AFM) order in the spin channel while differing in their valley configurations. 
For example, at $\theta = 3.89^\circ$ and $\frac{12}{\epsilon} = 0.4$, the energy differences among the quasi-degenerate AFM states are less than $0.2\text{meV}$. This energy difference is much smaller than the energy difference $\sim 1.5,\text{meV}$ between the AFM and ferromagnetic phases, indicating robust antiferromagnetic correlations. 
\cref{fig:dimer_hf} (a) and (b) show two representative antiferromagnetic (AFM) patterns at filling $\nu = 1$, which emerge as ground states in the Hartree-Fock calculations at $12/\epsilon = 0.2$ and $12/\epsilon = 0.6$, respectively. 
To establish the antiferromagnetic nature of the ground states shown in \cref{fig:dimer_hf} (a) and (b), we calculate 
\ba 
\label{eq:def_S_q_eta}
\langle \phi_{\qq,\eta}\rangle = \frac{1}{N}\sum_{\RR}\langle \bm{S}_{\RR,\eta}\rangle e^{-i\qq\cdot\RR}
\ea 
with expectation value taken with respect to the Hartree-Fock ground state. We illustrate the results in \cref{fig:afm_form_fac}. 
For both \cref{fig:dimer_hf}(a) and (b), we observe that $\langle \phi_{\qq,\eta} \rangle$ exhibits a peak at a finite momentum $\qq$, for $\eta = 0, 2$ in \cref{fig:afm_form_fac} (a), and for $\eta = 0$ in \cref{fig:afm_form_fac} (b). However, there is no signal at $\qq = 0$. These behaviors reflect the antiferromagnetic nature of the Hartree-Fock ground states in both cases. 
In addition, from \cref{fig:dimer_hf} (a) and (b), we also conclude that \cref{fig:dimer_hf}(b) develops a valley polarization, which is absent in \cref{fig:dimer_hf}(a). 
In our simulations, either \cref{fig:dimer_hf}(a) or \cref{fig:dimer_hf}(b) can emerge as the ground state in the strong coupling region, depending on the value of $12/\epsilon$, as detailed in the following paragraph. 
In the strong coupling region, the strong on-site Hubbard interaction $U$ also localizes the electrons into local moments, as evidenced by the charge excitation gap in our Hartree–Fock simulations [Fig.~\ref{fig:dimer_hf}(c)]. Below this gap, six nearly degenerate approximate flat bands emerge, consistent with the electron filling factor $2 \times 3 \times 1 = 6$ (where $2\times 3$ denotes the size of the supercell, and $1$ denotes the filling factor).

As we will demonstrate in \cref{sec:spin_strong_coupling}, the ground state properties in this region can be understood through an effective spin model, which leads to a large quasi-degeneracy on the mean-field level. 
In our momentum-space Hartree-Fock simulations, the AFM order (Fig.~\ref{fig:dimer_hf}(a)) is the ground state within the following interaction ranges for different twist angles: $0.08 < 12/\epsilon < 0.4$ for $\theta = 3.89^\circ$, $0.12 < 12/\epsilon < 0.48$ for $\theta = 4.41^\circ$, $0.2 < 12/\epsilon < 0.72$ for $\theta = 5.09^\circ$, $0.2 < 12/\epsilon <0.72$ for $\theta = 6.01^\circ$, and $0.32 < 12/\epsilon < 1.16$ for $\theta = 7.34^\circ$
The AFM VP order (antiferromagnetic spin order with valley polarization, see Fig.~\ref{fig:dimer_hf}(b)) is the ground state within the following interaction ranges for different twist angles: $0.4 < 12/\epsilon < 0.64$ for $\theta = 3.89^\circ$, $0.48 < 12/\epsilon < 0.68$ for $\theta = 4.41^\circ$, $0.72 < 12/\epsilon < 0.8$ for $\theta = 5.09^\circ$, $0.72 < 12/\epsilon < 0.8$ for $\theta = 6.01^\circ$. We do not observe the AFM VP order at $7.34^\circ$. We also note that the two types of ground states are quasi-degenerate, for example, with an energy difference of less than $0.2 \text{meV}$ in the case of $\theta = 3.89^\circ$ and $\frac{12}{\epsilon} = 0.4$. Since the Hartree-Fock method underestimates the quantum fluctuations, it cannot accurately determine the true ground state in this region. However, certain exact results based on the spin model can be obtained in the strong coupling regime, as discussed later in \cref{sec:spin_strong_coupling}.

\begin{figure}
    \centering
\includegraphics[width=1.0\linewidth]{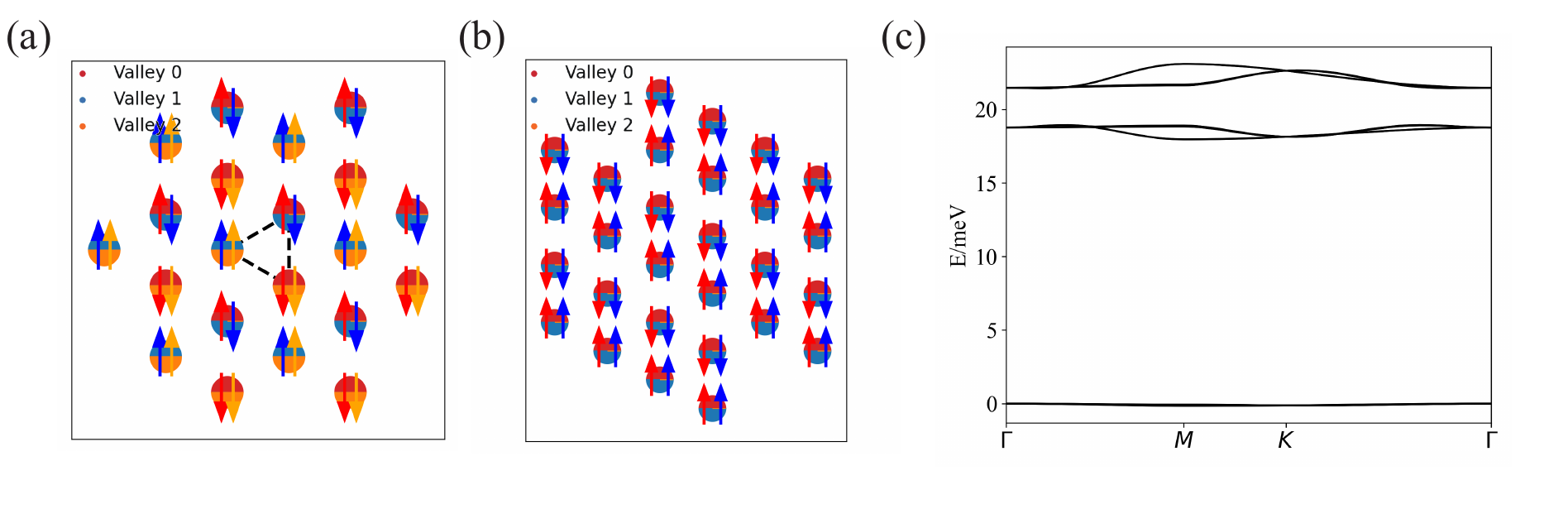}
    \caption{Two possible valley and spin distributions of the ground states for $\theta=3.89^\circ$ AA stacked bilayer SnSe${}_2$ with filling number $\nu=2$ at the AFM phase. The colors indicate the valley components, with valleys $\eta = 0$, $1$, and $2$ shown in red, blue, and orange, respectively. (a) $12/\epsilon=0.24$ (b) $12/\epsilon=0.4$.   The directions of the arrows represent the spin direction. The dashed line in (a) labels the super unit cell. (c) The Hartree-Fock spectrum for $\theta=3.89^\circ$ AA stacked bilayer SnSe${}_2$ with $\nu=2$ and $12/\epsilon=0.24$ with real-space mean-field solution. A large charge gap is observed. Below the gap, there are 6 quasi-degenerate approximate flat bands with band width $0.2$meV, which is consistent with the electron filling, that is filling $\nu=2$ within the unit-cell of 3 sites. 
    }
    \label{fig:nu=2HF}
\end{figure}

At filling $\nu = 2$, in the strong coupling region, the ground states within the AFM phase can be categorized into two distinct types of valley-spin configurations, depending on the exact interaction strength, as shown in Fig.~\ref{fig:nu=2HF}(a) and (b). 
The first configuration features a $\sqrt{3} \times \sqrt{3}$ supercell, where electrons within the same valley form AFM order along $C_{3z}^{\eta}\mathbf{a}_{M,2}$ direction \cref{fig:main:spin_config}. 
In the second configuration, the valley degrees of freedom are polarized into two valleys, with electrons in each valley exhibiting antiferromagnetic correlations along the direction of their dominant hopping. 
In our momentum-space Hartree-Fock simulations, the AFM order without valley polarization (Fig.~\ref{fig:nu=2HF}(a)) is the ground state within the following interaction ranges for different twist angles: $0.08 < 12/\epsilon < 0.28$ for $\theta = 3.89^\circ$, $0.12 < 12/\epsilon < 0.36$ for $\theta = 4.41^\circ$, $0.2 < 12/\epsilon < 1.08$ for $\theta = 5.09^\circ$, $0.24 < 12/\epsilon < 1.04$ for $\theta = 6.01^\circ$, and $0.76 < 12/\epsilon < 1.28$ for $\theta = 7.34^\circ$. 
The AFM VP order (antiferromagnetic spin order with valley polarization, see Fig.~\ref{fig:nu=2HF}(b)) is the ground state within the following interaction ranges for different twist angles: $0.28 < 12/\epsilon < 1.16$ for $\theta = 3.89^\circ$, $0.36 < 12/\epsilon < 1.12$ for $\theta = 4.41^\circ$, and  $1.08 < 12/\epsilon < 1.12$ for $\theta = 5.09^\circ$, and $0.44 < 12/\epsilon < 0.76$ for $\theta = 7.34^\circ$. We do not observe the AFM VP order at $\theta=6.01^\circ$. In this regime, similar to the filling $\nu=1.0$, the strong on-site Hubbard interaction $U$ localizes the electrons into local moments, and a charge excitation gap is observed in our Hartree–Fock simulations as shown in Fig.~\ref{fig:nu=2HF}(c). Below this gap, six nearly degenerate approximate flat bands emerge.
The number of the approximate flat bands is consistent with the electron filling factor with $6=2\times 3$, with $2$ denoting the filling factor per original unit cell and $3$ characterizing the size of the supercell. 
The AFM VP state (\cref{fig:nu=2HF} (b)) 
 occurs with a stronger interaction with $\theta=3.89^\circ, 4,41^\circ$ and $5.09^\circ$. 
We note that the two types of ground states, that is, AFM without valley polarization and AFM with valley polarization, are quasi-degenerate, for example, with an energy difference of less than $0.03 \text{meV}$ in the case of $\theta = 3.89^\circ$ and $\frac{12}{\epsilon} = 0.4$. Thus, we do not distinguish these two states in the phase diagram but attribute them both to the AFM state, indicating that they both develop antiferromagnetic spin order. Since the Hartree-Fock method underestimates the quantum fluctuations, it cannot accurately determine the true ground state in this region. However, we can get exact ground states based on the spin model which is obtained in the strong coupling regime, as discussed later in \cref{sec:spin_strong_coupling}.

\begin{figure}
    \centering
\includegraphics[width=.7\linewidth]{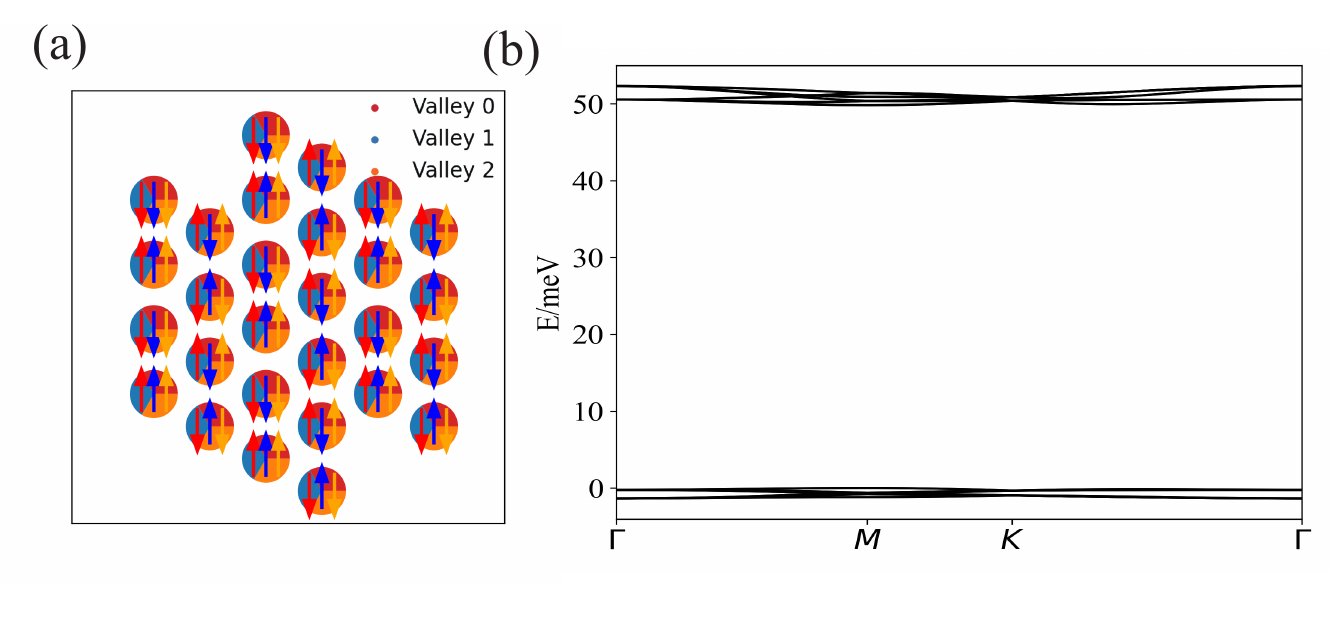}
    \caption{(a) Valley and spin distributions of the ground states for $\theta=3.89^\circ$ AA stacked bilayer SnSe${}_2$. 
    The colors indicate the valley components, with valleys $\eta = 0$, $1$, and $2$ shown in red, blue, and orange, respectively.
    Here, the filling number $\nu=3$ and $6/\epsilon=0.6$, which is in the AFM phase. The directions of the arrows represent the spin direction. (c) The Hartree-Fock spectrum for $\theta=3.89^\circ$ AA stacked bilayer SnSe${}_2$ with $\nu=3$ $12/\epsilon=0.6$. A large charge gap is observed. Below the gap, there are 12 quasi-degenerate approximate flat bands with band width 1meV which is consistent with the electron filling.}
    \label{fig:nu=3HF}
\end{figure}

At filling $\nu = 3$, in the strong coupling region, the ground state is illustrated in Fig.~\ref{fig:nu=3HF}(a). At this filling, each site is occupied by three electrons, one from each valley. Electrons within a given valley $\eta$ develop AFM order. This AFM order is observed within the following interaction ranges for different twist angles: $0.08 < 12/\epsilon < 7.84$ for $\theta = 3.89^\circ$, $0.12 < 12/\epsilon < 3.56$ for $\theta = 4.41^\circ$, $0.2 < 12/\epsilon < 1.84$ for $\theta = 5.09^\circ$, $0.28 < 12/\epsilon < 1.4$ for $\theta = 6.01^\circ$, and $0.6 < 12/\epsilon < 1.48$ for $\theta = 7.34^\circ$.

Finally, as the interaction strength increases further from the upper boundary of AFM phases, the system enters the flat band region. The system favors valley-polarized ferromagnetic states at $\nu = 1, 2, 3$ (blue region in \cref{fig:phase_diagram_AA_mom} and \cref{fig:phase_diagram_AA}), which are described as follows:
\begin{align}
\label{eq:gnd_at_ultra_strong}
    &\nu=1:\quad  \prod_{\vR} \cre{d}{\vR,\eta,s}|0\rangle \nonumber\\ 
    &\nu=2:\quad  \prod_{\vR}\cre{d}{\vR,\eta,s} \cre{d}{\vR,\eta',s'}|0\rangle,\quad \eta\ne \eta' \nonumber\\ 
    &\nu=3:\quad  \prod_{\vR}\cre{d}{\vR,0,s_0}
    \cre{d}{\vR,1,s_1}
    \cre{d}{\vR,2,s_2}|0\rangle 
\end{align}
In this limit, due to the strong interaction strength, the dispersion of the system can be treated as negligible. The Heisenberg coupling is also highly suppressed, placing the system in the flat-band limit. Due to the nonzero quantum geometry of the lowest bands~\cite{calugaru2024mtwist}, we expect the valley-polarized ferromagnetic state to be energetically favored, which has also been confirmed within the Hartree-Fock simulations. Similar ferromagnetic states have also been observed and studied in twisted bilayer graphene, as discussed in Ref.~\cite{PhysRevB.103.205414}.

Comparing the mean-field phase diagrams for $\nu=1,2,3$, we observe that the AFM (antiferromagnetic spin order) phase, as marked by red in Fig.~\ref{fig:phase_diagram_AA_mom}, becomes more extended with a higher electron filling. 
This can be understood at the mean-field level. 
We consider the following generic interactions 
\ba 
H_{int} = \sum_{\RR, \Delta \RR, \Delta \rr_1,\Delta\rr_2,i_1j_1i_2j_2} \frac{1}{2}V(\Delta\RR,\Delta\rr_1, i_1,j_1,\Delta\rr_2, i_2,j_2):d_{\RR, i_1}^\dag d_{\RR+\Delta\rr_1 , j_1}  d_{\RR +\Delta\RR,i_2}^\dag d_{\RR+\Delta\RR+\Delta\rr_2,j_2}:,
\ea 
where $i$ and $j$ are the electron flavors $(\eta,s)$. We next perform a Hartree-Fock decoupling of the interactions and define 
\ba 
O_{\Delta\RR,ij}  = \frac{1}{N}\sum_{\RR} \langle  d_{\RR,i}^\dag d_{\RR+\Delta\RR,j} \rangle
=\frac{1}{N}\sum_{\kk} \langle d_{\kk,i}^\dag d_{\kk,j}\rangle e^{i\kk\cdot(\RR+\rr_j-\rr_i) }
\ea 
where the expectation values are now taken with respect to the mean-field ground states. 

This leads to 
\ba 
H_{int}^{MF} = &
 \sum_{\RR, squ\Delta \RR, \Delta \rr_1,\Delta\rr_2,i_1j_1i_2j_2} \frac{1}{2}V(\Delta\RR,\Delta\rr_1, i_1,j_1,\Delta\rr_2, i_2,j_2)\nonumber\\ 
 &\bigg[ 
O_{\Delta  \rr_1, i_1j_1}
 d_{\RR +\Delta\RR,i_2}^\dag d_{\RR+\Delta\RR+\Delta\rr_2,j_2}  
 +O_{\Delta\rr_2 ,i_2j_2} d_{\RR, i_1}^\dag d_{\RR+\Delta\rr_1 , j_1} \nonumber\\ 
 &- 
O_{\Delta\RR+\Delta\rr_2,i_1j_2}d_{\RR+\Delta\RR,i_2}^\dag d_{\RR+\Delta\rr_1,j_1} 
-  O_{\Delta\rr_1-\Delta\RR,i_2,j_1} d_{\RR,i_1}^\dag d_{\RR+\Delta\RR+\Delta\rr_2,j_2}  \nonumber\\ 
&- O_{\Delta \rr_1, i_1j_1}O_{\Delta\rr_2,i_2j_2} 
+O_{\Delta\RR+\Delta\rr_2,i_1j_2}O_{\Delta\rr_1-\Delta\RR,i_2,j_1}  \bigg] 
\ea 
We can rewrite the Hartree-Fock Hamiltonian in a more compact form by introducing
\ba 
M_{(\Delta\xx_1,i_1,j_1),(\Delta\xx_2,i_2,j_2)} = \frac{1}{2} \sum_{\Delta\RR} \bigg[ V(\Delta\RR,\Delta\xx_1,i_1,j_1,\Delta \xx_2,i_2,j_2) - V(\Delta\RR, \Delta \xx_2 +\Delta\RR, i_1  , j_2 , \Delta \xx_1 - \Delta\RR, i_2 , j_1 )
\bigg] 
\ea 
The Hartree-Fock Hamiltonian now reads
\ba 
H_{int}^{MF} = &
 \sum_{\RR,\Delta \xx_1,\Delta\xx_2,i_1j_1i_2j_2}M_{(\Delta \xx_1,i_1,j_1),(\Delta \xx_2,i_2,j_2) }\nonumber\\ 
 &\bigg[ 
 O_{\Delta  \xx_1, i_1j_1}
 d_{\RR ,i_2}^\dag d_{\RR+\Delta \xx_2,j_2}   
 +O_{\Delta\xx_2 ,i_2j_2} d_{\RR, i_1}^\dag d_{\RR+\Delta\xx_1 , j_1} 
 - O_{\Delta \xx_1,i_1 j_1} 
 O_{\Delta \xx_2,i_2 j_2} 
  \bigg].
\ea 
When $\Delta \xx_1$ or $\Delta \xx_2=0$, the Hartree-Fock Hamiltonian gives density-hopping terms as
\begin{align}
H_{dh}^{MF}=\sum_{\RR,\Delta x,i_1,j_1,i_2,j_2} \left[M_{(0,i_1,j_1),(\Delta x,i_2,j_2)}n_{i_1}\delta_{i_1,j_1}d_{\RR ,i_2}^\dag d_{\RR+\Delta \xx,j_2}   
 +M_{(\Delta x,i_1,j_1),(0,i_2,j_2)}n_{i_2} \delta_{i_2,j_2}d_{\RR, i_1}^\dag d_{\RR+\Delta\xx , j_1} \right],\label{app:eq:density_hopping_mf}
\end{align}
which effectively enhanced the electron hopping amplitude $t$. Moreover, the antiferromagnetic order arises from the superexchange mechanism, with a spin–spin coupling $J\sim\frac{t^2}{U}$, where $U$ is the onsite Hubbard interaction. Consequently, the density-hopping terms serve to stabilize the AFM phase. Since the strength of the effective hopping in \cref{app:eq:density_hopping_mf} is proportional to the particle number $n_{i}$, we expect that larger filling has a stronger tendency towards AFM order. 

The phase diagram of the momentum-space in  \cref{fig:phase_diagram_AA_mom} and that of the real-space in \cref{fig:phase_diagram_AA} exhibit similar structure but with different phase boundaries locations
. As discussed above, the interactions will stabilize the AFM phase. In particular, including the full long‐range interaction in the momentum‐space calculation shifts the balance in favor of the AFM phase and pushes the onset of ferromagnetism to stronger coupling. The phase boundary is sensitive to the specific interactions due to the large quasi-degeneracy of the mean-field states near the phase boundary.

\section{Spin model in the strong coupling limit}
\label{sec:spin_strong_coupling}
In this appendix, we discuss the spin model and the corresponding ground states in the strong coupling limit. 
In this limit, the Heisenberg coupling dominates, and the system tends to form antiferromagnetic order.

\subsection{Ground state in the strong coupling limit at $\nu=1$}\label{sec:nu1spinliquid}

\subsubsection{Effective spin model at $\nu=1$}
\label{sec:spin_mode_nu_1}
We now derive the effective spin model at $\nu = 1$, starting from the electron model with only the dominant density-density interactions:
\begin{align}
    H_{V} =\frac{1}{2} \sum_{\vR,\Delta\vR, \eta,\eta'}V_{\eta\eta'}(\Delta\vR) n_{\vR,\eta} n_{\vR+\Delta\vR,\eta'},
    \quad \quad n_{\vR,\eta}=\sum_s n_{\vR,\eta,s}
\end{align}
Since the Wannier orbitals of the three valleys are nearly positioned at the same locations, the interactions exhibit an approximate $U(6)$-valley symmetry. The Hamiltonian can therefore be decomposed into a $U(6)$-valley symmetric part and a weak symmetry-breaking part: 
\begin{align}
    &H_V = H_{U(6)} +H_{\delta V} \nonumber\\ 
    & H_{U(6)} =\frac{1}{2} \sum_{\vR,\Delta\vR}V(\Delta\vR) n_{\vR} n_{\vR+\Delta\vR},
    \quad \quad n_{\vR}=\sum_{\eta,s} n_{\vR,\eta,s},\quad V(\Delta\vR) =\frac{1}{9}\sum_{\eta\eta'}V_{\eta\eta'}(\Delta\vR) \nonumber\\ 
    & H_{\delta V}
    =\frac{1}{2} \sum_{\vR,\Delta\vR, \eta,\eta'}\delta V_{\eta\eta'}(\Delta\vR) n_{\vR,\eta} n_{\vR+\Delta\vR,\eta'},
    \quad \quad \delta V_{\eta\eta'}(\Delta\vR) = V_{\eta\eta'}(\Delta\vR)-V(\Delta\vR)
\end{align}
Numerically, we find, at $\theta=3.89^\circ$, 
\begin{align}
\label{eq:sym_break_delta_V}
\frac{\text{max}_{\Delta\vR,\eta\eta'}|\delta V_{\eta\eta'}(\Delta\vR)|}
{|V(0)|} = 2.1\%.
\end{align}
For the hopping term,  we consider the strongest hopping introduced in~\cref{eq:nn_t}
\begin{align}
    H_t =\sum_{\vR,\eta,s} \sum_{\Delta\vR = \pm C_{3z}^\eta \bm{a}_{M,2}}t \cre{d}{\vR,\eta,s}\des{d}{\vR+\Delta\vR,\eta,s} 
\end{align}

We separate the Hamiltonian $H=H_t+H_V$ into two parts
\begin{align}
\label{eq:SW_ham_h0_h1}
    &H = H_0 +H_1 \nonumber\\ 
    &H_0 = H_{U(6)},\quad H_1=H_t +H_{\delta V} 
\end{align}
and derive the effective Hamiltonian by treating $H_1$ perturbatively. 

The ground state of $H_0=H_{U(6)}$ describes an atomic problem. Due to the large onsite Hubbard interaction, the ground state of $H_0$ at filling $\nu=1$ can be written as
\begin{align}
\label{eq:low_energy_state_nu_1}
   |\{\eta_\vR,s_\vR\}_\vR \rangle =  \prod_{\vR}\cre{d}{\vR, \eta_{\vR},s_{\vR}}|0\rangle 
\end{align}
where each site $\vR$ is filled by one electron with valley $\eta_{\vR}$ and spin $s_{\vR}$ that can be taken arbitrarily. 

We also introduce the projection operator that only keeps low-energy states with $1$ electron per moiré unit cell and is defined as
\begin{align}
P_L^{\nu=1} 
= \sum_{ \{\eta_{\vR},s_{\vR} \}_\vR  } | \{\eta_\vR,s_\vR\}_\vR \rangle \langle \{ \eta_\vR,s_\vR\}_\vR | 
\end{align}
where $\sum_{ \{\eta_\vR,s_\vR\}_\vR }$ sums over all possible configurations of $\{\eta_\vR,s_\vR\}_{\vR}$. 

Then the effective Hamiltonian at $\nu=1$ reads
\begin{align}
    H_J^{\nu=1} = P_L^{\nu=1}\bigg[ H_0 +H_1
    \bigg]  P_L^{\nu=1}  
    - P_L^{\nu=1} H_1 \sum_{|H\rangle } \frac{|H\rangle \langle H|}{E_H} H_1 P_L^{\nu=1}
\end{align}
where $|H\rangle$ denotes the high-energy states created by acting $H_1$ on the low-energy states defined in~\cref{eq:low_energy_state_nu_1}. $E_H$ is the corresponding excitation energy. 
We observe that states in~\cref{eq:low_energy_state_nu_1} are also the eigenstates of $H_{\delta V}$ (because $H_{\delta V}$ describes density-density interactions). In addition, since $H_t$ is a hopping term that moves one electron from one unit cell to another unit cell which violates the constrains of $1$ electron per unit cell, we expect
\begin{align}
   P_L^{\nu=1} H_t P_{L}^{\nu=1} =0 
\end{align}
We conclude that the effective Hamiltonian can be simplified as
\begin{align}
    H_J^{\nu=1} = P_L^{\nu=1} 
    \bigg[ H_{\delta V}- H_t\sum_{|H\rangle } \frac{|H\rangle \langle H|}{E_H} H_t\bigg] P_L^{\nu=1}
\end{align}

$H_t$ creates electron and holon at two different sites within the same valley and spin sectors, the high-energy states created by acting $H_t$ on $|\{\eta_{\vR'}, s_{\vR'}\}_{\vR'}\rangle $ can be written as
\begin{align}
\label{eq:charge_0_nu_1}
   &\cre{d}{\tilde{\vR},\eta_{\tilde{\vR}+\tilde{\Delta\vR}},s_{\tilde{\vR}+\Delta\vR}}  \des{d}{\tilde{\vR}+\tilde{\Delta\vR},\eta_{\tilde{\vR}+\tilde{\Delta\vR}},s_{\tilde{\vR}+\tilde{\Delta\vR}}}  |\{\eta_{\vR'},{s_{\vR'}}\}_{\vR'}\rangle = \cre{d}{\vR,\eta_{\vR+\Delta\vR},s_{\vR+\Delta\vR}}\prod_{\vR'}\cre{d}{\vR', \eta_{\vR'},s_{\vR'}}|0\rangle  \nonumber\\ 
   &\tilde{\Delta\vR} =\pm  C_{3z}^\eta \bm{a}_{M,2} ,\quad 
   ({\eta}_{\tilde{\vR}+\tilde{\Delta\vR}},{s}_{\tilde{\vR}+\tilde{\Delta\vR}}) \ne ({\eta}_{\tilde{\vR}},{s}_{\tilde{\vR}})
\end{align}
Here, we also require $({\eta}_{\tilde{\vR}+\tilde{\Delta\vR}},{s}_{\tilde{\vR}+\tilde{\Delta\vR}}) \ne ({\eta}_{\tilde{\vR}},{s}_{\tilde{\vR}})$. Since $\cre{d}{\tilde{\vR},\eta_{\tilde{\vR}+\tilde{\Delta\vR}},s_{\tilde{\vR}+\Delta\vR}}  \des{d}{\tilde{\vR}+\tilde{\Delta\vR},\eta_{\vR+\Delta\vR},s_{\vR+\Delta\vR}}$ moves one electron with valley $\eta_{\tilde{\vR}+\tilde{\Delta\vR}}$ and spin $s_{\tilde{\vR}+\tilde{\Delta\vR}}$ from site $\tilde{\vR}+\tilde{\Delta\vR}$ to $\tilde{\vR}$. 
If $({\eta}_{\tilde{\vR}+\tilde{\Delta\vR}},{s}_{\tilde{\vR}+\tilde{\Delta\vR}}) = ({\eta}_{\tilde{\vR}},{s}_{\tilde{\vR}})$, then site $\tilde{\vR}$ is already occupied by one electron with valley $\eta_{\tilde{\vR}+\tilde{\Delta\vR}}$ spin $s_{\tilde{\vR}+\tilde{\Delta\vR}}$. 
Then, we cannot create one more electron at site $\tilde{\vR}$ with valley $\eta_{\tilde{\vR}+\tilde{\Delta\vR}}$ spin $s_{\tilde{\vR}+\tilde{\Delta\vR}}$. Therefore, we will have 
$\cre{d}{\tilde{\vR},\eta_{\tilde{\vR}+\tilde{\Delta\vR}},s_{\tilde{\vR}+\tilde{\Delta\vR}}}  \des{d}{\tilde{\vR}+\tilde{\Delta\vR},\eta_{\tilde{\vR}+\tilde{\Delta\vR}},s_{\tilde{\vR}+\tilde{\Delta\vR}}}  |\{\eta_{\vR'},s_{\vR'}\}_{\vR'}\rangle =0$
if $({\eta}_{\tilde{\vR}+\tilde{\Delta\vR}},{s}_{\tilde{\vR}+\tilde{\Delta\vR}}) 
=({\eta}_{\tilde{\vR}},{s}_{\tilde{\vR}})$. 

The excitation energy, defined with respect to the unperturbed Hamiltonian $H_0$, is 
\begin{align}
    [H_0,  \cre{d}{\tilde{\vR},\eta_{\tilde{\vR}+\tilde{\Delta\vR}},s_{\tilde{\vR}+\Delta\vR}}  \des{d}{\tilde{\vR}+\tilde{\Delta\vR},\eta_{\tilde{\vR}+\tilde{\Delta\vR}},s_{\tilde{\vR}+\tilde{\Delta\vR}}}]|\{\eta_{\vR'},{s_{\vR'}}\}_{\vR'}\rangle  = E_H    \cre{d}{\tilde{\vR},\eta_{\tilde{\vR}+\tilde{\Delta\vR}},s_{\tilde{\vR}+\Delta\vR}}  \des{d}{\tilde{\vR}+\tilde{\Delta\vR},\eta_{\tilde{\vR}+\tilde{\Delta\vR}},s_{\tilde{\vR}+\tilde{\Delta\vR}}} |\{\eta_{\vR'},{s_{\vR'}}\}_{\vR'}\rangle 
\end{align}
We now calculate $E_H$ explicitly. First we observe that $H_0$ only includes density-density interactions. The energy of a given state characterized by $\{\eta_{\vR'},s_{\vR'}\}_{\vR'}$ is just 
\begin{align}
    &H_V |\{\eta_{\vR'},s_{\vR'}\}_{\vR'}\rangle = E_{\{\eta_{\vR},s_\vR\}_\vR}^{H_V}|\{\eta_{\vR'},s_{\vR'}\}_{\vR'}\rangle \nonumber\\ 
    &E_{\{\eta_{\vR'},s_{\vR'}\}_{\vR'}}^{H_V} = \frac{1}{2}\sum_{\vR',\Delta\vR}V(\Delta\vR) n_{\vR'}n_{\vR'}
\end{align}
where 
\begin{align}
    n_{\vR',\eta} =  1
\end{align}
characterizes the number of electrons at site $\vR'$ which is $1$ for the states we considered. We can then calculate the energy of $ \cre{d}{\tilde{\vR},\eta_{\tilde{\vR}+\tilde{\Delta\vR}},s_{\tilde{\vR}+\Delta\vR}} \des{d}{\tilde{\vR}+\tilde{\Delta\vR},\eta_{\tilde{\vR}+\tilde{\Delta\vR}},s_{\tilde{\vR}+\tilde{\Delta\vR}}}  |\{\eta_{\vR'},{s_{\vR'} }\}_{\vR'}\rangle $. We notice that the number of electrons at site $\vR'$ now changes to 
\begin{align}
    \tilde{n}_{\vR'}= n_{\vR'} - \delta_{\vR',\tilde{\vR}+\tilde{\Delta\vR}} + \delta_{\vR',\tilde{\vR}}
\end{align}
where we have one less electrons at site $\tilde{\vR}+\tilde{\Delta\vR}$ and one more electron at site $\tilde{\vR}$. 
The excitation energy will then depends on the change of the particle numbers 
\begin{align}
\label{eq:charge_0_nu_1_en}
    E_H = &\frac{1}{2}\sum_{\vR',\Delta\vR}
    V_{}(\Delta\vR) 
    \bigg[ -n_{\vR'}n_{\vR'+\Delta\vR} + \tilde{n}_{\vR'}\tilde{n}_{\vR'+\Delta\vR}\bigg] \nonumber\\ 
    =& \frac{1}{2}\sum_{\vR',\Delta\vR}
    V_{}(\Delta\vR) 
    \bigg[ 
    (-\delta_{\vR',\tilde{\vR}+\Delta\vR}
    +\delta_{\vR',\tilde{\vR}})n_{\vR'+\Delta\vR}
    + n_{\vR'} (-\delta_{\vR'+\Delta\vR,\tilde{\vR}+\Delta\vR}
    +\delta_{\vR'+\Delta\vR,\tilde{\vR}}) \nonumber\\
    & + 
     (-\delta_{\vR',\tilde{\vR}+\tilde{\Delta\vR}}
    +\delta_{\vR',\tilde{\vR}})(-\delta_{\vR'+\Delta\vR,\tilde{\vR}+\tilde{\Delta\vR}}
    +\delta_{\vR'+\Delta\vR,\tilde{\vR}})
    \bigg] 
    \nonumber\\ 
    =& 0 + 0 +\frac{1}{2}
    \sum_{\vR',\Delta\vR}V(\Delta\vR)
    \bigg[ -\delta_{\vR',\tilde{\vR}+\tilde{\Delta\vR}}\delta_{\Delta\vR,\tilde{\vR}-\vR'}
    -\delta_{\vR',\tilde{\vR}}
    \delta_{\Delta\vR, \tilde{\vR}+\tilde{\Delta\vR}-\vR'} +\delta_{\vR',\tilde{\vR}+\tilde{\Delta\vR}}\delta_{\Delta\vR,\tilde{\vR}+\tilde{\Delta\vR}-\vR'}
    +\delta_{\vR',\tilde{\vR}}
    \delta_{\Delta\vR,\tilde{\vR}-\vR'}
    \bigg]  \nonumber\\
    =& \frac{1}{2}
    \bigg[- V(\tilde{\Delta\vR})-V(-\tilde{\Delta\vR})
    + V(0) + V(0) 
    \bigg] \nonumber\\ 
    =& V(0) -V(\tilde{\Delta\vR})
\end{align}
where we have also used the fact that $V(\tilde{\Delta\vR}) = V(-\tilde{\Delta\vR})$ (see also the discussion near \cref{app:eq:ddinteraction}). We also comment that $E_H$ are the same for any choices of $\tilde{\Delta\vR} =C_{3z}^{\eta}\bm{a}_{M,2}$ due to the $C_{3z}$ and effective $U(6)$ symmetry. 

Taking the states in~\cref{eq:charge_0_nu_1} to be the high-energy states, we define the projector of the high-energy state as 
\begin{align}
   P_{H}^{\nu=1}= \sum_{H} |H\rangle \langle H|
\end{align}
Written explicitly, the effective coupling generated by $H_t$ is
\begin{align}
 &-P_L^{\nu=1} 
    \bigg[  H_t\sum_{|H\rangle } \frac{|H\rangle \langle H|}{E_H} H_t\bigg] P_L^{\nu=1}    \nonumber\\ 
    =&-\frac{t^2}{E_H} 
        \sum_{\vR,\vR',\eta,s,\eta',s'}\sum_{\Delta\vR = \pm C_{3z}^{\eta}\bm{a}_{M,2},\Delta\vR' = \pm C_{3z}^{\eta'}\bm{a}_{M,2}} \nonumber\\ 
        &
        P_L^{\nu=1}\cre{d}{\vR,\eta,s} \des{d}{\vR+\Delta\vR,\eta,s} 
  P_H^{\nu=1}
          \cre{d}{\vR'+\Delta\vR',\eta',s'} \des{d}{\vR',\eta',s'} 
        P_L^{\nu=1}
\end{align}
A non-zero contribution requires that any operators between two $P_L^{\nu=1}$ operators preserve the number of electrons for each unit cell. Since $P_H^{\nu=1}$ is a projection operator that preserves the electron number on each valley and site, we require that 
\begin{align}
    &\tilde{\vR} = \vR+\Delta\vR =\vR'+\Delta\vR' \nonumber\\ 
    &\tilde{\vR}+\Delta\tilde{\vR} = \vR = \vR' \, .
\end{align}
Since $\Delta\vR = \pm C_{3z}^{\eta}\bm{a}_{M,2},\Delta\vR' = \pm C_{3z}^{\eta'}\bm{a}_{M,2},\tilde{\Delta\vR} = \pm C_{3z}^\eta\bm{a}_{M,2}$, we also have $\eta=\eta''=\eta'$. Therefore, we find 
\begin{align}
\label{eq:perturb_PddPddP}
 &-P_L^{\nu=1} 
    \bigg[  H_t\sum_{|H\rangle } \frac{|H\rangle \langle H|}{E_H} H_t\bigg] P_L^{\nu=1}    \nonumber\\ 
    =&-\frac{t^2}{E_H} 
        \sum_{\vR,\eta,s,s',s''}\sum_{\Delta\vR = \pm C_{3z}^{\eta}\bm{a}_{M,2}} \nonumber\\ 
        &
        P_L^{\nu=1}\cre{d}{\vR,\eta,s} \des{d}{\vR+\Delta\vR,\eta,s} 
    P_H^{\nu=1}
          \cre{d}{\vR+\Delta\vR,\eta,s'} \des{d}{\vR,\eta,s'} 
        P_L^{\nu=1}
\end{align}
We now investigate 
\ba 
  P_H^{\nu=1}
          \cre{d}{\vR+\Delta\vR,\eta,s'} \des{d}{\vR,\eta,s'} 
        P_L^{\nu=1}
\ea 
We aim to show that 
\ba
 P_H^{\nu=1}
          \cre{d}{\vR+\Delta\vR,\eta,s'} \des{d}{\vR,\eta,s'} 
        P_L^{\nu=1} =
          \cre{d}{\vR+\Delta\vR,\eta,s'} \des{d}{\vR,\eta,s'} 
        P_L^{\nu=1}
\ea 
We first note that 
\ba 
 P_H^{\nu=1}
          \cre{d}{\vR+\Delta\vR,\eta,s'} \des{d}{\vR,\eta,s'} 
        P_L^{\nu=1}
        = \sum_{H,L}|H\rangle \langle H|\cre{d}{\vR+\Delta\vR,\eta,s'} \des{d}{\vR,\eta,s'} 
        |L\rangle \langle L| 
\ea 
where $|H\rangle$ and $|L\rangle$ denote the high-energy and low-energy states, respectively. For any given low-energy state $|L\rangle$, $\cre{d}{\vR+\Delta\vR,\eta,s'} \des{d}{\vR,\eta,s'} $ either annihilates it or map the low-energy state to a specific high-energy state. We use $|\tilde{L}\rangle$ to denote the state that is not annihilated by $\cre{d}{\vR+\Delta\vR,\eta,s'} \des{d}{\vR,\eta,s'} $. In addition, we let $|H_{\tilde{L}}\rangle$ to denote the high-energy state created by $\cre{d}{\vR+\Delta\vR,\eta,s'} \des{d}{\vR,\eta,s'}$. In other words, we let 
\ba 
 |H_{\tilde{L}}\rangle =\cre{d}{\vR+\Delta\vR,\eta,s'} \des{d}{\vR,\eta,s'}|\tilde{L}\rangle
\ea 
Written in a compact formula we find 
\ba 
\label{eq:PddP_L_expand}
P_H^{\nu=1}
          \cre{d}{\vR+\Delta\vR,\eta,s'} \des{d}{\vR,\eta,s'} 
        P_L^{\nu=1}= \sum_{H,L}|H\rangle \langle H|\cre{d}{\vR+\Delta\vR,\eta,s'} \des{d}{\vR,\eta,s'} 
        |L\rangle \langle L| 
        = \sum_{ \tilde{L},H}|H\rangle \langle H| H_{\tilde{L}}\rangle  \langle \tilde{L}|
        = \sum_{\tilde{L}} |H_{\tilde{L}} \rangle \langle \tilde{L}|
\ea 
where we use $\tilde{L}$ to denote the low-energy state that is not annihilated by $\cre{d}{\vR+\Delta\vR,\eta,s'} \des{d}{\vR,\eta,s'}$. Similarly, we find 
\ba 
\label{eq:ddP_L_expand}
          \cre{d}{\vR+\Delta\vR,\eta,s'} \des{d}{\vR,\eta,s'} 
        P_L^{\nu=1}=
        \sum_{L} \cre{d}{\vR+\Delta\vR,\eta,s'} \des{d}{\vR,\eta,s'} |L\rangle \langle L| 
        = \sum_{\tilde{L}} |H_{\tilde{L}}\rangle \langle \tilde{L}|
\ea 
Combining \cref{eq:PddP_L_expand,eq:ddP_L_expand}, we conclude that 
\ba 
P_H^{\nu=1}
          \cre{d}{\vR+\Delta\vR,\eta,s'} \des{d}{\vR,\eta,s'} 
        P_L^{\nu=1}= \cre{d}{\vR+\Delta\vR,\eta,s'} \des{d}{\vR,\eta,s'} 
        P_L^{\nu=1}
\ea 
Then \cref{eq:perturb_PddPddP} can be written as
\ba 
&-P_L^{\nu=1} 
    \bigg[  H_t\sum_{|H\rangle } \frac{|H\rangle \langle H|}{E_H} H_t\bigg] P_L^{\nu=1}    \nonumber\\ 
    =&-\frac{t^2}{E_H} 
        \sum_{\vR,\eta,s,s'}\sum_{\Delta\vR = \pm C_{3z}^{\eta}\bm{a}_{M,2}} 
        P_L^{\nu=1}\cre{d}{\vR,\eta,s} \des{d}{\vR+\Delta\vR,\eta,s} 
    P_H^{\nu=1}
          \cre{d}{\vR+\Delta\vR,\eta,s'} \des{d}{\vR,\eta,s'} 
        P_L^{\nu=1} \nonumber\\
        =& 
        -\frac{t^2}{E_H} 
        \sum_{\vR,\eta,s,s'}\sum_{\Delta\vR = \pm C_{3z}^{\eta}\bm{a}_{M,2}} 
        P_L^{\nu=1}\cre{d}{\vR,\eta,s} \des{d}{\vR+\Delta\vR,\eta,s} 
          \cre{d}{\vR+\Delta\vR,\eta,s'} \des{d}{\vR,\eta,s'} 
        P_L^{\nu=1} \nonumber\\
        =& 
        -\frac{t^2}{E_H} 
        \sum_{\vR,\eta,s,s'}\sum_{\Delta\vR = \pm C_{3z}^{\eta}\bm{a}_{M,2}} 
        P_L^{\nu=1}\cre{d}{\vR,\eta,s} \bigg[
        \delta_{s,s'} -  \cre{d}{\vR+\Delta\vR,\eta,s'} 
        \des{d}{\vR+\Delta\vR,\eta,s} \bigg] 
         \des{d}{\vR,\eta,s'} 
        P_L^{\nu=1}  \nonumber\\ 
        =&\frac{t^2}{E_H} 
        \sum_{\vR,\eta,s,s'}\sum_{\Delta\vR = \pm C_{3z}^{\eta}\bm{a}_{M,2}} 
        P_L^{\nu=1}\cre{d}{\vR,\eta,s}  
         \des{d}{\vR,\eta,s'} \cre{d}{\vR+\Delta\vR,\eta,s'} 
        \des{d}{\vR+\Delta\vR,\eta,s} 
        P_L^{\nu=1} 
        -\frac{t^2}{E_H}
        \sum_{\vR}\sum_{\Delta\vR = \pm C_{3z}^{\eta}\bm{a}_{M,2}}
        P_{L}^{\nu=1} \sum_{\eta,s} \hat{d}_{\RR,\eta s}^\dag \hat{d}_{\RR,\eta,s}P_L^{\nu=1} \nonumber\\ 
        & 
        =\frac{t^2}{E_H} 
        \sum_{\vR,\eta,s,s'}\sum_{\Delta\vR = \pm C_{3z}^{\eta}\bm{a}_{M,2}} 
        P_L^{\nu=1}\cre{d}{\vR,\eta,s}  
         \des{d}{\vR,\eta,s'} \cre{d}{\vR+\Delta\vR,\eta,s'} 
        \des{d}{\vR+\Delta\vR,\eta,s} 
        P_L^{\nu=1} -\frac{2t^2}{E_H}N
\ea 
where we can safely remove $P_H^{\nu=1}$, and $N$ denotes the number of unit cells.
We omitted the constant term, which leads to the following effective coupling
\ba 
\label{eq:ham_nu_1_after_SW}
\frac{t^2}{E_H} 
        \sum_{\vR,\eta,s,s'}\sum_{\Delta\vR = \pm C_{3z}^{\eta}\bm{a}_{M,2}} 
        P_L^{\nu=1}\cre{d}{\vR,\eta,s}  
         \des{d}{\vR,\eta,s'} \cre{d}{\vR+\Delta\vR,\eta,s'} 
        \des{d}{\vR+\Delta\vR,\eta,s} 
        P_L^{\nu=1} 
\ea

For convenience, we define the spin operator as
\begin{align}
   S^\mu_{\vR,\eta} =\sum_{ss'}\cre{d}{\vR,\eta,s}\frac{\sigma^\mu_{ss'}}{2}\des{d}{\vR,\eta,s'}
\end{align}
We have the following relation for the spin operators (Fierz identity)
\begin{align}
\label{eq:rel_SS_dddd}
    &\sum_\mu  S^\mu_{\vR,\eta}
     S^\mu_{\vR+\Delta\vR,\eta}=\sum_\mu \sum_{s_1s_1',s_2s_2'}\cre{d}{\vR,\eta,s_1}\frac{\sigma^\mu_{s_1s_1'}}{2}\des{d}{\vR,\eta,s_1'}
     \cre{d}{\vR+\Delta\vR,\eta,s_2}\frac{\sigma^\mu_{s_2s_2'}}{2}\des{d}{\vR+\Delta\vR,\eta,s_2'} \nonumber\\ 
     =&\frac{1}{2} \sum_{ss'}
     \cre{d}{\vR,\eta,s} 
   \des{d}{\vR,\eta,s'}\cre{d}{\vR+\Delta\vR,\eta,s'}\des{d}{\vR+\Delta\vR,\eta,s}
   -\frac{1}{4}
   \sum_{ss'}
     \cre{d}{\vR,\eta,s} 
   \des{d}{\vR,\eta,s}\cre{d}{\vR+\Delta\vR,\eta,s'}\des{d}{\vR+\Delta\vR,\eta,s'}
\end{align}
Combining~\cref{eq:ham_nu_1_after_SW,eq:rel_SS_dddd}, we obtain the effective spin-spin coupling $\nu=1$
\begin{align}
  J\sum_{\vR,\eta,s,s'}\sum_{\Delta\vR = \pm C_{3z}^{\eta}\bm{a}_{M,2}}
        P_L^{\nu=1}
       \bigg[\bm{S}_{\vR,\eta}\cdot \bm{S}_{\vR+\Delta\vR,\eta} +\frac{1}{4}n_{\vR,\eta}n_{\vR+\Delta\vR,\eta}\bigg] 
    P_L^{\nu=1}
\end{align} which is consistent with the Heisenberg exchange coupling derived for multi-orbital Hubbard models, where the valley index plays the role of an orbital degree of freedom. 
In addition, we note that for the single-orbital Hubbard model at half-filling in the strong-coupling limit, the density–density interaction term becomes a constant, as each site is constrained to have exactly one electron from a fixed orbital. 
with Heisenberg coupling strength
\begin{align}
    J= 2t^2/E_H
\end{align}

The full Hamiltonian now reads
\begin{align} 
\label{eq:def_spin_model_nu_1}
    H_{J}^{\nu=1} = &   P_L^{\nu=1}\bigg\{ 
    \frac{1}{2} \sum_{\vR,\Delta\vR, \eta,\eta'}\delta V_{\eta\eta'}(\Delta\vR) n_{\vR,\eta} n_{\vR+\Delta\vR,\eta'}
    \nonumber\\ 
    & +J\sum_{\vR,\eta,s,s'}\sum_{\Delta\vR = \pm C_{3z}^{\eta}\bm{a}_{M,2}}
       \bigg[\bm{S}_{\vR,\eta}\cdot \bm{S}_{\vR+\Delta\vR,\eta} +\frac{1}{4}n_{\vR,\eta}n_{\vR+\Delta\vR,\eta}\bigg] 
       \bigg\} 
    P_L^{\nu=1}
\end{align}

We  observe that we have coupling in both spin channel $\bm{S}_{\vR,\eta}$ and density channel $n_{\vR,\eta}$. For the spin channel, each spin operator $S^\mu_{\vR,\eta}$ at valley $\eta$ only couples to the spin operator at the same valley. For valley $\eta$, the effective spin coupling is one-dimensional antiferromagnetic coupling along direction $C_{3z}^\eta \bm{a}_{M,2}$. 
For the couplings in the density channel, we have two types of density-density interactions. One has interaction strength $\delta V_{\eta\eta'}(\Delta\vR)$, which comes from projecting the Coulomb repulsions to the Wannier orbital; the other one has interaction strength $J$ and comes from the Schrieffer-Wolff transformation. Since the Coulomb repulsion is proportional to $1/\epsilon$, then $\delta V_{\eta\eta'}(\Delta\vR) \propto 1/\epsilon$ and $J\propto \epsilon$. Therefore, types of interactions have different behaviors as we tune $\epsilon$, and we thus separate the density-density interactions into two parts. 
In the limit where $\delta V_{\eta\eta'}(\Delta\vR) = 0$, we only have 1D coupling for both spin and density channels. As a result, the effective spin model $H_J^{\nu=1}$ possesses an additional $C_{2y}$ symmetry defined as
\begin{align}
    &C_{2y} \bm{S}_{\vR,\eta=0} C_{2y}^{-1} = 
    \bm{S}_{C_{2y}(\vR+\rr_0)-\rr_0,\eta=0} \nonumber\\ 
    &C_{2y} \bm{S}_{\vR,\eta=1} C_{2y}^{-1} = 
    \bm{S}_{C_{2y}(\vR+\rr_1)-\rr_2,\eta=2} \nonumber\\ 
    &C_{2y} \bm{S}_{\vR,\eta=2} C_{2y}^{-1} = 
    \bm{S}_{C_{2y}(\vR+\rr_2)-\rr_1,\eta=1}.
\end{align} 
In addition, each valley and each chain possesses a $U(2)$ symmetry, as discussed in near \cref{eq:additional_U2_symmetry}.


\subsubsection{Ground states of the effective spin model at $\nu=1$}
\label{sec:gnd_state_spin_nu_1}
To solve the $H_J^{\nu=1}$ (\cref{eq:def_spin_model_nu_1}), we utilize the fact that the following density operator is a good quantum number
\begin{align}
    n_{\vR,\eta} =  \sum_{s} \cre{d}{\vR, \eta,s}\des{d}{\vR,\eta,s}
\end{align} 
In other words, 
\begin{align}
    [n_{\vR,\eta}, H_J^{\nu=1}] = 0 
\end{align}
Therefore, we could assume the eigenstates of the $H_J^{\nu=1}$ are also the eigenstates of the density operator $n_{\vR,\eta}$. We could characterize the eigenstate $|\psi\rangle$ of $H_J^{\nu=1}$ via the $\{N_{\vR,\eta}\}$ which are defined as 
\begin{align}
   n_{\vR,\eta} |\psi\rangle = N_{\vR,\eta} |\psi\rangle 
\end{align}
with $N_{\vR,\eta} \in \{0,1\}$.
In addition, we also require there to be only one electron per unit cell which indicates
\begin{align}
    \sum_\eta N_{\vR,\eta}=1,\quad \text{for all }\vR
\end{align}

For a given set of quantum numbers $\{N_{\vR,\eta}\}$ with 
\begin{align}
    N_{\vR,\eta} \in \{0,1\} ,\quad \sum_{\eta}N_{\vR,\eta} = 1 \, 
\end{align}
with the corresponding Hilbert space defined as 
\begin{align}
    \mathcal{H}_{ \{N_{\vR,\eta}\}} = \bigg\{ |\psi\rangle \bigg| \forall (\vR,\eta), n_{\vR,\eta}|\psi\rangle = N_{\vR,\eta}|\psi\rangle \bigg\} \, ,
\end{align}
 we now show that the lowest-energy state within $\mathcal{H}_{ \{N_{\vR,\eta}\}}$ 
 can be obtained. 
For convenience, we consider the system with open boundary conditions. 

\begin{figure}
    \centering
\includegraphics[width=0.8\linewidth]{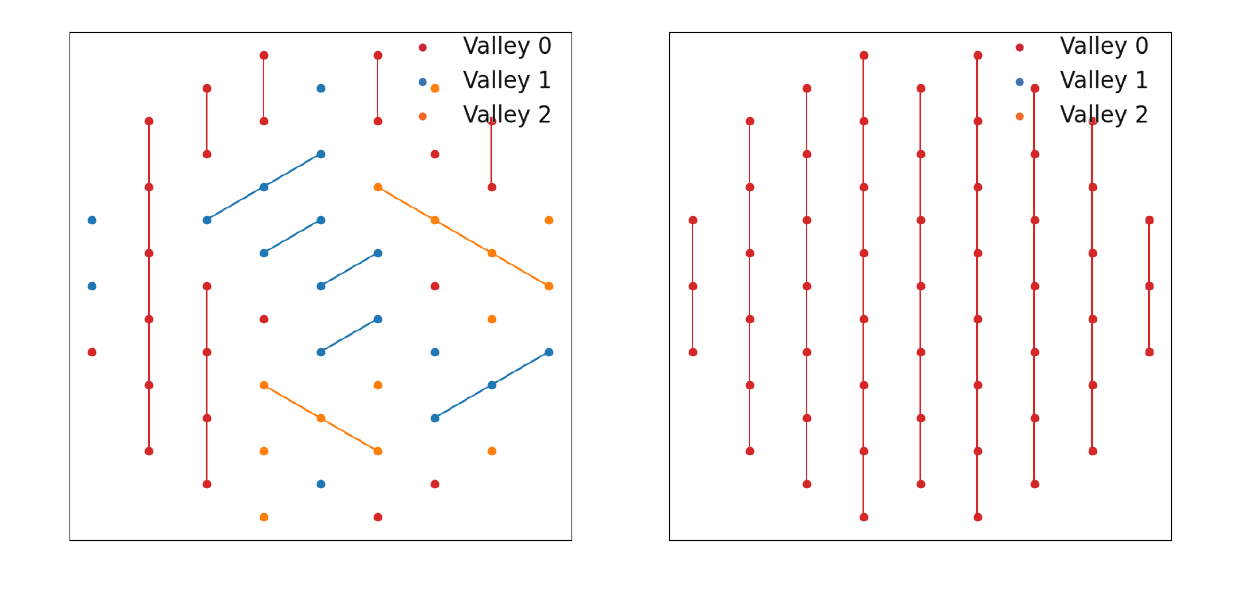}
    \caption{A random $\{N_{\vR,\eta}\}$ configuration (left) and a valley-polarized configuration with $N_{\vR,\eta} = \delta_{\eta,0}$ (right) are illustrated. Each dot represents a unit cell of the system, with red, orange, and blue indicating that the unit cell contains one electron in valley 0, 1, or 2, respectively. The lines connecting dots of the same colors represent spin chains formed in the respective valleys.}
    \label{fig:spin_config}
\end{figure}
To obtain the ground state, we first decomposed our Hamiltonian into two parts: the density-density interactions $H_{Charge}^{\nu=1}$ and the spin-spin $H_{spin}^{\nu=1}$ interactions
\begin{align}
\label{eq:H_J_nu_1_sep}
    &H_J^{\nu=1} = H_{Charge}^{\nu=1} + H_{Spin}^{\nu=1} \nonumber\\ 
    & H_{Charge}^{\nu=1}=
    \frac{1}{2} \sum_{\vR,\Delta\vR, \eta,\eta'}
    \bigg[ \delta V_{\eta\eta'}(\Delta\vR)  +\frac{J\delta_{\Delta\vR , \pm C_{3z}^\eta \bm{a}_{M,2}}\delta_{\eta,\eta'}}{2}\bigg] P_L^{\nu=1}n_{\vR,\eta}n_{\vR+\Delta\vR,\eta'}P_L^{\nu=1}\nonumber\\ 
    &
     H_{Spin}^{\nu=1} =J\sum_{\vR,\eta,s,s'}\sum_{\Delta\vR = \pm C_{3z}^{\eta}\bm{a}_{M,2}}
         P_L^{\nu=1}\bm{S}_{\vR,\eta}\cdot \bm{S}_{\vR+\Delta\vR,\eta} 
    P_L^{\nu=1}
\end{align}
Here, we also comment that the density operator and spin operator commute with each other
\begin{align}
    [n_{\vR,\eta}, \bm{S}_{\vR',\eta'}] = 0
\end{align}
which suggests 
\begin{align}
    [H_{Charge}^{\nu=1}, H_{Spin}^{\nu=1}]  =0 \,.
\end{align}
We now aim to find the state $|\psi\rangle \in \mathcal{H}_{\{N_{\RR,\eta}\}_{\vR,\eta} }$ that minimize the energy of the system
\begin{align}
    E = \langle \psi| H_{Charge}^{\nu=1} |\psi\rangle 
    +\langle \psi| H_{Spin}^{\nu=1} |\psi\rangle 
\end{align}
The energy contribution from the $H_{Charge}^{\nu=1}$ term for a given configuration $\{N_{\vR,\eta}\}$ can be obtained directly, where $\{N_{\vR,\eta}\}$ denotes the set of all $N_{\vR,\eta}$ obtained by varying $\vR$ and $\eta$
\begin{align}
    E_{Charge} = \langle \psi |  H_{Charge}^{\nu=1} |\psi\rangle =  
    \frac{1}{2} \sum_{\vR,\Delta\vR, \eta,\eta'}
    \bigg[ \delta V_{\eta\eta'}(\Delta\vR)  +\frac{J\delta_{\Delta\vR , \pm C_{3z}^\eta \bm{a}_{M,2}}\delta_{\eta,\eta'}}{2}\bigg] N_{\vR,\eta}N_{\vR+\Delta\vR,\eta'}
\end{align}
with $|\psi\rangle \in \mathcal{H}_{ \{N_{\vR,\eta}\}}$.

We now discuss the energy contribution from the spin-coupling term $H_{Spin}^{\nu=1}$. For a given configuration  $\{N_{\vR,\eta}\}$, we first classify the unit cells into different groups. 
Here, we note that each unit cell can be regarded as a ``site" of the system. Although a randomly generated configuration may break translational symmetry, we continue to use the terminology of unit cells, each labeled by a lattice vector \(\vR\).

We notice that, if we only consider $H_{Spin}^{\nu=1}$, spins at $\vR_1,\vR_2$  interact with each other if and only if there exists $\eta$ with $\vR_1-\vR_2= \pm C_{3z}^{\eta}\bm{a}_{M,2}$ and $N_{\vR_1,\eta}=N_{\vR_2,\eta}=1$. Based on this fact, we could use the following rules to classify the unit cells into different groups
\begin{itemize}
    \item Unit cell at $\vR$ and unit cell at $\vR + \Delta\vR$ belong to the same group if there exists $\eta \in \{0,1,2\}$ such that $\Delta\vR = \pm C_{3z}^\eta \bm{a}_{M,2}$ and $N_{\vR+\Delta\vR,\eta}=N_{\vR,\eta}=1$. 
    \item If $\vR$ and $\vR'$ belong to the same group, and $\vR'$ and $\vR''$ belong to the same group, then $\vR$ and $\vR''$ also belong to the same group. 
\end{itemize}
Based on the above rules, for given $\{ N_{\vR,\eta}\}$, we could classify the unit cells into several groups. Only spin operators within the same group interact with each other through $H_{Spin}^{\nu=1}$. 
Two unit cells from different groups are then decoupled if only $H_{Spin}^{\nu=1}$ is considered, since they are not interacting with each other directly and are not coupled through other sites. 

One particular example for a randomly generated $\{ N_{\vR,\eta}\}_{\vR,\eta}$ has been shown in~\cref{fig:spin_config} (left), in which the unit cells that belong to the same group are connected by lines. 
We use $\mathcal{S} = \{(\vR_1,\eta),(\vR_2,\eta),...\} \in \mathcal{N}_{\{N_{\vR,\eta}\}} $ to denote the group formed by a series of unit cells. The number of unit cells of each group will then depend on the values of $\{N_{\vR,\eta}\}$. 
We also comment that a single site could also form a group as shown in ~\cref{fig:spin_config} (left). 
In addition, we also let $\mathcal{N}_{\{N_{\vR,\eta}\}}$ denote the set of all groups of unit cells we constructed for $\{N_{\vR,\eta}\}$. In other words, for any given configuration $\{N_{\vR,\eta}\}$, we classify the unit cells into different groups. The set of all these groups is denoted by $\mathcal{N}_{\{N_{\vR,\eta}\}}$. 
Since unit cells from different groups are decoupled, we can solve each group of unit cells separately. 
Each group of unit cells ($\mathcal{S}$) then corresponds to a spin chain with open boundaries if we only consider $H_{Spin}^{\nu=1}$.

For a spin chain with the open boundary and length $n$, the exact ground state energy could be obtained with Bethe ansatz or DMRG. Without loss of generality, we assume its ground-state energy is 
\begin{align}
    nE_{n} 
\end{align}
where $E_n$ denotes the average energy of each site. 
Therefore the energy contribution from $H_{Spin}^{\nu=1}$ reads
\begin{align}
    E_{Spin} = \sum_{ \mathcal{S} \in \{N_{\vR,\eta}\} } n_{\mathcal{S}}\text{ }E_{n_{\mathcal{S}}}
\end{align}
where $n_{\mathcal{S}}$ denotes the number of unit cells of the spin chain described by set $\mathcal{S}$. $E_{Spin}$ is the summation of the ground state energies for all spin chains we constructed.

Since $[H_{Charge}^{\nu=1},H_{Spin}^{\nu=1}]=0$, the lowest energy we can realized with the state $|\psi\rangle \in \mathcal{H}_{\{N_{\vR,\eta}\}}$ is 
\begin{align}
    E = E_{Charge} +E_{Spin} 
\end{align}
The corresponding state that minimizes the energy for a given configuration of $\{N_{\vR,\eta}\}$ is just the product of the ground states of all the spin chains
\begin{align}
\label{eq:gnd_state_product_nu_1}
    |\{N_{\vR,\eta}\}\rangle = \prod_{\mathcal{S} \in \mathcal{N}_{\{N_{\vR,\eta}\}} }| \text{Ground state of spin chain labeled by $\mathcal{S}$ }\rangle  
\end{align}

In order to obtain the final ground state, we need to determine the configuration $\{N_{\vR,\eta}\}$ that minimizes the energy. This can be done in two limits, either $\delta V_{\eta\eta'}(\Delta\vR)=0$ or $|\Delta V_{\eta\eta'}(\Delta\vR)| \gg J$. 

We focus on filling $\nu=1$. 
We first start from the case with
\begin{align}
    \delta V_{\eta\eta'}(\Delta\vR) = 0 
\end{align}
In this situation, we notice that the energy contributions from $H_{Charge}^{\nu=1}$ (\cref{eq:H_J_nu_1_sep})
 of a spin chain with open boundary conditions and length $n$ is 
\begin{align}
    E_{Charge}^n=2\frac{J}{4}(n-1)=\frac{J}{2}(n-1)
\end{align}
Then the total energy of the system in the limit of $\delta V_{\eta\eta'}(\Delta\vR)=0$ can be written as
\begin{align}
\label{eq:def_energy_of_spin_chain}
    E = \sum_{\mathcal{S} \in \mathcal{N}_{ \{N_{\vR,\eta}\}} } n_{\mathcal{S}} \tilde{E}_{n_{\mathcal{S}}} ,\quad \tilde{E}_{n_{\mathcal{S}}} = E_{n_{\mathcal{S}}} + \frac{(n_{\mathcal{S}}-1)J}{2n_{\mathcal{S}}}
\end{align}
where $\tilde{E}_{n}$ denotes the average energy per unit cell for a spin chain with length $n$. 
Here, $\mathcal{S}$ represents a specific spin chain with length $n_{\mathcal{S}}$ associated with the given charge configuration $\{N_{\vR,\eta}\}$. The set of all such spin chains associated to the charge configuration $\{N_{\vR,\eta}\}$ is denoted by $\mathcal{N}_{\{N_{\vR,\eta}\}}$.

\begin{figure}
    \centering
    \includegraphics[width=0.7\linewidth]{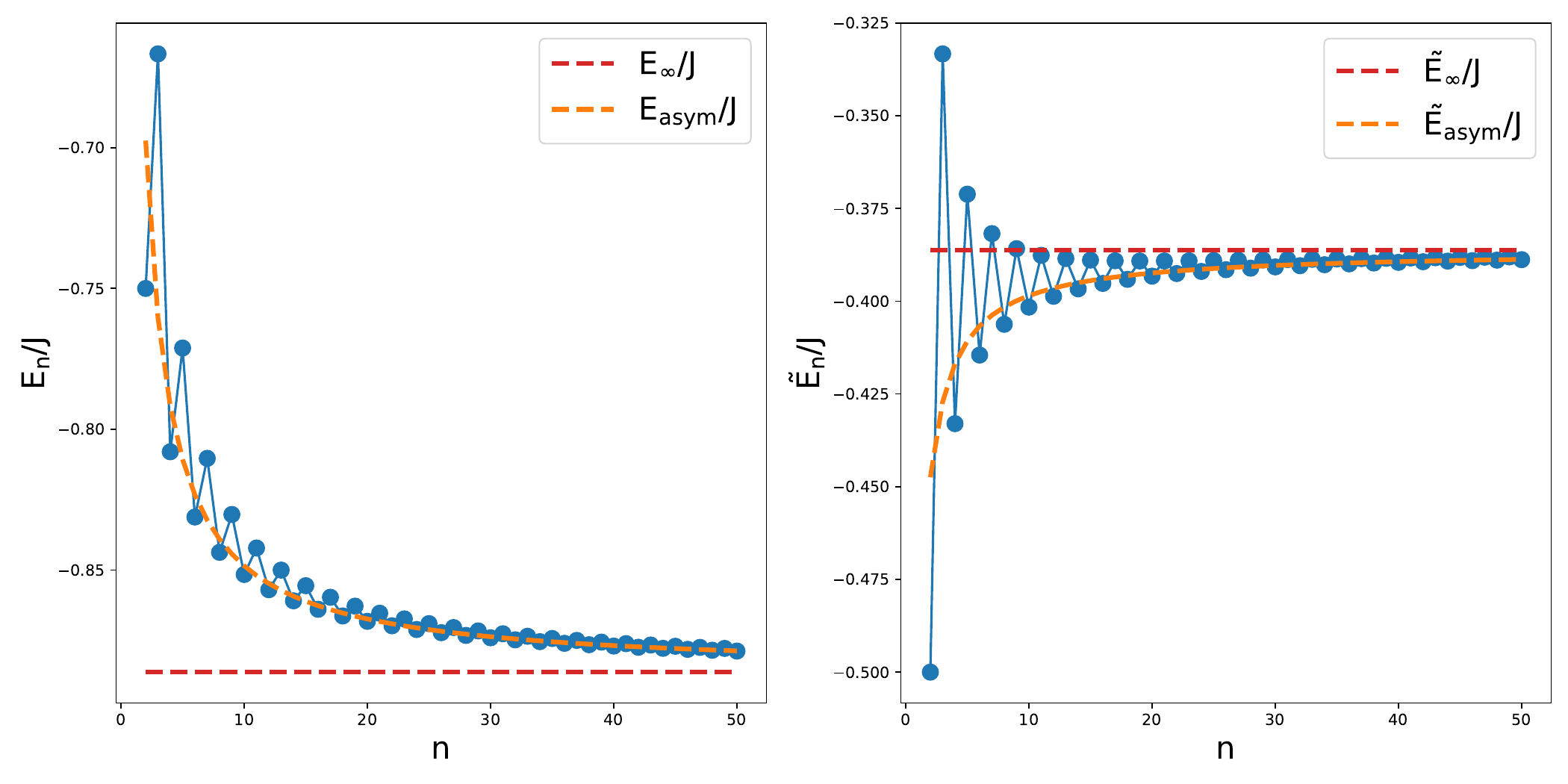}
    \caption{Ground state energy of the open boundary spin chain with length $n$ (calculated from DMRG). The red dashed lines mark the energies for the spin chain with infinite length. The orange dashed line shows the asymptotic behaviors (\cref{eq:energy_infinit_spin_chain}) of ground state energies at large $n$ limit. }
    \label{fig:gnd_energy_spin_chain}
\end{figure}

For $n=1$, $\tilde{E}_{1}=0$. For $n=2$, we find 
\begin{align}
\label{eq:energy_length_2_spin_chain}
    E_{2}/J = -\frac{3}{4},\quad  \tilde{E}_{2}/J= -\frac{1}{2}
\end{align}
The analytical formula of ground state energy at large $n$ can also be obtained from Bethe ansatz\cite{PhysRev.112.309,des1966anisotropic,alcaraz1987surface,batchelor1990surface} which gives 
\begin{align}
\label{eq:energy_infinit_spin_chain}
   E_{n}/J= -0.886 + \frac{0.378}{n} +O(\frac{1}{n^2}). ,\quad \tilde{E}_{n}/J=-0.386 - \frac{0.122}{n}
    +O(\frac{1}{n^2}).
\end{align}
For generic $n$, the numerical value of $E_{n}, \tilde{E}_{n}$ can be obtained using methods such as the Bethe ansatz~\cite{PhysRev.112.309,des1966anisotropic,alcaraz1987surface} and DMRG~\cite{SCHOLLWOCK201196}. In~\cref{fig:gnd_energy_spin_chain}, we perform DMRG calculations and show the evolution of $E_{n}$ and $\tilde{E}_{n}$ as a function $n$

\begin{figure}
    \centering
    \includegraphics[width=0.5\linewidth]{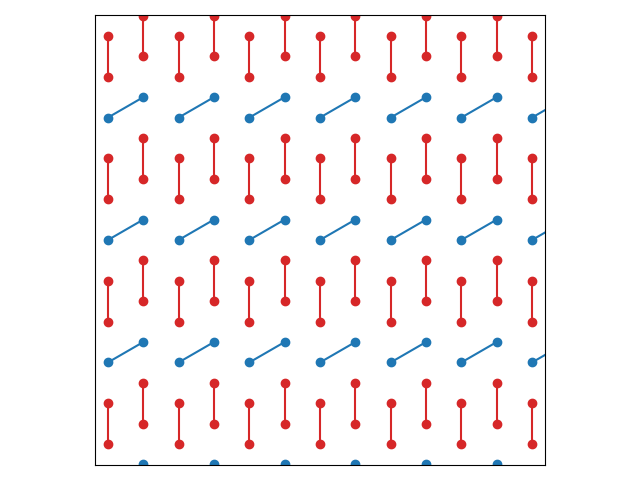}
    \caption{Ground state configuration at the limit of $\delta V_{\eta\eta'}(\Delta\vR)=0$ (\cref{eq:H_J_nu_1_sep}).
     The red, blue, and orange dots indicate that the corresponding sites are occupied by valley indices 0, 1, and 2, respectively. The solid lines in different colors indicate spin-singlet formation within different valleys. }
    \label{fig:Spin_chain_VBL}
\end{figure}

By combining numerical simulations and analytical results in~\cref{eq:energy_infinit_spin_chain}, we conclude that 
\begin{align}
\label{eq:min_E_is_min_E2}
\text{min}_{n}\tilde{E}_{n} = \tilde{E}_2 
\end{align} 
This indicates that the energy per unit cell is minimized when the spin operators form spin chains of length two. Consequently, in the limit $\delta V_{\eta\eta'}(\Delta\vR) = 0$, the total energy of the system is minimized when all open-boundary spin chains have length two. 
In other words, the ground state of the system is realized by a configuration of $\{n_{\vR,\eta}\}$ where all the unit cells form a length-2 spin chain. One specific configuration of the ground state has been shown in~\cref{fig:Spin_chain_VBL} and can be written as
\begin{align}
\label{eq:gnd_config_nu_1_VBL}
    &N_{\vR,\eta} = \delta_{\eta,0},\quad \vR=  
    n_1(2\bm{a}_{M,1}+\bm{a}_{M,2}) +3n_2\bm{a}_{M,2} + \bm{a}_{M,2} \nonumber\\ 
    & N_{\vR,\eta} = \delta_{\eta,0},\quad \vR =  
    n_1(2\bm{a}_{M,1}+\bm{a}_{M,2}) +3n_2\bm{a}_{M,2} + 2\bm{a}_{M,2} \nonumber\\ 
    & N_{\vR,\eta} = \delta_{\eta,0},\quad \vR =  
    n_1(2\bm{a}_{M,1}+\bm{a}_{M,2}) +3n_2\bm{a}_{M,2} +\bm{a}_{M,1}+ 2\bm{a}_{M,2} \nonumber\\ 
    & N_{\vR,\eta} = \delta_{\eta,0},\quad \vR =  
    n_1(2\bm{a}_{M,1}+\bm{a}_{M,2}) +3n_2\bm{a}_{M,2} +\bm{a}_{M,1} +3\bm{a}_{M,2} \nonumber\\ 
     &N_{\vR,\eta} = \delta_{\eta,1},\quad \vR =  
    n_1(2\bm{a}_{M,1}+\bm{a}_{M,2}) +3n_2\bm{a}_{M,2}  \nonumber\\ 
     &N_{\vR,\eta} = \delta_{\eta,1},\quad \vR =  
    n_1(2\bm{a}_{M,1}+\bm{a}_{M,2}) +3n_2\bm{a}_{M,2} + \bm{a}_{M,1} +\bm{a}_{M,2}
\end{align}
with $n_1,n_2 \in \mathbb{Z}$. 

\begin{figure}
    \centering
    \includegraphics[width=1.0\linewidth]{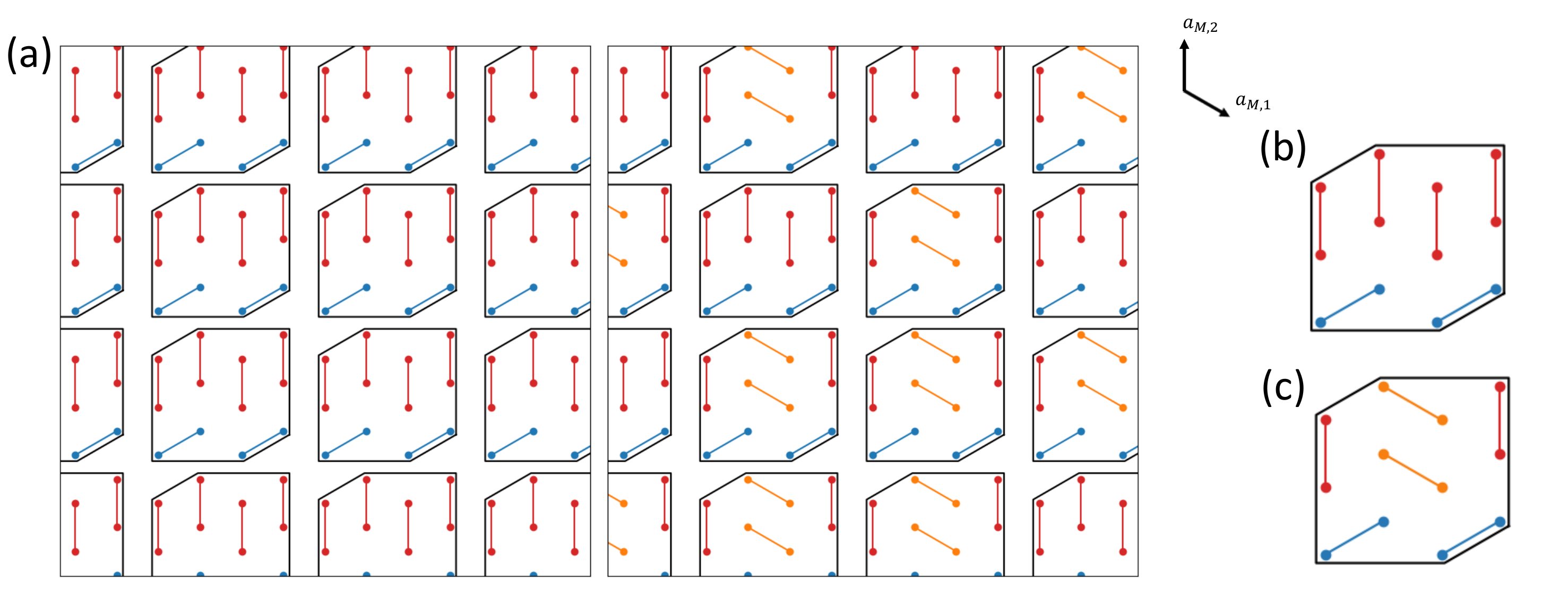}
    \caption{ (a) Two ground state configurations at the limit of $\delta V_{\eta\eta'}(\Delta\vR)=0$ (\cref{eq:H_J_nu_1_sep}). 
    The red, blue, and orange dots indicate that the corresponding sites are occupied by valley indices 0, 1, and 2, respectively. The solid lines in red, blue, and orange indicate spin-singlet formation within valley 0, 1, 2, respectively.
The black solid lines encircle the 12 sites that form the supercell.
Each supercell can independently adopt either of the two configurations shown in (c) and (d), and all resulting states are valid ground states.
    }
    \label{fig:Spin_chain_VBL_SC}
\end{figure}


We note that, the ground states are highly degenerate. However, a given configuration of the ground state can still develop long-range order in spin and valley channels, such as the configuration illustrated in \cref{fig:Spin_chain_VBL}.
We observe that the ground states are the product states of the ground states of the length-2 spin chain. The ground state of a length-2 spin chain is a spin-singlet state. Therefore, the ground states of the whole system could be understood as product states of the spin-singlet states.

In the following, we show the existence of multiple ground states and establish a lower bound on the ground-state entropy.
Starting from the ground state shown in \cref{fig:Spin_chain_VBL}, we can construct additional ground states, which can be characterized by a 12-site supercell. 
Each supercell is labeled by integers $n,m$, and includes the following 12 sites (as marked by solid black lines in \cref{fig:Spin_chain_VBL_SC} (a)) 
\begin{align}
\label{eq:VBL_12_supercell}
     &\vR_1^{n,m} = 2n(2\bm{a}_{M,1}+\bm{a}_{M,2}) + m(3\bm{a}_{M,2}) + \bm{a}_{M,1}+2\bm{a}_{M,2}, \quad \nonumber\\ 
    &\vR_2^{n,m} =2n(2\bm{a}_{M,1}+\bm{a}_{M,2}) + m(3\bm{a}_{M,2}) + \bm{a}_{M,1}+3\bm{a}_{M,2},\quad  \nonumber\\ 
    &\vR_3^{n,m} =(2n+1)(2\bm{a}_{M,1}+\bm{a}_{M,2}) + m(3\bm{a}_{M,2}) + \bm{a}_{M,2}, \quad \nonumber\\ 
    &\vR_4^{n,m}=(2n+1)(2\bm{a}_{M,1}+\bm{a}_{M,2}) + m(3\bm{a}_{M,2}) + 2\bm{a}_{M,2} \nonumber\\ 
    & \vR_5^{n,m} = 2n(2\bm{a}_{M,1}+\bm{a}_{M,2}) + m(3\bm{a}_{M,2})  \nonumber\\
    & \vR_6^{n,m} = 2n(2\bm{a}_{M,1}+\bm{a}_{M,2}) + m(3\bm{a}_{M,2}) + \bm{a}_{M,1} + \bm{a}_{M,2}  \nonumber\\
    & \vR_7^{n,m} = (2n+1)(2\bm{a}_{M,1}+\bm{a}_{M,2}) + m(3\bm{a}_{M,2})  \nonumber\\
    & \vR_8^{n,m} = (2n+1)(2\bm{a}_{M,1}+\bm{a}_{M,2}) + m(3\bm{a}_{M,2}) + \bm{a}_{M,1} + \bm{a}_{M,2}  \nonumber\\
    & \vR_9^{n,m} = 2n(2\bm{a}_{M,1}+\bm{a}_{M,2}) + m(3\bm{a}_{M,2})  + \bm{a}_{M,2} \nonumber\\
    & \vR_{10}^{n,m} = 2n(2\bm{a}_{M,1}+\bm{a}_{M,2}) + m(3\bm{a}_{M,2})  + 2 \bm{a}_{M,2}\nonumber\\ 
     & \vR_{11}^{n,m} = (2n+1)(2\bm{a}_{M,1}+\bm{a}_{M,2}) + m(3\bm{a}_{M,2})  + \bm{a}_{M,1} +2\bm{a}_{M,2}\nonumber\\
    & \vR_{12}^{n,m} = (2n+1)(2\bm{a}_{M,1}+\bm{a}_{M,2}) + m(3\bm{a}_{M,2})  +  \bm{a}_{M,1} +3\bm{a}_{M,2}
\end{align}
Note that the labeling used in \cref{eq:VBL_12_supercell} differs from that in \cref{eq:gnd_config_nu_1_VBL}. 

Each supercell can independently adopt one of two configurations (as shown in \cref{fig:Spin_chain_VBL_SC} (c) and (d)), and all such resulting states remain ground states.  
Written explicitly, we have the following two configurations
\begin{align}
    & N_{\vR_{1}^{n,m},\eta} =  N_{\vR_{2}^{n,m},\eta} = 
     N_{\vR_{3}^{n,m},\eta} = 
      N_{\vR_{4}^{n,m},\eta} =\delta_{\eta,\eta^{n,m}} \nonumber\\ 
    &N_{\vR_{5}^{n,m},\eta} =  N_{\vR_{6}^{n,m},\eta} = 
     N_{\vR_{7}^{n,m},\eta} = 
      N_{\vR_{8}^{n,m},\eta} = \delta_{\eta,1} \nonumber\\ 
    &N_{\vR_{9}^{n,m},\eta} =  N_{\vR_{10}^{n,m},\eta} = 
     N_{\vR_{11}^{n,m},\eta} = 
      N_{\vR_{12}^{n,m},\eta} = \delta_{\eta,0}
\end{align}
where $\eta^{n,m}$ could be either $0$ or $2$, characterizing the two possible configurations for supercell labeled by $n,m$ (see \cref{fig:Spin_chain_VBL_SC} (c) and (d)). 
In practice, we fix the valley indices of the sublattice sites $\vR_{5,\dots,12}^{n,m}$. 
However, within each supercell labeled by $n,m$, the valley indices of the sites $\vR_{1,2,3,4}^{n,m}$, which we denote by $\eta^{n,m}$, are all the same and take values either $0$ or $2$. 
However, the valley number $\eta^{n,m}$ can vary from one supercell to another. The resulting states are all ground states, since each corresponds to a product state of length-2 spin-singlet states. 

We now discuss the ground-state degeneracy arising from the two configuration choices shown in \cref{fig:Spin_chain_VBL_SC} (c) and (d). For a system with $N$ moir\'e unit cells, there are $N/12$ supercells, each of which can independently adopt either configuration (\cref{fig:Spin_chain_VBL_SC} (c) or (d)). This yields a total ground-state degeneracy of $2^{N/12}$. 
Consequently, this leads to a lower bound on the zero-temperature entropy per moiré unit cell, given by
\begin{align}
    S \geq \frac{k_B}{N} \log 2^{N/12} = k_B \frac{\log 2}{12}\approx 0.059k_B.
\end{align} 

\begin{figure}
    \centering
    \includegraphics[width=0.7\linewidth]{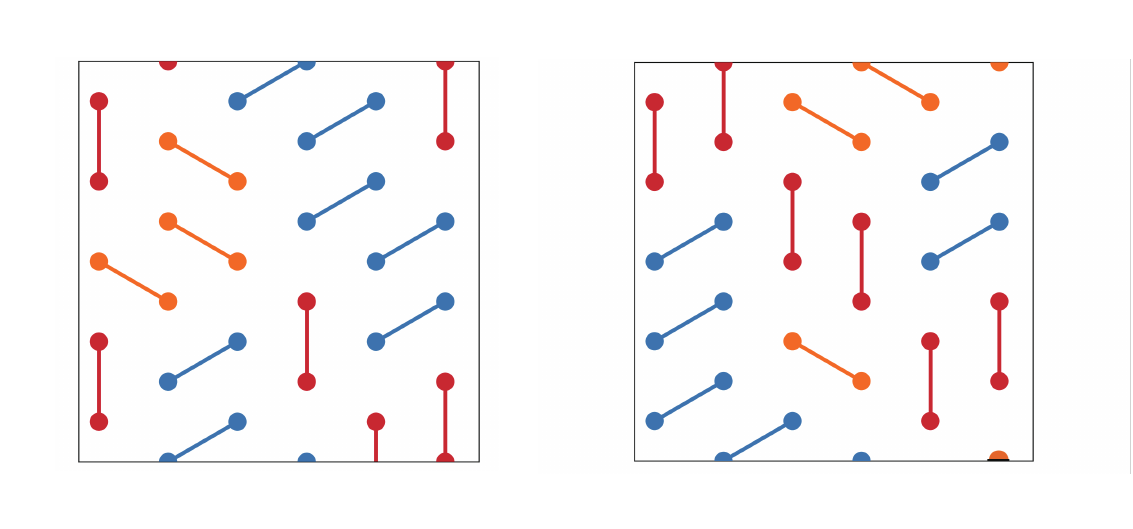}
    \caption{Two random ground state configurations of filling $\nu=1$ without translational symmetry.
    }
    \label{fig:random_config}
\end{figure}

Finally, we note that many other ground-state configurations exist. Two randomly generated ground state configurations are also given in \cref{fig:random_config}. 
In \cref{sec:entropy_of_dimer}, we provide a detailed discussion on the entropy per moir\'e unit cell of this dimer phase (where all spins tend to form length-2 spin chain or dimer). Approximately, the entropy is around 
$S \sim 0.307k_B$, which is much larger than the lower bound we found here. 
We also comment that, such degeneracy of the ground states may be lifted if high-order terms in the $t/V$ are included.

We now discuss the effect of non-zero $\delta V_{\eta\eta'}(\Delta\vR)$. 
In the case where $\delta V_{\eta,\eta'}(\Delta\vR) = 0$, we have shown that configurations in which every moir\'e unit cell forms a length-2 spin chain minimize the energy. 
With a finite $\delta V_{\eta\eta'}$, the dimer states are not necessarily ground states, as they may not be energetically favored by $\delta V_{\eta,\eta'}$. 
Numerically, we compare the energies of all charge configurations at filling $\nu=1$ with translational symmetry defined by a $3 \times 3$ supercell. 
We find that the energy of the $\delta V_{\eta\eta'}(\Delta\vR)$ term 
\begin{align}
    E_{\delta V} = \frac{1}{2} \sum_{\vR,\Delta\vR, \eta,\eta'}
\delta V_{\eta\eta'}(\Delta\vR)  N_{\vR,\eta}N_{\vR+\Delta\vR,\eta'}
\end{align}
will be minimized by developing valley-polarized states characterized by
\begin{align}
    N_{\vR,\eta} = \delta_{\eta,\eta_0},\quad \forall \vR 
    \label{eq:valley_pol_nu_1_N_config}
\end{align}
with a given valley number $\eta_0$ (see \cref{fig:spin_config} right panel, for $\eta_0=0$). 
We note that $\delta V_{\eta\eta'}(\Delta\vR)$ is always repulsive ($\delta V_{\eta\eta'}(\Delta\vR) > 0$), as it originates from the projected Coulomb interaction. This term captures the charge repulsion between electrons, and thus tends to favor more uniform charge distributions. Note that the Wannier centers of the three valleys are close to each other but do not exactly coincide. As a result, one natural way to achieve a more uniform charge distribution is to develop a valley-polarized state, as we have also confirmed numerically. 

We note that the $\delta V_{\eta\eta'}(\Delta\vR)$ term energetically favors valley polarization, while the Heisenberg-type terms proportional to $J$ in Eq.~(\cref{eq:def_spin_model_nu_1}) favor the dimer phase. These two contributions compete: if the energy gain from valley polarization due to $\delta V_{\eta\eta'}(\Delta\vR)$ exceeds the energy cost from the $J$-terms, the system tends to develop valley polarization. However, since the exact ground state at finite $\delta V_{\eta\eta'}(\Delta\vR)$ is hard to determine analytically, it is difficult to find the threshold value of $\delta V_{\eta\eta'}(\Delta\vR)$ at which the valley-polarized state becomes the ground state. We expect, with large enough $|\delta V_{\eta\eta'}(\Delta\vR)|$ ($\gg J$), system tends to develop a valley-polarized state. 
After the valley polarization, the system forms infinite spin chains, as illustrated in the right panel of \cref{fig:spin_config}. 
The ground state of an infinite spin chain is a quantum paramagnetic state, where long-range spin order is suppressed by strong quantum fluctuations.

In summary, the system exhibits different ground states depending on the interaction strength. In the limit of small $\delta V_{\eta\eta'}(\Delta\vR)$, it forms a highly degenerate dimer phase. In contrast, for large $\delta V_{\eta\eta'}(\Delta\vR)$, the system favors a valley-polarized paramagnetic state. 

\subsubsection{Comparison between exact solutions of spin model and Hartree-Fock simulations} 
\label{sec:compare_hf_exact}
In this section, we discuss the consistency between our exact results in the spin model and our HF results. We first note that, the spin coupling $J$ and the density-density interaction $\delta V_{\eta,\eta'}(\Delta\vR)$ scale differently as we tune $\epsilon$. In particular, we have 
\begin{align}
    J \propto \epsilon,\quad \delta V_{\eta,\eta'}(\Delta\vR) \propto \frac{1}{\epsilon}. 
\end{align}
Therefore, within the strong coupling region, a relatively small $1/\epsilon$ corresponds to the limit of small $\delta V_{\eta,\eta'}(\Delta\vR)$, while a relatively large $1/\epsilon$ corresponds to the limit of dominant $\delta V_{\eta,\eta'}(\Delta\vR)$. Here, we also comment that $\delta V_{\eta,\eta'}(\Delta\vR)$ is much smaller than the on-site Hubbard interactions (see \cref{eq:sym_break_delta_V}). Even though we focus on the strong coupling limit (with large on-site Hubbard interaction), the energy contributions from $\delta V_{\eta\eta'}(\Delta\vR)$ could still be much smaller than the Heisenberg term that is proportional to $J$, when a relatively large $\epsilon$ is considered. 
Within the Hartree-Fock simulations, we indeed observe that a valley polarization develops at a relatively large $12/\epsilon$ with $12/\epsilon \gtrsim 0.68 $ for $\theta= 3.89^\circ$ (see also \cref{sec:HFphase_diagram}). 

We now discuss the case where the Heisenberg coupling dominates and the system does not develop valley polarization. 
In this region, HF simulations have found an AFM ground state without valley polarization. 
For the spin model (\cref{eq:def_spin_model_nu_1}), 
exact solutions demonstrate a degenerate dimer phase in this limit ($\delta V_{\eta\eta'}(\Delta\vR)) \approx 0$). We first notice that the dimer state of the spin models is formed by the following spin-singlet state 
\begin{align}
    |\uparrow,\downarrow\rangle - |\downarrow,\uparrow\rangle  \, .
\end{align}
Such spin-singlet formation cannot be accurately captured by the HF approximation, within which the ground states are represented by Slater determinant states. However, the antiferromagnetic interactions that drive spin-singlet formation will favor an antiferromagnetic order, which has been observed in HF simulations. 
\\

To better understand the Hartree-Fock results in the strong coupling regime, we now analyze the corresponding spin model by treating spin operators as classical spins. This approach is expected to yield consistent results with Hartree-Fock simulation of the electronic model, as both approximations similarly underestimate quantum fluctuations of the spin degrees of freedom. 
The energy of the spin model can be approximately written as (at $\delta V_{\eta\eta'}(\Delta\vR)=0$ limit)
    \begin{align} \label{app:eq:hfeffectivespin}
    E_J^{MF} = &  +J\sum_{\vR,\eta,s,s'}\sum_{\Delta\vR = \pm C_{3z}^{\eta}\bm{a}_{M,2}}
       \bigg[\langle \bm{S}_{\vR,\eta}\rangle \cdot \langle \bm{S}_{\vR+\Delta\vR,\eta} \rangle +\frac{1}{4}\langle n_{\vR,\eta}\rangle \langle n_{\vR+\Delta\vR,\eta}\rangle \bigg] 
\end{align}
whee $\langle \bm{S}_{\vR,\eta}\rangle$ and $\langle n_{\vR,\eta}\rangle$ denote the expectation value of the spin operator and density operator, respectively, of site $\vR$ and valley $\eta$.
We focus on the filling $\nu=1$ where each site is filled with one electron. If site $\vR$ is filled with one electron of valley $\eta$, we have $|\langle \bm{S}_{\vR,\eta'} \rangle | =\delta_{\eta,\eta'}1/2$ and $\langle n_{\vR,\eta'}\rangle =\delta_{\eta,\eta'}$. $2\langle \bm{S}_{\vR,\eta'}\rangle$ is then a unit vector describing the orientation of the spin. We then aim to find the spin and charge configurations, such that the energy $E_J^{MF}$ is minimized. 

The spin and charge configurations that minimize the $E_J^{MF}$ can be obtained by minimizing each term of $E_J^{MF}$. If a given configuration minimizes each term of the $E_J^{MF}$, then it minimizes the $E_J^{MF}$.  
Considering a bond $(\vR,\eta, \vR+\Delta\vR,\eta)$ (where $\Delta\vR = \pm C_{3z}^\eta \bm{a}_{M,2}$) with 
\begin{align}
\label{eq:h_bond_strong_coupling}
   E_{\vR,\eta,\vR+\Delta\vR,\eta}= J\bigg[\langle \bm{S}_{\vR,\eta}\rangle \cdot \langle \bm{S}_{\vR+\Delta\vR,\eta} \rangle +\frac{1}{4}\langle n_{\vR,\eta}\rangle \langle n_{\vR+\Delta\vR,\eta}\rangle \bigg] 
\end{align}
Since each site is filled with one electron, there will be four possibilities: 
(1) $\langle n_{\vR,\eta}\rangle =\langle n_{\vR+\Delta\vR,\eta}\rangle = 1$;
(2) 
$\langle n_{\vR,\eta}\rangle =\langle n_{\vR+\Delta\vR,\eta}\rangle = 0$;
(3) $\langle n_{\vR,\eta}\rangle =1,\langle n_{\vR+\Delta\vR,\eta}\rangle = 0$; (4)
$\langle n_{\vR,\eta}\rangle =0,\langle n_{\vR+\Delta\vR,\eta}\rangle = 1$. 
We note that the total filling of each site is $1$. Therefore, each valley can be either empty or singly occupied. 
We now discuss each situation. 
\begin{itemize}
    \item  (1) $\langle n_{\vR,\eta}\rangle =\langle n_{\vR+\Delta\vR,\eta}\rangle = 1$. Due to the antiferromagnetic coupling between spin operator, we can minimize the energy by taking 
\begin{align}
    \langle S^\mu_{\vR,\eta}\rangle  = -\langle S^\mu_{\vR+\Delta\vR,\eta}\rangle 
    = \frac{1}{2}\bm{e}^\mu 
\end{align}
with $\bm{e}^{\mu}$ an arbitrary unit vector. In other words, a short-range antiferromagnetic correlation developed between site $\vR$ and $\vR+\Delta\vR$.
We find the corresponding energy contributions are $E_{\Delta\vR,\eta,\vR+\Delta\vR,\eta} = 0$. 
\item 
(2) $\langle n_{\vR,\eta}\rangle =\langle n_{\vR+\Delta\vR,\eta}\rangle = 0$. Then we also have $\langle \bm{S}_{\vR,\eta}\rangle =\langle \bm{S}_{\vR,\eta}\rangle  =0 $. The corresponding energy reads $E_{\Delta\vR,\eta,\vR+\Delta\vR,\eta} = 0$. 
\item (3) $\langle n_{\vR,\eta}\rangle=1 ,\langle n_{\vR+\Delta\vR,\eta}\rangle = 0$. Then we have $\langle \bm{S}_{\vR+\Delta\vR,\eta} \rangle =0$. For $\langle \bm{S}_{\vR,\eta}\rangle$, we can let $\langle \bm{S}_{\vR,\eta}\rangle=\bm{e}^\mu/2$ with $\bm{e}^\mu$ an arbitrary unit vector. The corresponding energy reads $E_{\Delta\vR,\eta,\vR+\Delta\vR,\eta} = 0$. 
\item 
(4) we consider $\langle n_{\vR,\eta}\rangle=0 ,\langle n_{\vR+\Delta\vR,\eta}\rangle = 1$. This is equivalent to the case of $\langle n_{\vR,\eta}\rangle=1 ,\langle n_{\vR+\Delta\vR,\eta}\rangle = 0$.
\end{itemize}
In summary, we conclude that the smallest value of $E_{\vR,\eta,\RR+\Delta\vR,\eta}$ (\cref{eq:h_bond_strong_coupling})
is $0$. 
If a spin and charge configurations are chosen such that for all $(\vR,\eta,\RR+\Delta\vR,\eta)$ (with $\Delta\vR = \pm C_{3z}^\eta \bm{a}_{M,2}$), $E_{\vR,\eta,\RR+\Delta\vR+\eta} =0$, the corresponding spin and charge configurations are the ground state in the classical limit (where the Hamiltonian of the spin model has been treated at the mean-field level). 
Indeed, the antiferromagnetic ground state (\cref{fig:dimer_hf} (a)) 
observed in the Hartree-Fock (HF) simulations minimizes $E_J^{MF}$.
The nature of the antiferromagnetic spin order (\cref{fig:dimer_hf} (a)) that appeared in the Hartree-Fock simulation has been discussed near \cref{eq:def_S_q_eta} and \cref{fig:afm_form_fac} (a). 
However, similar to the exact solutions of the spin model in the strong-coupling limit, there could be multiple configurations that minimize $E_J^{MF}$, which then leads to degenerate ground states in the Hartree-Fock simulations. However, since Hartree-Fock simulations include all the long-range interactions and hopping that are not included in \cref{eq:h_bond_strong_coupling}, the degeneracy will be lifted, and the ground state shown in \cref{fig:dimer_hf} (a) is the one with the lowest energy for relatively small $1/\epsilon$ ($12/\epsilon \lesssim 0.68$ at $\theta=3.89^\circ$). 

For relatively large $1/\epsilon$ in the strong coupling limit, HF simulations have found an AFM ground state with valley polarization. In the spin model, a valley-polarized spin liquid state is observed in this limit. Again, the HF simulation cannot capture the spin liquid phase. However, the HF simulation does capture the antiferromagnetic order induced by the Heisenberg coupling (\cref{fig:dimer_hf} (b)). 
The nature of the antiferromagnetic spin order (\cref{fig:dimer_hf} (b)) that appeared in the Hartree-Fock simulation has been discussed near \cref{eq:def_S_q_eta} and \cref{fig:afm_form_fac} (b). 


 Finally, in the flat-band regime, which corresponds to the blue region of the Hartree-Fock phase diagram in \cref{fig:phase_diagram_AA_mom}, the effective Heisenberg coupling is strongly suppressed due to the large interactions. 
The system can be treated as a flat-band system with zero kinetic energy.
The spin model cannot correctly describe the system since other interactions, such as ferromagnetic spin-spin coupling (\cref{eq:spin_spin_coupling}) from the projected Coulomb repulsions, instead of the second-order processes become relevant.  As we discussed near \cref{eq:gnd_at_ultra_strong}, a ferromagnetic valley polarized state appears in such a limit.

\begin{figure}
    \centering
    \includegraphics[width=0.5\linewidth]{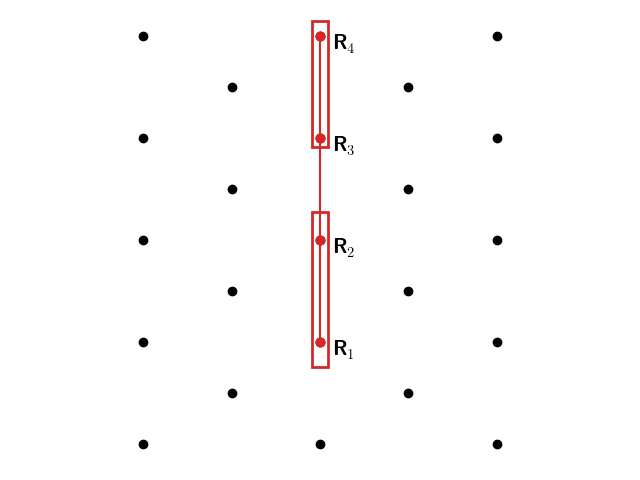}
    \caption{Violation of the dimer constraints. If sites $\RR_1$ and $\RR_2$ form a dimer, then sites $\RR_1,\RR_2$ are all filled by valley $0$. 
    If sites $\RR_3$ and $\RR_4$ form a dimer, 
    then sites $\RR_3,\RR_4$ are filled by valley $0$.
    Since $\RR_1,\RR_2,\RR_3,\RR_4$ are all formed by valley $0$, then instead of forming two dimers, $\RR_1,\RR_2,\RR_3,\RR_4$ will form a length-4 chain. 
    }
    \label{fig:violation_dimer_constraint}
\end{figure}

\subsection{Entropy of the dimer phase at $\nu=1$}
\label{sec:entropy_of_dimer}
In this section, we provide a more detailed discussion on the ground state degeneracy (or zero-temperature entropy) of the dimer phase at $ \nu = 1 $ and $ \delta V_{\eta\eta'}(\Delta\vR) = 0 $. 
Our goal is to determine the number of configurations that satisfy the following constraints 
\begin{itemize}
    \item Each site forms a dimer (or a spin-singlet) with only one of its nearest neighbor sites.
    \item Dimer constraint:  
\refstepcounter{mybox}
\begin{tcolorbox}[colback=white, colframe=black, boxrule=0.5pt]
\label{box:dimer_constraint}
    \textbf{Box~\themybox.} For a linear arrangement of four consecutive sites labeled by $\vR_1,\vR_2,\vR_3,\vR_4$ that are directly adjacent to each other, if $\vR_1,\vR_2$ form a dimer, then $\vR_3,\vR_4$ cannot form a dimer. 
   \end{tcolorbox}
\end{itemize}
An illustration of the violation of the second constraint is shown in \cref{fig:violation_dimer_constraint}. If $\vR_1$ and $\vR_2$ form a dimer (highlighted by the red rectangle), and $\vR_3$ and $\vR_4$ form another dimer (also highlighted by a red rectangle), then all four sites, $\vR_1$, $\vR_2$, $\vR_3$, and $\vR_4$, belong to valley $0$. Then $\vR_1,\vR_2,\vR_3,\vR_4$ form a length-4 spin chain.

If only the first constraint is applied, our model reduces to the conventional dimer model on a triangular lattice \cite{kasteleyn1963dimer,kasteleyn1961statistics, PhysRevB.66.214513}, where the zero-temperature entropy can be calculated directly. 
The second constraint enforces our earlier statement that long-range spin chains (with lengths greater than 2) are prohibited (see discussion near \cref{eq:min_E_is_min_E2}). In this section, we present an analytical estimation of the ground-state degeneracy for our spin model at $ \delta V_{\eta\eta'}(\Delta\vR) = 0 $ and provide an exact numerical evaluation of the ground-state degeneracy for finite-size systems.

\subsubsection{Estimation of the entropy of dimer phase}
We first calculate the number of dimer configurations without the dimer constraint (\cref{box:dimer_constraint}). This is equivalent to calculating the ground state degeneracy of the conventional dimer model on the triangular lattice 
\cite{kasteleyn1963dimer, kasteleyn1961statistics, PhysRevB.66.214513}.
Even though our model features three valleys per site, each valley can form a dimer only with two of its nearest neighbors, specifically along the direction $C_{3z}^\eta \bm{a}_{M,2}$ for valley $\eta$. In the conventional dimer model on a triangular lattice, each site can form a dimer with any one of its six nearest neighbors. Therefore, when all three valleys are combined, and in the absence of dimer constraint (\cref{box:dimer_constraint}), our model becomes equivalent to the conventional dimer model.

For the conventional dimer model, a dimer configuration is defined via the following requirement
\begin{itemize}
    \item Within each dimer configuration, each site must be connected to exactly one of its nearest neighbor sites.
\end{itemize} 
The number of dimer configurations could be calculated by observing the equivalence between the dimer model and the free-fermion system, as discussed in Ref.~\cite{PhysRevB.66.214513}. Notice that in our model, the valley on each site is determined by the direction of the bond it connects to. We also review the mapping between the dimer model and the free-fermion system in \cref{sec:dimer_of_triangular}.

\begin{figure}
    \centering
    \includegraphics[width=0.6\linewidth]{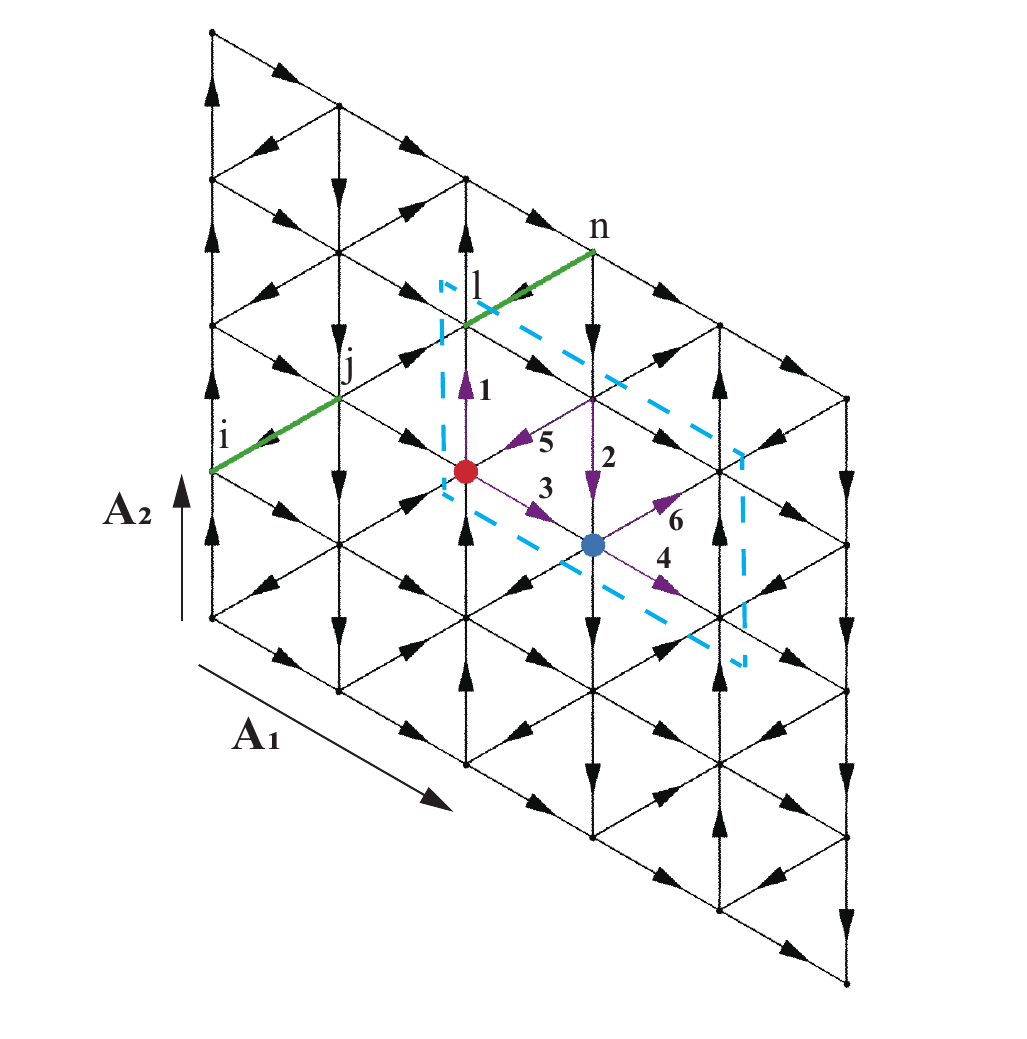}
    \caption{Illustration of the dimer model on the triangular lattice. 
     We label the doubled unit cell introduced in \cref{eq:define_super_lattice_vector_for_dimer} with blue dashed lines.
    The six bonds introduced in \cref{eq:introduce_of_bond_op},
    for the unit cell marked by blue dashed lines,
    are marked in purple with indices characterizing six bonds. 
    For the ground-state configuration of our model, two bonds marked by green cannot both form a spin-singlet bond due to the dimer constraint (\cref{box:dimer_constraint}) we imposed.
    The arrows characterize the matrix $M_{(\RR,i), (\RR',j)}$ (in \cref{eq:define_S_fermion_for_dimer}) and are discussed below \cref{eq:define_S_fermion_for_dimer}. 
    }
    \label{fig:illustration_dimer_model}
\end{figure}

We follow the procedure introduced in Refs.~\cite{kasteleyn1963dimer, kasteleyn1961statistics, PhysRevB.66.214513}. The mapping can only be established by considering an enlarged unit cell, since the corresponding fermionic system only has translational symmetry with respect to a doubled unit cell.
We let the new lattice vectors of the enlarged unit cell be 
\begin{align}
\label{eq:define_super_lattice_vector_for_dimer}
    \bm{A}_{1} = 2\bm{a}_{M,1},\quad \bm{A}_2 = \bm{a}_{M,2} 
\end{align} 
Two sublattices within each enlarged unit cell are located at 
\begin{align}
    \bm{r}_1 =1/2 \bm{A}_1 ,\quad \bm{r}_2 = \bm{0}
\end{align}
For each site, we could introduce a fermionic operator 
\begin{align}
    \psi_{\RR,\alpha} 
\end{align}
where $\RR \in \mathbb{Z}\bm{A}_1+\mathbb{Z}\bm{A}_2$ denotes the positions of the enlarged unit cell, and $\alpha =1,2$ denotes sublattice index. 
The fermionic operators satisfy 
\begin{align}
     \psi_{\RR,\alpha} =  \psi_{\RR,\alpha}^\dag,\quad \{ \psi_{\RR,\alpha},\psi_{\RR',\gamma}\} = 0 
     \label{eq:def_const_psi_commute}
\end{align}
The number of dimer configurations $ \Omega_{\text{dimer}} $ can be obtained by calculating the partition function $ Z $ of a free fermion system \cite{kasteleyn1963dimer, kasteleyn1961statistics}
\begin{align}
  \Omega_{dimer} = Z = \int D[\psi_{\RR,\alpha}]e^{-S}
  \label{eq:Omega_dimer_equal_Z}
\end{align}
where the action of the fermionic system reads
\begin{align}
\label{eq:define_S_fermion_for_dimer}
    S =- \frac{1}{2}\sum_{\RR,\alpha,\RR',\gamma}\psi_{\RR,\alpha} M_{(\RR,\alpha),(\RR',\gamma)} \psi_{\RR',\gamma}
\end{align} 
Here, the matrix $M_{(\RR,\alpha),(\RR',\gamma)}$ is defined according to \cref{fig:illustration_dimer_model}: on the triangular lattice, we assign arrows to links (nearest-neighbor bonds).
If two sites $(\RR,\alpha)$ and $(\RR',\gamma)$ are connected by an arrow pointing from $(\RR,\alpha)$ to $(\RR',\gamma)$, we let $M_{(\RR',\gamma),(\RR,\alpha)} = -M_{(\RR,\alpha),(\RR',\gamma)}=1$. 
Such a choice of $M$ matrix has been discussed in Refs.\cite{kasteleyn1963dimer, kasteleyn1961statistics, PhysRevB.66.214513}, and we are able to exactly solve the action once the $M$ matrix is fixed. 
As discussed in Refs.\cite{kasteleyn1963dimer, kasteleyn1961statistics, PhysRevB.66.214513}, the choice of $M$ matrix is not unique. However, certain requirements need to be satisfied by the $M$ matrix. We consider an arbitrary closed loop $L$ with length $|L|$ formed by a series of links 
\ba 
L: (\RR_1,\alpha_1)\rightarrow (\RR_2,\alpha_2) \rightarrow ... \rightarrow (\RR_{|L|},\alpha_{|L|})\rightarrow (\RR_1,\alpha_1)
\ea 
where $(\RR_i,\alpha_i)$ and $(\RR_{(i+1)\text{mod} |L|},\alpha_{(i+1)\text{mod}|L|})$ are nearest-neighbor sites. We require that, if the length of the closed loop $|L|$ is even, then 
\ba 
\prod_{i=1}^{|L|} M_{(\RR_i,\alpha_i), (\RR_{(i+1)\text{mod}|L|}, \alpha_{(i+1)\text{mod}|L|} } = -1 
\ea 
The $M$ matrix defined with respect to the \cref{fig:illustration_dimer_model} satisfies the above-mentioned requirement. 
In \cref{sec:dimer_of_triangular}, we review the proof of this exact mapping, and calculate the dimer-dimer correlation. Readers who are interested can refer to \cref{sec:dimer_of_triangular} first. 

For our current model, not all dimer configurations are ground state configurations due to the dimer constraint (\cref{box:dimer_constraint}).
 To obtain the mathematical formula of the dimer constrain, we attached the following variable to each nearest-neighbor bond  
\begin{align}
    \phi_{\RR,\alpha}
    \label{eq:introduce_of_bond_op}
\end{align}
with $\RR \in \mathbb{Z} \bm{A}_1 +\mathbb{Z}\bm{A}_2$ and $\alpha=1,2,\cdots,6$
(\cref{eq:define_super_lattice_vector_for_dimer}). $\RR$ labels the position of each enlarged unit cell, and $\alpha$ labels six nearest-neighbor bonds within each enlarged unit cell.  
The choices of the six bonds for a specific unit cell have been marked in purple in \cref{fig:illustration_dimer_model}.
For each dimer configuration $ c $, we let $ \phi_{\RR,\alpha}^c = 1 $ if the corresponding bond hosts a dimer (green lines in \cref{fig:illustration_dimer_model}), and $ \phi_{\RR,\alpha}^c = 0 $ otherwise. We note that the value of $\phi_{\RR,\alpha}^c$ is independent of the bond orientation. 
Then, only configurations satisfying 
\begin{align}
\label{eq:const_dimer}
    &\phi^c_{\RR,1}\phi^c_{\RR+2\bm{A}_2,1}= 0 ,\quad \phi^c_{\RR,2}\phi^c_{\RR+2\bm{A}_2,2}=0 \nonumber\\ 
    & \phi^c_{\RR,3}\phi^c_{\RR+\bm{A}_1,3}=0,\quad \phi^c_{\RR,4}\phi^c_{\RR+\bm{A}_1,4}=0\nonumber\\ 
    &\phi^c_{\RR,5}\phi^c_{\RR+\bm{A}_1+2\bm{A}_2,5}=0,\quad \phi^c_{\RR,6}\phi^c_{\RR+\bm{A}_1+2\bm{A}_2,6}=0
\end{align}
for all $\RR \in \mathbb{Z} \bm{A}_1 + \mathbb{Z} \bm{A}_2$ are ground states of our spin model. To see how the constraints work, we consider a linear arrangement of four consecutive sites labeled on edge $i,j$ and $l,n$ separately that are directly adjacent to each other (\cref{fig:illustration_dimer_model}), forming a straight line.
Here, we use indices $i,j,l,n$ to label both the position $\RR$ and sublattice index $\alpha$. 
We let $\phi_{ij}$ and $\phi_{ln}$ denote the dimer formed by $i,j$ and $l,n$ respectively.
In \cref{fig:illustration_dimer_model}, we show an illustration of $ \phi_{ij}, \phi_{ln} $, which are marked in green. To prevent the formation of long-range spin chains, we require at least one of the bonds to be equal to $0$. In other words, we require $\phi_{ij} \phi_{ln} = 0 $. This is exactly the constraints introduced in ~\cref{eq:const_dimer}. A similar violation of the valley constraint has also been shown in \cref{fig:violation_dimer_constraint}. 

The ground-state degeneracy of the model can then be described by 
\begin{align}
    \Omega = &\sum_{c} \prod_{\RR} \bigg\{ \bigg[ 1 - \phi^c_{\RR,1}\phi^c_{\RR+2\bm{A}_2,1}
 \bigg]
 \bigg[ 1 - \phi^c_{\RR,2}\phi^c_{\RR+2\bm{A}_2,2}
 \bigg]
 \bigg[ 1 - \phi^c_{\RR,3}\phi^c_{\RR+\bm{A}_1,3}
 \bigg] \nonumber\\ 
 &
 \bigg[ 1 -\phi^c_{\RR,4}\phi^c_{\RR+\bm{A}_1,4}
 \bigg]
 \bigg[ 1 - \phi^c_{\RR,5}\phi^c_{\RR+\bm{A}_1+2\bm{A}_2,5}
 \bigg]
 \bigg[ 1 - \phi^c_{\RR,6}\phi^c_{\RR+\bm{A}_1+2\bm{A}_2,6}
 \bigg]\bigg\} ,
\end{align} 
where only the configurations that satisfy constraints are counted. We next aim to express $\Omega$ via the path integral form using the action defined in \cref{eq:define_S_fermion_for_dimer}.

From the mapping between the fermionic model and the dimer model, we can replace $\phi_{\RR,i}$ by the following product of the fermionic fields (as discussed in Ref. \cite{kasteleyn1963dimer, kasteleyn1961statistics, PhysRevB.66.214513}, and briefly reviewed near  \cref{eq:multi_dimer_corr_fermion})
\begin{align}
     &\phi_{\RR,1} = \psi_{\RR,1} \psi_{\RR+\bm{A}_2,1} \nonumber\\ 
    &\phi_{\RR,2} = -\psi_{\RR,2}\psi_{\RR+\bm{A}_2,2} \nonumber\\ 
    &\phi_{\RR,3} = \psi_{\RR,1} \psi_{\RR,2} \nonumber\\ 
    &\phi_{\RR,4} =  \psi_{\RR,2} \psi_{\RR+\bm{A}_1,1} \nonumber\\
    &\phi_{\RR,5} =-\psi_{\RR,1} \psi_{\RR+\bm{A}_2,2} \nonumber\\ 
    &\phi_{\RR,6} = \psi_{\RR,2} \psi_{\RR+\bm{A}_1+\bm{A}_2,1}.
    \label{eq:def_const_phi}
\end{align} 
Notice that, the original electrons that form bond $\phi_{\RR,i}$ with $i=12/34/56$ are in valley 0/1/2, and we have neglected the valley index for convenience.  
For a given configuration where the dimer $\phi_{\RR,n}$ is occupied (or empty), the corresponding expectation value of the fermionic bilinear operator is 1 (or 0). 
For convenience, we could introduce $A_l$ 
\begin{align}
    A_{l= (\RR,n) } = 
    \begin{cases}
     \phi_{\RR,1}\phi_{\RR+2\bm{A}_2,1} & n=1\\ 
         \phi_{\RR,2}\phi_{\RR+2\bm{A}_2,2}& n=2\\ 
         \phi_{\RR,3}\phi_{\RR+\bm{A}_1,3}& n=3\\ 
          \phi_{\RR,4}\phi_{\RR+\bm{A}_1,4}& n=4\\ 
          \phi_{\RR,5}\phi_{\RR+\bm{A}_1+2\bm{A}_2,5}& n=5\\ 
        \phi_{\RR,6}\phi_{\RR+\bm{A}_1+2\bm{A}_2,6}& n=6
    \end{cases}
    \label{eq:def_const_A}
\end{align}
where index $l=(\RR,n)$ runs over all the constraints. $\prod_l(1-A_l)$ eliminate all the configurations that violate the second constraint. 
In other words, only configurations satisfying $\prod_l (1 - A_l) = 1$ are allowed.
For instance, consider $A_{l=(\RR,1)}$: the constraint is violated if both dimers $\phi_{\RR,1}$ and $\phi_{\RR+2\bm{A}_2,1}$ are occupied. In this case, we have $\phi_{\RR,1}= \phi_{\RR+2\bm{A}_2,1} = 1$, so $A_{(\RR,1)} = 1$, and the configuration is eliminated by the $1 - A_{(\RR,1)}$. 
Then, calculating the degeneracy $\Omega$ of the ground states of the dimer phase in our model \cref{eq:H_J_nu_1_sep} with $\delta V_{\eta\eta'}(\Delta\vR) = 0 $ is equivalent to calculating 
\begin{align}
    \frac{\Omega}{\Omega_{dimer}} =\frac{1}{Z}\int D[\psi]\prod_l(1-A_l) e^{-S} =  \langle \prod_l(1- A_l)\rangle 
\end{align}
where $\langle \rangle$ denotes the expectation value taken with respect to the action $S$ and $\Omega_{dimer}$ is the ground state degeneracy of the normal triangular lattice dimer model. 
Since $ S $ denotes a non-interacting fermionic system, the expectation value can be calculated via Wick's theorem. However, for a system with $ N $ sites, one needs to evaluate the expectation values of the product of at least $ N $ fermionic fields. 
For the product of $ N $ fermionic fields, 
when evaluating its expectation value with Wick's theorem, one needs to contract the fermionic fields in pairs. 
There will be approximately $ \sim \frac{N!}{(N/2)! 2^{N/2}} \sim N^{N/2} $ ways to perform such partitions/contractions. 
Therefore, we could only perform a direct exact  evaluation of $ \Omega/\Omega_{\text{dimer}} $ for small system sizes.

Fortunately, an estimation of $ \Omega/\Omega_{\text{dimer}} $ for an arbitrary system size can be performed. We observe the following exact relation
\begin{align}
    \prod_l (1-A_l) = &1 - \frac{1}{2}\sum_{l,l'} A_l A_{l'} +... \nonumber\\ 
    =&\sum_{n=0}^{N_l} \frac{(-1)^n}{n!}\bigg[ \sum_l A_l\bigg] ^n = \sum_{n=0}^{N_l} \frac{(-1)^n}{n!}\bigg[ A\bigg] ^n
\end{align}
where we have used the fact that $A_l A_l =0$ due to the anti-commutating property of the fermions (from~\cref{eq:def_const_psi_commute,eq:def_const_phi,eq:def_const_A}). We also define 
\begin{align}
    &N_l = \sum_l 1\nonumber\\ 
    &A = \sum_l A_l\, \quad  l\in(\RR,n).
\end{align}
In addition, since $ A_l A_l = 0 $ and we have at most $ N_l $ distinct $ A_l $ fields, 
\begin{align}
   \bigg[ \sum_l A_l\bigg]^m=0,\quad m> N_l
\end{align}
We can therefore extend the summation to infinity and find: 
\begin{align}
     \prod_l (1-A_l) = \sum_{n=0}^{N_l} \frac{(-1)^n}{n!}\bigg[ \sum_l A_l\bigg] ^n 
     =  \sum_{n=0}^{\infty } \frac{(-1)^n}{n!}\bigg[A\bigg] ^n 
     =\exp\bigg[ - A
     \bigg]
\end{align}
We now have 
\begin{align}
\label{eq:deg_to_exp_-A}
    \frac{\Omega}{\Omega_{dimer}} =\frac{1}{Z} \int D[\psi] e^{-A} e^{-S} = \langle e^{-A}\rangle 
\end{align} 
where $ A $ is the summation of the products of four-fermion operators (\cref{eq:def_const_A}). We can then conclude that calculating the degeneracy is equivalent to calculating the partition function of an effective interacting fermionic system with $A$ being quartic in $\psi$. We note that $A$ is quadratic in $\phi$ (see \cref{eq:def_const_A}), and $\phi$ is quadratic in the $\psi$ (see \cref{eq:def_const_phi}). Therefore, $A$ is quartic in the fermionic fields $\psi$. 
This result is exact so far.

We can use the cumulant expansion to approximately calculate such a partition function.
\begin{align}
    \frac{\Omega}{\Omega_{dimer}}  = \langle e^{-A}\rangle \approx 
    \exp\bigg[ -\langle A \rangle + 
    \frac{1}{2} \bigg[ \langle  A^2 \rangle -\langle A \rangle ^2 
    \bigg]  + ...
    \bigg] \label{eq:entropy_dimer_exact}
\end{align}
where we have truncated the cumulant expansion to the second order. The corresponding entropy is then:
\begin{align}
    S=  \frac{k_B}{N} \log(\Omega)\approx S_{approx} = k_B\frac{\log(\Omega_{dimer})}{N} 
    -\frac{k_B}{N} \langle A \rangle +\frac{k_B}{2N} \bigg[ \langle A^2  \rangle -\langle A\rangle^2 
    \bigg]
    \label{eq:entropy_dimer_approx}
\end{align}
with $N$ the number of triangular sites of the system. 
The first term 
$
   k_B \log(\Omega_{dimer})/N 
$
is the entropy of the conventional dimer model without the second constraint (\cref{sec:entropy_of_dimer}). 
$-k_B \langle A\rangle/N$ denotes the first-order contribution, and $k_B \bigg[\langle A^2 \rangle -\langle A\rangle^2 \bigg]/2N$  is the second-order contribution. In principle, we could also systematically refine the estimation of entropy by introducing high-order contributions. However, higher-order contribution is small as we will show in the following section, so we only consider the contribution up to second order.

\subsubsection{Numerical evaluations of the entropy}
\begin{figure}
    \centering
    \includegraphics[width=0.9\linewidth]{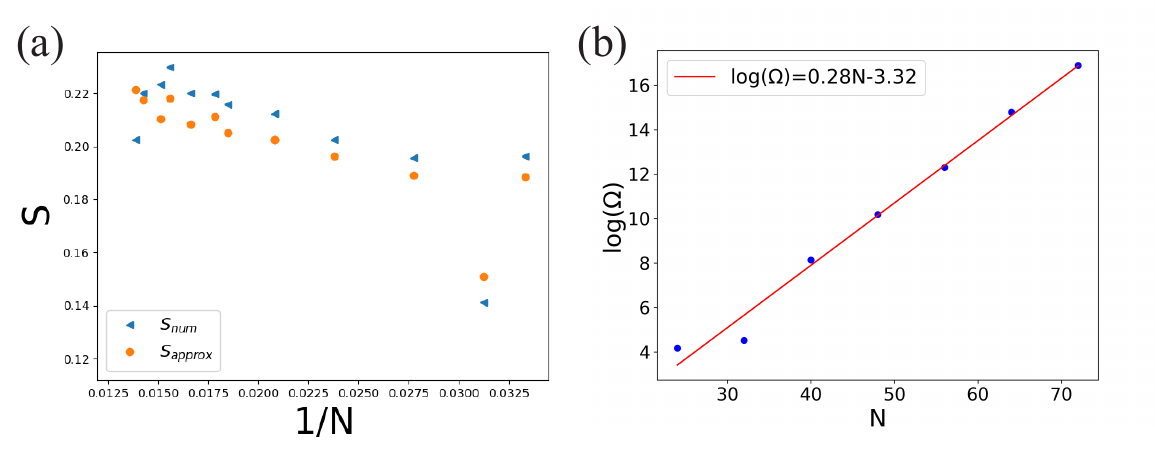}
    \caption{(a) Entropy $S=\frac{k_B}{N}\log(\Omega)$ of the open boundary system with $N$ sites. 
    The exact values and the geometry of the system have been provided in \cref{tab:dimer_config_num}.
    Blue dots represent the exact numerical values. The orange dots mark the approximate values (\cref{eq:entropy_dimer_approx}). (b) Linear scaling of $\log{\Omega}$ on system size $N$. 
    }
    \label{fig:dimer_entropy}
\end{figure}
We first numerically calculate the entropy of the system exactly for a finite-size system $S_{num}$. 
For small-size systems, the ground state degeneracy can also be calculated by exactly evaluating $\langle e^{-A}\rangle$ (\cref{eq:deg_to_exp_-A}) numerically
. However, to verify the validity of our approximate formula, we evaluate the degeneracy using \cref{eq:entropy_dimer_approx}. 
Then, we compare the exact numerical result with the approximate formula for this finite-size system
\begin{align}
\label{eq:entropy_dimer_approx_s12}
    &S_{approx} = \frac{\log(\Omega_{dimer})}{N} 
    -\frac{1}{N} \langle A \rangle +\frac{1}{2N} \bigg[ \langle A^2  \rangle -\langle A\rangle^2 
    \bigg].
\end{align}
The results are shown in~\cref{tab:dimer_config_num} and \cref{fig:dimer_entropy}.
In addition, we calculate the mean error of the approximate expression for the finite-size system
\begin{align}
&E_{S} =\text{mean}\frac{|S_{approx}-S_{num}|}{S_{num}} = 0.06
\end{align}
The approximate entropy leads to a quantitatively good result with $\sim 6\%$ deviations. 
We note that $S_{\text{approx}}$ can be either larger or smaller than the exact numerical result $S_{\text{num}}$, and therefore cannot serve as a strict upper or lower bound. Nonetheless, $S_{\text{approx}}$ remains a useful estimation of $S_{\text{num}}$.

\begin{table}[!ht]
    \centering
    \begin{tabular}{|l|l|l|l|l|l|}
    \hline
        $L_1$ & $L_2$ & $L_1L_2$  & $S_{approx}$ & $S_{num}$\\ \hline
6 & 3 & 18 & 0.162 & 0.174 \\
\hline
6 & 4 & 24 & 0.148 & 0.136 \\
\hline
6 & 5 & 30 & 0.189 & 0.196 \\
\hline
6 & 6 & 36 & 0.189 & 0.196 \\
\hline
6 & 7 & 42 & 0.196 & 0.202 \\
\hline
6 & 8 & 48 & 0.202 & 0.212 \\
\hline
6 & 9 & 54 & 0.205 & 0.216 \\
\hline
6 & 10 & 60 & 0.208 & 0.220 \\
\hline
6 & 11 & 66 & 0.210 & 0.223 \\
\hline
7 & 4 & 28 & 0.150 & 0.118 \\
\hline
7 & 6 & 42 & 0.196 & 0.202 \\
\hline
7 & 8 & 56 & 0.211 & 0.220 \\
\hline
7 & 10 & 70 & 0.217 & 0.220 \\
\hline
8 & 3 & 24 & 0.161 & 0.174 \\
\hline
8 & 4 & 32 & 0.151 & 0.141 \\
\hline
8 & 6 & 48 & 0.202 & 0.212 \\
\hline
8 & 7 & 56 & 0.211 & 0.220 \\
\hline
8 & 8 & 64 & 0.218 & 0.230 \\
\hline
8 & 9 & 72 & 0.221 & 0.203 \\
\hline 
    \end{tabular}
    \caption{ Ground state degeneracy for $L_{1}\times L_2$ lattice with open boundary condition (OBC). $S_{num}$ denotes the exact numerical value. $S_{approx}$ denotes approximate values. }
    \label{tab:dimer_config_num}
\end{table}

\subsubsection{Evaluating the entropy for the infinite-size system}
In this section, we provide a detailed calculation of the (\cref{eq:entropy_dimer_approx})
$S_{approx}$
for the infinite-size system.

To solve the action~\cref{eq:define_S_fermion_for_dimer} and the corresponding fermion correlators, we perform a Fourier transformation of the matrix $M$, and obtain the action in momentum space as 
\begin{align}
    S =& -\frac{1}{2}\sum_{\kk} \psi_{-\kk}^T 
    \cdot 
    \begin{bmatrix}
        2i\sin(k_2) &2i\sin(k_1/2) -2\cos(k_1/2+k_2) \\ 
      2i\sin(k_1/2) + 2 \cos(k_1/2+k_2)  & -2i \sin(k_2) 
    \end{bmatrix}\cdot \psi_\kk  \nonumber\\ 
    =& -\frac{1}{2}\sum_{\kk \in HBZ}
    \begin{bmatrix}
        \psi_\kk^T & \psi_{-\kk}^T
    \end{bmatrix}
    \begin{bmatrix}
      0   & A_{-\kk}  \\
      A_{\kk} & 0 
    \end{bmatrix}
     \begin{bmatrix}
        \psi_\kk \\ \psi_{-\kk}
    \end{bmatrix}\label{eq:dimer_action_matrix}
\end{align}
where $k_1=2k_x$ and $k_2=\sqrt{3}k_y/2-k_x/2$, with 
\begin{align}
    A_\kk =      \begin{bmatrix}
        2i\sin(k_2) &2i\sin(k_1/2) -2\cos(k_1/2+k_2) \\ 
      2i\sin(k_1/2) + 2 \cos(k_1/2+k_2)  & -2i \sin(k_2) 
    \end{bmatrix}
\end{align}
and $HBZ$ denotes half of the BZ. We choose the BZ to be the region $\{k_1\in [0, 2\pi], k_2\in [0, 2\pi] \}$. As discussed above (\cref{eq:dimer_pf_partition_relation}), the partition functions is given by
\begin{align}
    Z = \sqrt{ \prod_{\kk \in BZ}
    |\text{det}[A_\kk]|
    }, 
\end{align}
where
\begin{align}
    \log(Z) = \frac{1}{2}\sum_\kk \log\text{det}[A_\kk].
\end{align}
The entropy $S_{dimer}$ of the triangular dimer model without additional constratin is determined as  
\begin{align}
S_{dimer} = \frac{1}{2N}\log(Z)  = \frac{1}{4}\int d\kk \log 
\bigg| 4\sin^2(k_2) + 4 \sin(k_1/2)^2 + 4\cos^2(k_1/2+k_2)\bigg| ,
\end{align}
where $\int d\kk=\frac{1}{4\pi^2}\int^{2\pi}_0dk_1\int^{2\pi}_0dk_2$, and the prefactor $2N$ indicates we double the unit cell. One can numerically perform the integral and find $S_{dimer}=0.4286$, which is consistent with the result given in Ref.~\cite{PhysRevB.66.214513}.

 As discussed in \cref{eq:entropy_dimer_exact}, to obtain an estimation of the ground state degeneracy of our effective spin model \cref{eq:H_J_nu_1_sep}, we need to calculate the expectation values of the $A$ and $A^2$ operators (\cref{eq:entropy_dimer_approx}) with respect to the free fermion model (defined in \cref{eq:define_S_fermion_for_dimer}).  
We obtain $\langle A\rangle,\langle A^2\rangle$ via Wick's theorem. To do so, we first evaluate the two-point correlation functions
 \begin{align}
  G_{(\RR,\alpha),(\RR',\gamma)}= \langle  \psi_{\RR,i} \psi_{\RR',j}\rangle.
 \end{align}
Here $\alpha,\beta$ are sublattice indices of the doubled unit cell. Perform Fourier expansions with
 \begin{align}
G_{(\RR,\alpha),(\RR',\gamma)}= \frac{1}{N} \sum_\kk e^{i\kk\cdot(\RR+\rr_\alpha-\RR'-\rr_\gamma)}  \langle \psi_{\kk,\alpha} \psi_{-\kk,\gamma}\rangle,
 \end{align}
then $\langle\psi_{\kk,\gamma},\psi_{-\kk,\alpha}\rangle$ can be obtained:
\begin{align}\label{eq:dimer_green_func_momentum}
   &\langle \psi_{\kk,\alpha} \psi_{-\kk,\gamma}\rangle = \bigg[ M^{-1}(\kk)\bigg]_{(-\kk,\gamma), (\kk,\alpha)} \nonumber\\ 
   =&
    \frac{1}{\Delta(\kk)}\begin{bmatrix}
       i\sin(k_2) & i\sin(k_1/2)- \cos(k_1/2+k_2) \\ 
       i\sin(k_2/2) + \cos(k_1/2+k_2) & -i \sin(k_2)
   \end{bmatrix}_{\gamma\alpha},
\end{align}
 where $\Delta(\kk)=2\bigg[ \sin(k_1/2)^2 + \sin(k_2)^2 + \cos(k_1/2+k_2)^2 \bigg]$, and $M(\kk)$ is a $4\times4$ block matrix \cref{eq:dimer_action_matrix}
 \begin{align}
     M^{-1}(\kk) =   \begin{bmatrix}
      0   & A_{-\kk}  \\
      A_{\kk} & 0 
    \end{bmatrix}.
 \end{align}
In particular, 
\begin{align}
    G_{(\RR,1),(\RR',1)}&=-G_{(\RR,2),(\RR',2)}=\int d\kk e^{i\kk\cdot(\RR-\RR')}\frac{i\sin(k_2)}{\Delta(\kk)},\\
    G_{(\RR,1),(\RR',2)}&=\int d\kk e^{i\kk\cdot(\RR+\rr_1-\RR'-\rr_2)}\frac{i\sin(k_1/2)-\cos{(k_1/2+k_2)}}{\Delta(\kk)},\\
    G_{(\RR,2),(\RR',1)}&=\int d\kk e^{i\kk\cdot(\RR+\rr_2-\RR'-\rr_1)}\frac{i\sin(k_1/2)+\cos{(k_1/2+k_2)}}{\Delta(\kk)}.
\end{align}

Take $\RR'-\RR=m\bm{A}_1+ n \bm{A}_2$, the above integrals can be reduced to 
\begin{align}
        G_{(\RR,1),(\RR',1)}&=-G_{(\RR,2),(\RR',2)}=\int d\kk \frac{(-)^{m}\cos(k_2)\cos(nk_2)\cos(2mk_1)\sin(\frac{n}{2}\pi)}{2\bigg[ \cos(k_1)^2 + \cos(k_2)^2 + \cos(k_1+k_2)^2 \bigg]},\label{app:eq:green_func_integral11}\\
    G_{(\RR,1),(\RR',2)}&=\int d\kk \frac{(-)^{m}\left[\cos(k_1)\cos((2m+1)k_1+nk_2)\cos(\frac{n}{2}\pi)-\cos{(k_1+k_2)}\cos((2m+1)k_1+nk_2)\sin(\frac{n}{2}\pi)\right]}{2\bigg[ \cos(k_1)^2 + \cos(k_2)^2 + \cos(k_1+k_2)^2 \bigg]},\\
    G_{(\RR,2),(\RR',1)}&=\int d\kk \frac{(-)^{m}\left[-\cos(k_1)\cos((2m-1)k_1+nk_2)\cos(\frac{n}{2}\pi)-\cos{(k_1+k_2)}\cos((2m-1)k_1+nk_2)\sin(\frac{n}{2}\pi)\right]}{2\bigg[ \cos(k_1)^2 + \cos(k_2)^2 + \cos(k_1+k_2)^2 \bigg]}\label{app:eq:green_func_integral21},
\end{align}
where we have made the substitution of $k_1\to2k_1+\pi$ and $k_2\to k_2+\pi/2$.
We observe that $G_{(\RR,1),(\RR',1)}$ and the second term of $G_{(\RR,1),(\RR',2)}$ and $G_{(\RR,2),(\RR',1)}$ vanishes when $n$ is even, while the first term of $G_{(\RR,1),(\RR',2)}$ and $G_{(\RR,2),(\RR',1)}$ vanishes when $n$ is odd. Ref \cite{PhysRevB.66.214513} found that when $m=1$, the integrals \cref{app:eq:green_func_integral11}-\cref{app:eq:green_func_integral21} can be evaluated with the residue theorem and they proved that for $m=1$, the Green's function exponentially decays with the asymptotic behavior
\begin{align}
    |G_{(\RR,1),(\RR+n\bm{A}_1,1)}|\approx f(|n+1|)+f(|n-1|)),\\
    |G_{(\RR,1),(\RR+n\bm{A}_1,2)}|\approx(-1)^n(f(|n|)-f(|n-1|))
\end{align}
where
\begin{align}
    f(n)=\frac{e^{-\beta n}}{\sqrt{2n\pi3^{3/4}}}\cos(\alpha n+\frac{\pi}{8}),
\end{align}
with $\alpha=1.1960,\beta=\frac{1}{4}\ln\left[\frac{\sqrt{2}+3^{1/4}}{\sqrt{2}-3^{1/4}}\right]=0.83144$. For the Green's function with $m\ne1$, though Ref.~\cite{PhysRevB.66.214513} did not give the analytical expressions, they also proved the exponentially decay behavior for Green's function with field theory. Based on these results, we then numerically evaluated the Green function for a finite range $m, n\in[-12, 12]$, and neglected the terms with longer distance.

Using these fermionic Green's function, we can systematically compute higher‐order correlators such as $\langle A_l \rangle$ and $\langle A_l A_{l'}\rangle$, where $A_l$ is the four‐fermion operator defined in \cref{eq:def_const_A}.  Exploiting translational invariance, we focus on
\begin{align}
  A_{l=(\mathbf{0},n)}, 
  \quad
  A_{l'=(\mathbf{R},m)},
  \quad n,m = 1,2,\dots,6.
\end{align}
Using the previously evaluated Green’s functions, we apply Wick’s theorem to sum over all possible pairwise contractions contributing to
$ \langle A_{l=(\mathbf{0},n)} \rangle
  \quad\text{and}\quad
  \langle A_{l=(\mathbf{0},n)}\,A_{l'=(\mathbf{R},m)}\rangle$ and obtain their values.
We then calculate the global operators
\begin{align}
  A \;=\;\sum_{l=(\mathbf{R},n)}A_l
  \;=\;
  N \sum_{n=1}^6 A_{(\mathbf{0},n)},
\end{align}
and evaluate the cumulants
\begin{align}
  \langle A \rangle
  =N\sum_{n}\langle A_{(\mathbf{0},n)}\rangle,
  \quad
  \langle A^2\rangle
  =N\sum_{l=(\mathbf{0},n),l'=(\mathbf{R},m)}\langle A_l A_{l'}\rangle.
\end{align}
Finally, expanding the entropy to second order in these fluctuations yields
\begin{equation}
\label{eq:entropy_dimer_approx_s12_cal}
 S\approx  S_{\mathrm{approx}}
  =k_B\left(\frac{\log\bigl(\Omega_{\mathrm{dimer}}\bigr)}{N}
  -
  \frac{1}{N}\,\langle A \rangle
  +
  \frac{1}{2N}\Bigl[\langle A^2\rangle - \langle A\rangle^2\Bigr]\right)=0.307k_B.
\end{equation}
In this form, the first correction $-\langle A\rangle/N$ accounts for the mean four‐fermion contribution, while the second term $\bigl(\langle A^2\rangle - \langle A\rangle^2\bigr)/(2N)$ captures the leading fluctuation (variance) around the mean.


Here, we note that the second-order correction to the entropy from $\langle A^2\rangle- \langle A\rangle ^2$ is already very small $ 0.026k_B$. Meanwhile, the first-order correction to the entropy from $\langle A\rangle $ is around $0.096k_B$.  
We then conclude that the zero-temperature entropy of the spin model~\cref{eq:def_spin_model_nu_1} at $\delta V_{\eta\eta'}(\Delta\vR)=0$ and $\nu=1$ is around $0.307$. One can also perform linear fitting on the numerical results in~\cref{tab:dimer_config_num} to exclude the finite size effect. The result is shown in~\cref{fig:dimer_entropy} with $S=k_B\log{\Omega}/N\propto 0.28$ which is also quantitatively consistent with our obtained $S_{approx}$.

\subsubsection{Special case of $L_1=2$}
\begin{figure}
    \centering
\includegraphics[width=0.6\linewidth]{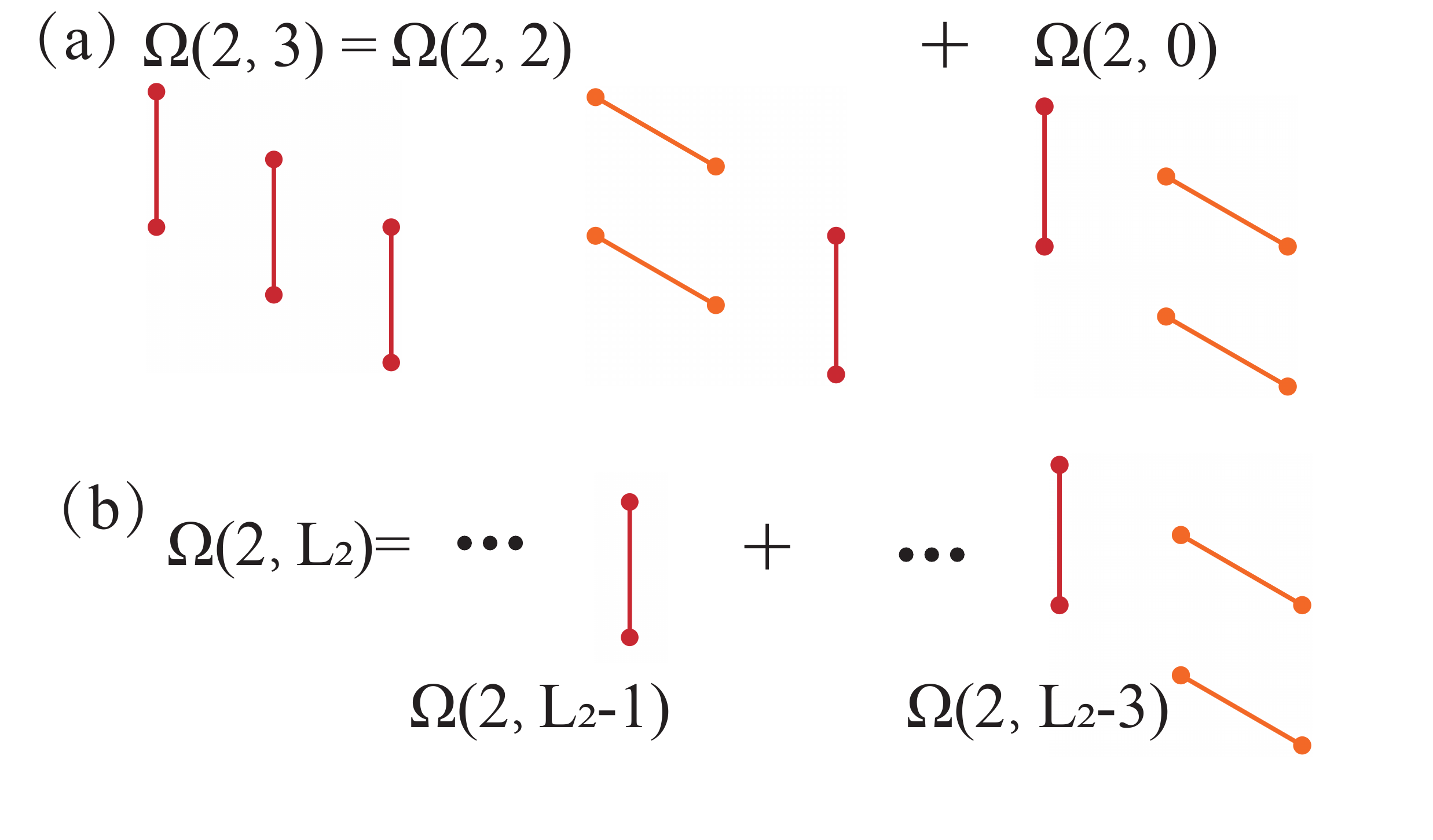}
    \caption{(a) All configurations of $2\times3$ that satisfy the dimer constraint (\cref{box:dimer_constraint}). (b) Illustration of the recurrence relation of $\Omega(2, L_2)$.}
    \label{fig:omega_2} 
\end{figure}

In the special case of a two‐leg ladder ($L_1 = 2$) with length $L_2>0$, the ground‐state degeneracy $\Omega(2,L_2)$ satisfies the simple recurrence
\begin{equation}
  \Omega(2, L_2) \;=\; \Omega(2, L_2 - 1) \;+\; \Omega(2, L_2 - 3)\,,
  \label{eq:recur_2}
\end{equation}
with initial values
\[
  \Omega(2,1) = 1,\quad \Omega(2,2) = 2,\quad \Omega(2,3) = 3.
\]
One can derive this relation directly—without invoking the mapping method introduced in the previous sections—by analyzing the possible terminations of a valid ground‐state configuration, as illustrated in \cref{fig:omega_2}(b).  Any such configuration on a ladder of length $L_2$ must end in exactly one of two ways:

\begin{enumerate}
  \item A single valley bond of type $\eta=0$, in which case the preceding part of the chain is any valid ground state of length $L_2-1$.  This yields the term $\Omega(2, L_2-1)$.
  \item A pair of valley bonds of type $\eta=2$.  However, by the constraints of our model, such a pair must be preceded immediately by an $\eta=0$ bond.  Stripping off this three‐bond segment reduces the problem to a ground state of length $L_2-3$, contributing $\Omega(2, L_2-3)$.
\end{enumerate}

Summing these two mutually exclusive possibilities gives exactly the recurrence in \eqref{eq:recur_2}. One can solve~\cref{eq:recur_2} with the initial condition above to obtain the exact expression of the ground-state degeneracy. $\Omega$ scales with $L_2$ with $\Omega \sim e^{\alpha L_2}$ where $\alpha$ is the root with the largest real part of the equation $\alpha^3-\alpha^2-1=0$. The numerical value of $\alpha$ is 1.47, which gives the entropy 
\begin{align}
    S=k_{N}\log \Omega / N\propto 0.382.
\end{align}

\subsubsection{Review of the dimer model on the triangular lattice}
\label{sec:dimer_of_triangular}
In this section, we review the mapping between the dimer model and the effective fermionic system, for the conventional dimer model without the second constraint introduced at the beginning of \cref{sec:entropy_of_dimer}. Further details on the dimer model can also be found in Refs.~\cite{kasteleyn1963dimer, kasteleyn1961statistics, PhysRevB.66.214513}.

First, we aim to prove that calculating the number of dimer configurations (without imposing the dimer constraint \cref{box:dimer_constraint})
is equivalent to calculating the partition function of a non-interacting fermionic system (\cref{eq:Omega_dimer_equal_Z}). 

We first discuss the Pfaffian of the matrix. For an antisymmetric $ n \times n $ matrix $M$, the Pfaffian is defined as 
\begin{align}
    \text{Pf}[M] = \sum_p {}'(-1)^p M_{p_1,p_2} M_{p_3,p_4}\dots
    \label{eq:def_pf_M}
\end{align}
where $\sum_p' $ denotes the summation over all the permutations with constraints $p_1 < p_2, p_3<p_4, p_{2i-1}<p_{2i}$, and $p_1<p_3<...<p_{n-1}$. $(-1)^p$ is the parity of the permutation.  
Equivalently, we could also define the Pfaffian via
\begin{align}
     \text{Pf}[M] =\frac{1}{(n/2)! 2^{n/2}} \sum_{p} (-1)^p M_{p_1,p_2} M_{p_3,p_4}\dots
\end{align}
$\sum_p$ denotes the summation over all the permutations. 
We aim to prove that, by choosing a specific matrix $M$, 
one can establish a one-to-one correspondence between the dimer configurations and the terms in the summation of
\cref{eq:def_pf_M}.

To establish this correspondence, we assume that each row/column index of the matrix corresponds to a lattice site.
Thus, the $ n \times n $ matrix describes the system with $ n $ sites. 
We further require $ M_{ij} \neq 0 $ if and only if the corresponding bonds are allowed to form a dimer. For our current model, $ M_{ij} \neq 0 $ if and only if the two sites labeled by $ i, j $ are nearest neighbors.
Therefore, each dimer configuration corresponds to a unique non-zero term in the summation of \cref{eq:def_pf_M}. 
However, $ (-1)^p M_{p_1,p_2} M_{p_3,p_4}\dots $ is not necessarily equal to 1 for all configurations. 
If one can construct $ M_{ij} $ such that $ (-1)^p M_{p_1,p_2} M_{p_3,p_4} \dots = 1 $ (or equivalently $ -1 $) for all the corresponding dimer configurations, then 
\begin{align}
    |\text{Pf}[M] |= \text{ number of dimer configurations}
\end{align}

One particular construction of the $ M $ matrix that satisfies our requirements has been discussed in Ref.~\cite{kasteleyn1963dimer, kasteleyn1961statistics, PhysRevB.66.214513}. We have shown it in \cref{fig:illustration_dimer_model}. 
Written explicitly, we have 
\begin{align}
    &M_{(\RR, 1),(\RR',1)} = \delta_{\RR', \RR+\bm{A}_2}-\delta_{\RR',\RR-\bm{A}_2} \nonumber\\ 
    &M_{(\RR, 2),(\RR', 2)} = -\delta_{\RR', \RR+\bm{A}_2}+\delta_{\RR',\RR-\bm{A}_2} \nonumber\\ 
    &M_{(\RR,1),(\RR',2)} = \delta_{\RR',\RR}
    -\delta_{\RR',\RR+\bm{A}_2}-\delta_{\RR',\RR-\bm{A}_1}-\delta_{\RR',\RR-\bm{A}_1-\bm{A}_2}
    \nonumber\\ 
    &M_{(\RR,2),(\RR',1)} = \delta_{\RR',\RR+\bm{A}_1}+\delta_{\RR',\RR+\bm{A}_1+\bm{A}_2}-\delta_{\RR',\RR} + \delta_{\RR',\RR-\bm{A}_2}
    \label{eq:def_M_mat}
\end{align}
Note that we have used $(\RR,\alpha)$ to label the positions of sites instead of the index $i,j$ in the above equation \cref{eq:def_M_mat}
(see also discussions near \cref{eq:define_super_lattice_vector_for_dimer}). In what follows, we will always use $i$ and $j$ to label the sites of the system.
.

Via $M$ defined in~\cref{eq:def_M_mat}, it is proved \cite{kasteleyn1963dimer, kasteleyn1961statistics} that calculating the numbers of dimer configurations $\Omega_{dimer}$ is equivalent to calculating the Pfaffian of the matrix $M_{ij}$. 
\begin{align}
    \Omega_{dimer} = \text{Pf}[M]
\end{align}

The Pfaffian of the matrix can be calculated using Grassmann numbers. We introduce fermionic fields to facilitate the computation
\begin{align}
    \psi_i,\quad i=1,...,n
\end{align}
with 
\begin{align}
   \{\psi_i,\psi_j\}=0 ,\quad \psi_i=[ \psi_i]^\dag 
\end{align}
We define 
\begin{align}
    Z= \int \prod_{i=n,n-1,...,1} d\psi_i e^{-S} ,\quad S= -\sum_{ij}\frac{1}{2}\psi_i M_{ij} \psi_j\label{eq:dimer_grassmann_action}
\end{align}
Using Grassmann number integral 
\begin{align}
    \int d\psi \psi = 1 
\end{align}
we find 
\begin{align}
    Z = &\frac{1}{2^{n/2} {n/2}!} \int \prod_i d\psi_i \bigg( \sum_{ij}\psi_iM_{ij}\psi_j\bigg)^{n/2}  \nonumber\\ 
    =&\frac{1}{2^{n/2} {n/2}!} \sum_p (-1)^p M_{p_1,p_2}... = \text{Pf}[M]\label{eq:dimer_pf_partition_relation}
\end{align}
Therefore, we can calculate the partition function of the effective fermionic system to obtain the ground state degeneracy.

We next show that, from the current Pfaffian technique, we can also calculate the multi-dimer correlations. We use $ \phi_{ij} $ to denote the dimer connecting two nearest-neighbor sites $ i $ and $ j $. For a given dimer configuration $ c $, we define $ \phi_{ij}^c = 1 $ if the configuration $ c $ contains the dimer $ ij $, and $ \phi_{ij}^c = 0 $ otherwise. The $m $-dimer correlators are defined as:
\begin{align}
  F_{i_1j_1,..,i_mj_m} = \frac{1}{Z} \sum_{c} \phi_{i_1j_1}^c \phi_{i_2j_2}^c ... \phi_{i_mj_m}^c 
\end{align} 
where $\sum_c$ runs over all the dimer configurations. We now aim to use the fermionic action to calculate the dimer correlators. 
Building on the one-to-one correspondence between dimer configurations and the terms in the summation of \cref{eq:def_pf_M},
 we can extract the $m$-dimer correlation functions by focusing on the terms on the right-hand side of \cref{eq:def_pf_M}
that include the product of $ M_{i_1j_1} $ (or $ M_{j_1i_1} $), ..., $ M_{i_mj_m} $ (or $ M_{j_m i_m} $). To extract such terms, we take the derivative with respect to $ M_{ij} $. Only terms containing $ M_{ij} $ in the product survive under the derivative. Thus, we have:
\begin{align}
    F_{i_1j_1,...,i_mj_m} = \frac{1}{Z} \bigg[ 
    \prod_{l=1}^m 
    \bigg(  M_{i_l ,j_l}\frac{d}{d M_{i_l,j_l}} + M_{j_l, i_l}\frac{d}{d M_{j_l,i_l}}
        \bigg) \bigg] 
    \text{Pf}[M] 
\end{align}
Since the derivative removes $ M_{i,j} $ (or $ M_{j,i} $), we introduce an additional $ M_{i,j} $ (or $ M_{j,i} $) such that all the terms that survive under the derivative retain the same sign. 
In addition, we notice that:
\begin{align}
    &\bigg[ 
    \prod_{l=1}^m 
    \bigg(  M_{i_l ,j_l}\frac{d}{d M_{i_l,j_l}} + M_{j_l, i_l}\frac{d}{d M_{j_l,i_l}}
        \bigg) \bigg] 
    \text{Pf}[M]  \nonumber\\ 
    =& \bigg[ 
    \prod_{l=1}^m 
    \bigg(  M_{i_l ,j_l}\frac{d}{d M_{i_l,j_l}} + M_{j_l, i_l}\frac{d}{d M_{j_l,i_l}}
        \bigg) \bigg]   \int D[\psi] e^{-S}  \nonumber\\ 
            =&\int D[\psi] \prod_{l=1}^mM_{i_l,j_l}\psi_{i_l}\psi_{j_l}e^{-S},
\end{align}
where $\psi_{i,l}\psi_{j,l}$ comes directly from the derivative of the action \cref{eq:dimer_grassmann_action}. Therefore, 
\begin{align}
    F_{i_1j_1,...,i_mj_m} =Z \langle \prod_{l=1}^mM_{i_l,j_l}\psi_{i_l}\psi_{j_l}\rangle  
    \label{eq:multi_dimer_corr_fermion}
\end{align} 
where the expectation value $ \langle \rangle $ is taken with respect to the action $ S $. 
$i_1,j_1,...,i_m,j_m$ are site indices where $i_l,j_l$ labels the $l$-th dimer we are interested in. 
We thus conclude that calculating the dimer correlators is equivalent to calculating the fermion correlators, which can be done using Wick's theorem.

\subsection{Ground state in the strong coupling limit at $\nu=2$}\label{app:sec:vbsnu=2}

\subsubsection{Effective spin model at $\nu=2$} 
We now derive the effective spin model at $\nu=2$ in the strong coupling limit. As we discussed in~\cref{sec:spin_mode_nu_1} and around~\cref{eq:SW_ham_h0_h1}, we separate our Hamiltonian into
\begin{align}
    &H = H_0 +H_1 \nonumber\\ 
    &H_0 = H_{U(6)},\quad H_1=H_t +H_{\delta V} 
\end{align}
We treat $H_1$ as perturbation and derive an effective spin model. 

For $H_0$, the ground states at $\nu=2$ is 
\begin{align}
\label{eq:low_energy_state_nu_2}
    \prod_{\vR}\cre{d}{\vR, \eta^1_{\vR},s^1_{\vR}}\cre{d}{\vR, \eta^2_{\vR},s^2_{\vR}}|0\rangle 
\end{align}
where each site $\vR$ is filled by two electrons with valley spin indices $(\eta^1_{\vR},s_{\vR}^1)$ and $(\eta^2_{\vR},s_{\vR}^2)$ respectively and $(\eta^1_{\vR},s_{\vR}^1) \ne (\eta^2_{\vR},s_{\vR}^2)$. 
We introduce the projection operator that retains only the low-energy states with $2$ electrons per unit cell, defined as:
\begin{align}
P_L^{\nu=2} 
= \sum_{ \{(\eta^1_{\vR},s^1_{\vR}),(\eta^2_{\vR},s^2_{\vR}) \}, (\eta^1_{\vR},s^1_{\vR})\ne (\eta^2_{\vR},s^2_{\vR})}
\prod_{\vR}\cre{d}{\vR, \eta^1_{\vR},s^1_{\vR}}\cre{d}{\vR, \eta^2_{\vR},s^2_{\vR}}|0\rangle \langle 0 |
\prod_{\vR}\hat{d}_{\vR, \eta^2_{\vR},s^2_{\vR}}\hat{d}_{\vR, \eta^1_{\vR},s^1_{\vR}} 
\end{align}
Then the effective Hamiltonian at $\nu=2$ reads
\begin{align}
    H_J^{\nu=2} =& P_L^{\nu=2}\bigg[ H_0 +H_1
    \bigg] P_L^{\nu=2}  
    - P_L^{\nu=2} H_1 \sum_{|H\rangle } \frac{|H\rangle \langle H|}{E_H} H_1 P_L^{\nu=2}\nonumber\\ 
    =&  P_L^{\nu=2}\bigg[ H_{\delta V}
    \bigg] P_L^{\nu=2}  
    - P_L^{\nu=2} H_t \sum_{|H\rangle } \frac{|H\rangle \langle H|}{E_H} H_t P_L^{\nu=2}
\end{align}

Note that $H_{\delta V}$ describes density-density interactions and does not mix $\ket{L}$ with $\ket{H}$.
Since $H_t$ is a hopping term that will create electron and hole states at two different sites within the same valley and spin sectors, the high-energy states created by $H_t$ can be written as:
\begin{align}
\label{eq:charge_0_nu_2}
   &\cre{d}{\vR,\eta_{\vR+\Delta\vR},s_{\vR+\Delta\vR}}  \des{d}{\vR+\Delta\vR,\eta_{\vR+\Delta\vR},s_{\vR+\Delta\vR}}  \prod_{\vR'}\cre{d}{\vR', \eta^1_{\vR'},s^1_{\vR'}}
   \cre{d}{\vR', \eta^2_{\vR'},s^2_{\vR'}}
   |0\rangle  \nonumber\\ 
   &
  \Delta\vR =\pm  C_{3z}^\eta \bm{a}_{M,2} \nonumber\\ 
  &
   ({\eta}_{\vR+\Delta\vR},{s}_{\vR+\Delta\vR}) \ne ({\eta}^1_{\vR},{s}^1_{\vR}),\quad 
    ({\eta}_{\vR+\Delta\vR},{s}_{\vR+\Delta\vR}) \ne ({\eta}^2_{\vR},{s}^2_{\vR}) \nonumber\\ 
    &
   ({\eta}_{\vR+\Delta\vR},{s}_{\vR+\Delta\vR}) = ({\eta}^1_{\vR+\Delta},{s}^1_{\vR+\Delta \vR})\text{  or  }
    ({\eta}_{\vR+\Delta\vR},{s}_{\vR+\Delta\vR}) =({\eta}^2_{\vR+\Delta \vR},{s}^2_{\vR+\Delta \vR})
\end{align}
The excitation energy is
\begin{align}
\label{eq:charge_0_nu_2_en}
    E_H = V(0) -V(\Delta\vR)
\end{align}
Taking the states in \cref{eq:charge_0_nu_2}, we introduce the projection operator to the high-energy state as 
\begin{align}
P_L^{\nu=2} = 
    \sum_{H} |H\rangle \langle H|
\end{align}

Written explicitly, the effective coupling generated by $H_t$ is
\begin{align}
 &P_L^{\nu=2} 
    \bigg[  H_t\sum_{|H\rangle } \frac{|H\rangle \langle H|}{E_H} H_t\bigg] P_L^{\nu=2}    \nonumber\\ 
    =&-\frac{t^2}{2E_H} 
        \sum_{\vR,\eta,s,s',s''}\sum_{\Delta\vR = \pm C_{3z}^{\eta}\bm{a}_{M,2}} 
        P_L^{\nu=2}\cre{d}{\vR,\eta,s} \des{d}{\vR+\Delta\vR,\eta,s} 
   P_H^{\nu=2}
          \cre{d}{\vR+\Delta\vR,\eta,s'} \des{d}{\vR,\eta,s'} 
        P_L^{\nu=2}
        \label{eq:derive_spin_model_nu_2}
\end{align}
Following the same procedure near \cref{eq:ham_nu_1_after_SW}, we obtain the following spin-spin coupling:
\begin{align}
  J\sum_{\vR,\eta,s,s'}\sum_{\Delta\vR = \pm C_{3z}^{\eta}\bm{a}_{M,2}}
        P_L^{\nu=2}
       \bigg[\bm{S}_{\vR,\eta}\cdot \bm{S}_{\vR+\Delta\vR,\eta} +\frac{1}{4}n_{\vR,\eta}n_{\vR+\Delta\vR,\eta}\bigg] 
    P_L^{\nu=2}
\end{align}
with Heisenberg coupling 
\begin{align}
    J= 2t^2/E_H
\end{align}

The full Hamiltonian now reads
\begin{align} 
\label{eq:def_spin_model_nu_2}
    H_{J}^{\nu=2} = &   P_L^{\nu=2}\bigg\{ 
    \frac{1}{2} \sum_{\vR,\Delta\vR, \eta,\eta'}\delta V_{\eta\eta'}(\Delta\vR) n_{\vR,\eta} n_{\vR+\Delta\vR,\eta'}
    \nonumber\\ 
    & +J\sum_{\vR,\eta,s,s'}\sum_{\Delta\vR = \pm C_{3z}^{\eta}\bm{a}_{M,2}}
       \bigg[\bm{S}_{\vR,\eta}\cdot \bm{S}_{\vR+\Delta\vR,\eta} +\frac{1}{4}n_{\vR,\eta}n_{\vR+\Delta\vR,\eta}\bigg] 
       \bigg\} 
    P_L^{\nu=2}
\end{align}
which takes a similar form to the effective Hamiltonian at $\nu = 1$ (\cref{eq:def_spin_model_nu_1}), except that we project the Hamiltonian onto the subspace with 2 electrons per unit cell. 

In addition, in the limit where $\delta V_{\eta\eta'}(\Delta\vR) = 0$, the spin model in the 1D limit possesses an additional spinless $C_{2y}$ symmetry defined via
\begin{align}
    &C_{2y} \bm{S}_{\vR,\eta=0} C_{2y}^{-1} = 
    \bm{S}_{C_{2y}(\vR+\rr_0)-\rr_0,\eta=0} \nonumber\\ 
    &C_{2y} \bm{S}_{\vR,\eta=1} C_{2y}^{-1} = 
    \bm{S}_{C_{2y}(\vR+\rr_1)-\rr_2,\eta=2} \nonumber\\ 
    &C_{2y} \bm{S}_{\vR,\eta=2} C_{2y}^{-1} = 
    \bm{S}_{C_{2y}(\vR+\rr_2)-\rr_1,\eta=1}
\end{align}
There is also a $U(2)$ symmetry for each valley and each chain attached to each valley (\cref{eq:additional_U2_symmetry}).

\subsubsection{Ground states of the effective spin model at $\nu=2$}
\label{sec:gnd_state_spin_nu_2}
We now discuss the ground state of the spin model at $\nu=2$. 
We have the following good quantum numbers
\begin{align}
    n_{\vR,\eta} =  \sum_{s} \cre{d}{\vR,\eta,s}\des{d}{\vR,\eta,s}
\end{align} 
We use $N_{\vR,\eta}$ to characterize the eigenvalue of $n_{\vR,\eta}$. We only consider states $|\psi\rangle $ that are eigenstates of $n_{\vR,\eta}$ 
\begin{align}
    n_{\vR,\eta}|\psi\rangle = N_{\vR,\eta} |\psi\rangle 
\end{align}
However, unlike the $\nu = 1$ case, we now have the following constraints indicating that there are two electrons per unit cell.
\begin{align}
    \sum_\eta N_{\vR,\eta}=2 
\end{align}

For a given configuration $\{N_{\vR,\eta}\}$, we first classify the unit cells $\vR$ into different groups. We apply the following rule to group the unit cells:
\begin{itemize}
    \item For each unit cell $\vR$ (or each site $\vR$) 
    filled with electrons from valley $\eta_1,\eta_2$ respectively ($N_{\vR,\eta_1} = N_{\vR,\eta_2} = 1$), it belongs to the same group as the unit cell $\vR + \Delta \vR$ if and only if at least one of the following two conditions is satisfied 
\begin{align}
    (1) \quad & : \quad \Delta \vR = \pm C_{3z}^{\eta_1} \bm{a}_{M,2} \quad \text{and} \quad N_{\vR + \Delta \vR, \eta_1} = 1 \nonumber \\
    (2) \quad & : \quad \Delta \vR = \pm C_{3z}^{\eta_2} \bm{a}_{M,2} \quad \text{and} \quad N_{\vR + \Delta \vR, \eta_2} = 1 \nonumber
\end{align}
\end{itemize}
Here, similarly to $\nu=1$ case (see \cref{sec:gnd_state_spin_nu_1}), the system can be described by a series of open-boundary spin chains for each valley as shown in \cref{fig:spin_config_nu2} (g). However, unlike $\nu=1$ case, each unit cell contributes to two open-boundary spin chains instead of one, since each cell is filled by two electrons. As discussed in~\cref{sec:gnd_state_spin_nu_1}, if a configuration exists in which all spin chains have length 2, then the ground state at $\delta V_{\eta\eta'}(\Delta \vR) = 0$ corresponds to this configuration.

\begin{figure}
    \centering
\includegraphics[width=0.4\linewidth]{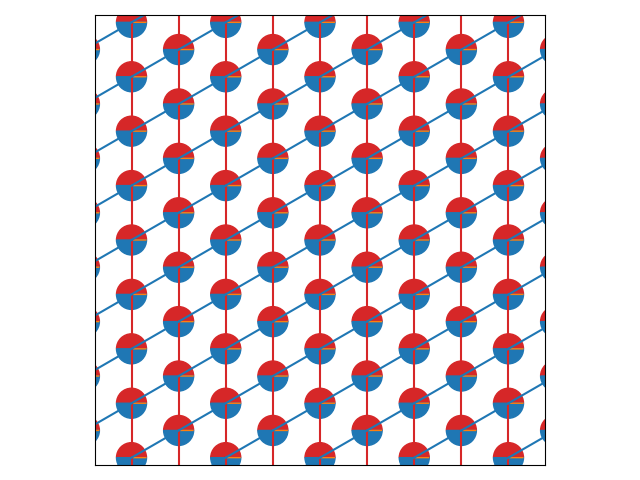}
    \caption{Configuration of the valley polarization states at $\nu=2$.  
    Each dot represents a unit cell of the system. Each unit cell is filled with two electrons of two valleys which have been marked by two colors. The dots connected by the same color denote the spin chain formed by spins of the same valley. 
    }
    \label{fig:spin_config_nu2_vp} 
\end{figure}

With an infinite lattice, we find that there are 6 degenerate ground states at $ \delta V_{\eta\eta'}(\Delta\vR) = 0 $ that are all valence-bond solid (VBS) states with $C_{3z}$ symmetry. We have shown the ground-state patterns in \cref{fig:spin_config_nu2} (a-f) (see \cref{sec:VBS_nu_2} for further discussions). 
Similarly to the $ \nu = 1 $ case, when $ \delta V_{\eta\eta'}(\Delta\vR)  $ is dominant, we have valley-polarized states with a 1D spin liquid developed on top of it. One illustration of the VP quantum spin liquid state has been shown in \cref{fig:spin_config_nu2_vp}.

\subsubsection{VBS at $\nu=2$}\label{sec:VBS_nu_2}
\begin{figure}
    \centering
\includegraphics[width=1.0\linewidth]{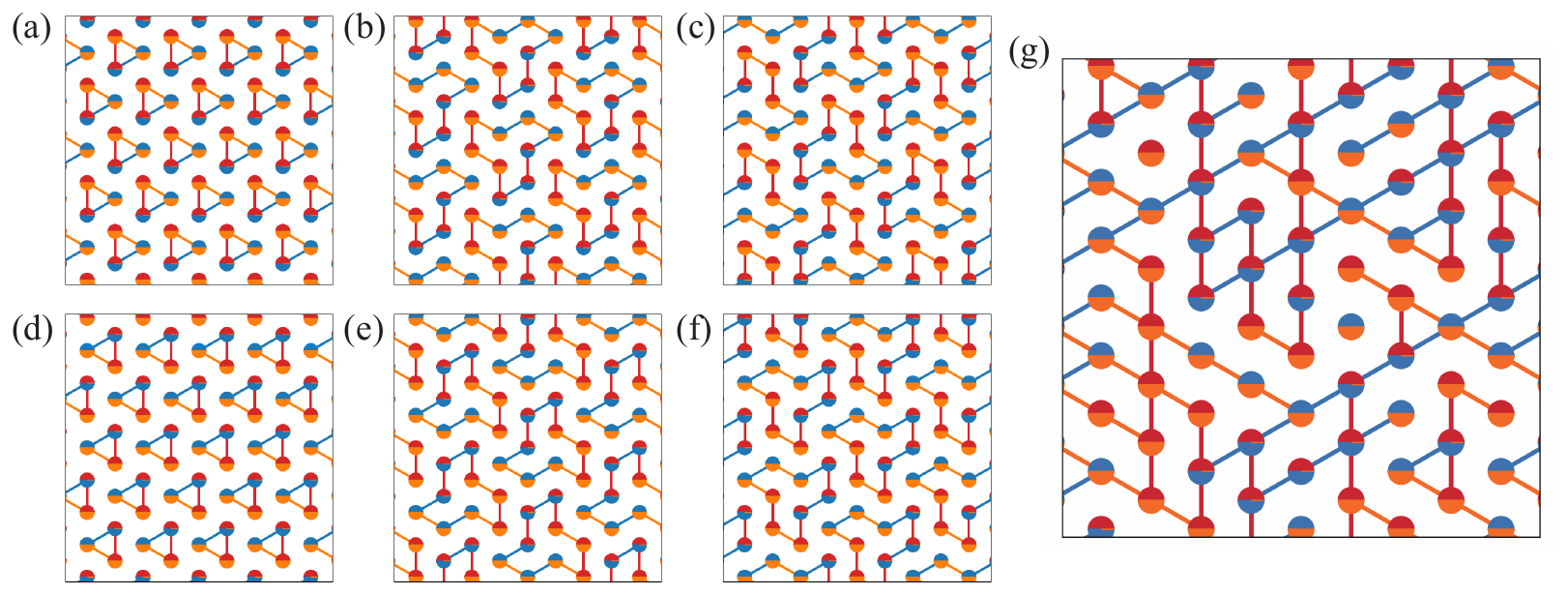}
    \caption{All the ground state (a-f) configurations at $\delta V_{\eta\eta'}(\Delta\vR)=0$ and a random configuration for illustrate (g). Each dot represents a unit cell of the system. Each unit cell is filled with two electrons of two valleys which have been marked by two colors. The dots connected by the same color denote the spin chain formed by spins of the same valley.}
    \label{fig:spin_config_nu2} 
\end{figure}

We now analyze the ground state manifold of the spin model defined in \cref{eq:def_spin_model_nu_2} at $\delta V_{\eta\eta'}(\Delta\vR)=0$ limit.
We identify six valence-bond solid (VBS) ground states, each characterized by a distinct spin dimerization pattern, as shown in \cref{fig:spin_config_nu2}(a–f). 
All other ground-state configurations can be generated by performing a translational transformation.
These configurations strictly satisfy the geometric constraint requiring all spin chains to exhibit a uniform length of two unit cells, thereby minimizing the system energy.
Notably, the patterns in \cref{fig:spin_config_nu2}(a) and (d) manifest $\sqrt{3}\times\sqrt{3}$ translational symmetry breaking, while configurations (b), (c), (e), and (f) display $3\times3$ translational symmetry breaking. 
Furthermore, configurations (a), (b), and (c) are related to (d), (f), and (e), respectively, by a $C_{2y}$ symmetry. Configurations (b) and (c) are connected by a $C_{2x}$ symmetry. Consequently, there are only two symmetry-inequivalent ground state classes: (a) and (b). 
We now rigorously demonstrate the completeness of this classification in an infinite lattice by proving the absence of additional ground states beyond these symmetry-related configurations. We leave the discussion of the boundary condition at the end of this section.

\begin{figure}
    \centering
\includegraphics[width=0.9\linewidth]{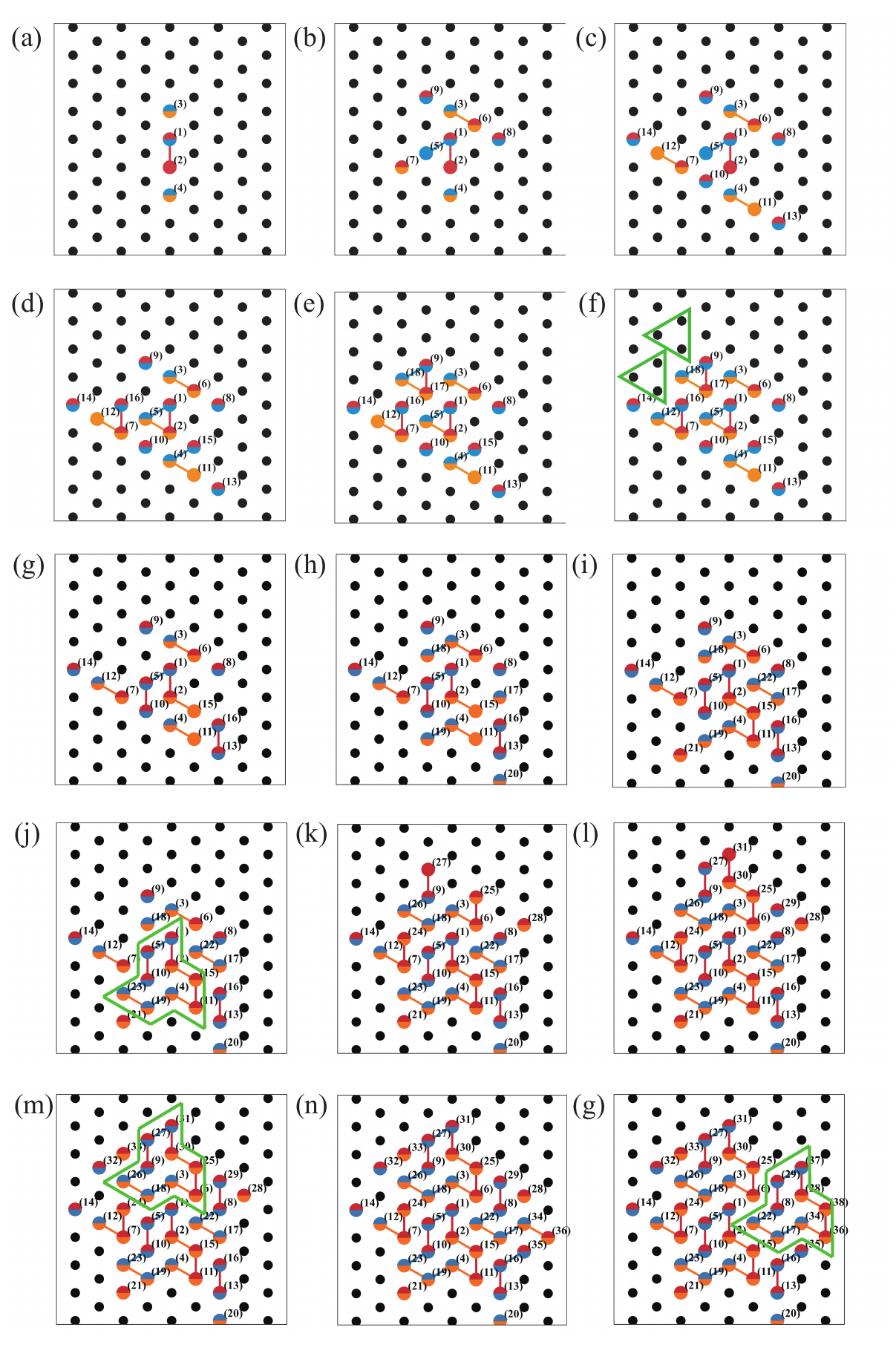}
    \caption{Illustration of the proof of ground state configurations at filling $\nu=2$. Red, blue, orange: valley indices 0, 1, 2. Colored sites indicate valley-specific electron occupation; inter-site bonds mark spin-singlet pairs of shared-valley electrons.}
    \label{fig:spin_config_nu2_prove} 
\end{figure}

 The proof is outlined as follows, with an illustration provided in~\cref{fig:spin_config_nu2_prove}. We start from a single site and gradually construct the ground state by filling electrons on the nearby sites.

 We start from an arbitrary lattice site $\vR$ (denoted as site 1). 
 Due to symmetry, we can assume, without loss of generality, that two electrons occupy the valleys $ \eta = 0 $ (red) and $ \eta = 1 $ (blue) at lattice site 1. We then need to place an electron of valley $ \eta = 0 $ on either site $ \vR + \bm{a}_{M,2} $ or site $ \vR - \bm{a}_{M,2} $ to form a dimer with site 1. Both choices are symmetry-equivalent. We choose to place an electron of valley $ \eta = 0 $ at site $ \vR - \bm{a}_{M,2} $ (site 2). 
To satisfy the ground-state conditions, the sites $\vR + \bm{a}_{M,2}$ (site 3) and $\vR - 2\bm{a}_{M,2}$ (site 4) must each be occupied by electrons not in valleys 0, thus in valleys 1 and 2, respectively, as illustrated in~\cref{fig:spin_config_nu2_prove} (a). Under these constraints, the electron in valley $\eta = 1$ at site $\vR$ (site 1) can pair with the electron in valley $\eta = 1$ located at site 
$\vR \pm \bm{a}_{M,1} \pm \bm{a}_{M,2}$. The choice of $+$ sign will lead to a configuration corresponding to configurations (a)-(c) in~\cref{fig:spin_config_nu2}, while the choice of $-$ sign will lead to configurations (d)-(f) in~\cref{fig:spin_config_nu2}. For illustration, we consider the $+$ sign case, specifically choosing site $\vR - \bm{a}_{M,1} - \bm{a}_{M,2}$ (site 5), forming a bond between sites 1 and 5. This bond further constrains the occupation of sites $\vR + \bm{a}_{M,1} + \bm{a}_{M,2}$ (site 6) and $\vR - 2\bm{a}_{M,1} - 2\bm{a}_{M,2}$ (site 7) by electrons in valleys 0 (red) and 2 (orange) so that no chains of valley 1 with length longer than 2 is formed, as shown in~\cref{fig:spin_config_nu2_prove} (b). 

At this stage, a bond in valley 2 (orange) must connect sites 3 and 6, otherwise the valley 2 would form a chain with length larger than 2, and this further fixes the electrons at sites 8 and 9, as depicted in~\cref{fig:spin_config_nu2_prove} (b).
A bond must be formed between the two adjacent sites if they are in the same color and the edge connecting them is also in the corresponding direction. To ensure no spin chain exceeds a length of 2, it follows that the site 10 between sites 4 and 7 must be occupied by electrons in valleys 0 and 1. Furthermore, the valley $\eta = 2$ (orange) electrons at sites 4 and 7 must pair with the valley $\eta = 2$ (orange) electrons at sites 11 and 12 because every electron must be paired with another one, respectively, thereby determining the electron configurations at sites 13 and 14 which cannot be valley 2 (orange), as illustrated in~\cref{fig:spin_config_nu2_prove} (c).

At this point, the constraints imposed by the ground-state configuration cannot determine additional electron positions, leaving freedom to assign the remaining electron at site 2 and its associated dimer. There are three options: the electron could belong to valley $\eta = 2$ (orange) and pair with either site 5 or 15, or it could belong to valley $\eta = 1$ and pair with site 10. These choices correspond to configurations (d), (e), and (f) in~\cref{fig:spin_config_nu2}.  Since there are two types of translational symmetry breaking in the six VBS state: $\sqrt{3}\times\sqrt{3}$ and $3\times3$. Here, we select the first and the second option for illustration of both cases, since they represent the two symmetry in-equivalent cases as discussed before.

In the first option, i.e., the remaining electron at site 2 in ~\cref{fig:spin_config_nu2}(c) belongs to the valley $\eta = 2$ (orange) and an orange bond connects site 2 and site 5. This configuration fixes the electrons at sites 15 and 16 (not orange), creating dimers between sites 7 and 16, and sites 4 and 15, as shown in~\cref{fig:spin_config_nu2_prove} (d). Additionally, since valley $\eta = 1$ (blue) electrons occupy sites 3 and 16, the two electrons at site 17 must belong to valleys $\eta = 0$ and $\eta = 2$, as depicted in~\cref{fig:spin_config_nu2_prove} (e). With the electrons at site 17 fixed, the valley $\eta = 1$ (blue) electron at site 16 can only pair with the valley $\eta = 1$ (blue) electron at site 12. Then site 12 must have an electron in valley 1 (blue). This configuration results in the $\sqrt{3} \times \sqrt{3}$ unit cell shown in~\cref{fig:spin_config_nu2_prove} (f). Starting from the configuration \cref{fig:spin_config_nu2_prove} (f), we now show that the valley numbers of all other sites can be uniquely determined. We notice that starting from the current configurations of sites 1,2,5, i.e., starting from ~\cref{fig:spin_config_nu2_prove} (d), we have uniquely determined the configurations of sites 9,18,17, and sites 7,12,16, which have the same valley configurations as 1,2,5. From valley configurations of $9,18,17$, we could uniquely generate the configurations of the sites marked by green triangular in \cref{fig:spin_config_nu2_prove} (f). This procedure can be repeated until all the sites of the system are filled.

In the second option, i.e., the remaining electron at site 2 in ~\cref{fig:spin_config_nu2}(c) belongs to the valley $\eta = 2$ (orange) and an orange bond connects site 2 and site 15. This fixes the electrons at site 16 in valleys 0 (red) and 1 (blue), and the other electron on site 5 to be valley 0 (red). Subsequently, valley $\eta=0$ (red) bond connects site 13 and 16, as shown in~\cref{fig:spin_config_nu2_prove} (g). The bond between sites 13 and 16, as well as between sites 5 and 10, constrains the electrons at sites 17–20. Next, two valley $\eta=1$ (blue) bonds should connect site 3 to 18 and site 4 to 19, as shown in~\cref{fig:spin_config_nu2_prove} (h). The bond between site 4 and 19 further constrains the electrons at sites 15 and 21 to valleys $\eta=0$ (red) and 2 (orange). 

Since sites 6 and 15 host valley $\eta=0$ electrons, then electrons at site 22 (located between them) must be in valleys $\eta=1$ (blue) and 2 (orange). Consequently, a valley 0 (red) bond, a valley 1 (blue) bond, and a valley 2 (orange) bond form between sites 11 and 15, sites 8 and 22, and sites 17 and 22, respectively, as illustrated in~\cref{fig:spin_config_nu2_prove} (i). Given that valley $\eta = 0$ (red) electrons occupy sites 7 and 21, the two electrons at site 23 between them must belong to valleys $\eta = 1$ (blue) and $\eta = 2$ (orange). This results in a valley 1 (blue) bond between sites 10 and 23, and a valley 2 bond between sites 19 and 23, as shown in~\cref{fig:spin_config_nu2_prove} (j). Then one $3\times3$ unit cell is determined in this step and it is shown in green contours in ~\cref{fig:spin_config_nu2_prove} (j). 

Next, we demonstrate that the entire lattice can only be tiled with this $3\times3$ unit cell under the ground state constraints. To proceed, we first notice that the blue bond between 3 and 18 in ~\cref{fig:spin_config_nu2_prove} (j) constrains the electrons at site 24 and 25 to be in valley 0 (red) and 2 (orange), necessitating two valley 0 (red) bonds between sites 7 and 24 and between sites 6 and 25 as shown in ~\cref{fig:spin_config_nu2_prove} (k). The bond between sites 7 and 24 then constrains the electrons at site 26 to valleys 1 (blue) and 2 (orange), pairing them with electrons in the same valleys at sites 9 and 18, respectively, as depicted in~\cref{fig:spin_config_nu2_prove} (k). Also, site 29 cannot be valley 1 (blue), otherwise the blue bonds will form a chain longer than length 2.

We then notice that since valley $\eta = 2$ (orange) electrons occupy sites 25 and 28, the two electrons at site 29 between them must belong to valleys $\eta = 0$ (red) and $\eta = 1$ (blue). In addition, site 30 cannot contain valley 1 (blue) electrons, otherwise a length-3 spin chain will be formed by sites 9, 26 and 30. 
Hence, site 30 contains valley 0 (red) and 2 (orange). The valley 2 electron at site 30 pairs with that at site 25, while the other electron at site 30, determined by the bond between sites 9 and 26, belongs to valley 0. It must pair with another valley 0 (red) electron on site 31, as shown in~\cref{fig:spin_config_nu2_prove} (l). 
The bond between sites 25 and 30 constrains the remaining electron at site 27 to valley 1. 
To determine its pairing, we note that the bond between sites 18 and 26 fixes the electrons at site 32 to be in valley 0 (red) and 1 (blue). Further, since valley $\eta = 1$ (blue) electrons occupy sites 27 and 32, the two electrons at site 33 must belong to valleys $\eta = 0$ (red) and $\eta = 2$ (orange), leading to a valley 1 bond between sites 27 and 31. We identify another unit cell displaced by 3$a_{M,2}$ from the original one, as shown in~\cref{fig:spin_config_nu2_prove} (m).

Next, we try to identify another $3\times3$ unit cell with a displacement $3\bm{a}_{M,0}+3\bm{a}_{M,1}$. Notice that the bond between 17 and 22 fixes the electrons at site 35 to be in valley 0 and 1, and the valley 1 (blue) electron at this site pairs with the one at site 16. Furthermore, the valley 0 electrons at sites 28 and 35 prevent the electrons at site 34 (located between them) from occupying valley 0 (red), causing its electrons to pair with electrons in matching valleys at sites 17 and 36, as depicted in~\cref{fig:spin_config_nu2_prove} (n).

The bond between 17 and 34 fixes the electrons at site 38 to be not in valley 1 (blue), generating a valley 0 bond between site 36 and 28. Additionally, the valley 0 (red) and 1 (blue) electrons at sites 28 and 29 pair with the same-valley electrons at site 37. This completes the identification of another unit cell, as shown in~\cref{fig:spin_config_nu2_prove} (g). The entire lattice can thus be uniquely constructed using this $3\times3$ unit cell.

Our systematic analysis reveals that precisely six distinct configurations [\cref{fig:spin_config_nu2}(a-f)] satisfy the ground state constraints. These states exhibit spontaneous translational symmetry breaking with periodicity of either $\sqrt{3}\times\sqrt{3}$ or $3\times3$, lattice units. To numerically verify possible ground state configurations, we implemented an optimized backtracking search algorithm with constraint propagation. This depth-first search approach employs two key optimizations: (1) dynamic prioritization of lattice sites based on their remaining configuration possibilities, and (2) maintenance of a backtracking stack for efficient decision-tree navigation. The algorithm proceeds as follows: 

Initialization begins with the central lattice site, enumerating all its possible configurations that are consistent with local bonding constraints and add it to the backtracking list. At each step, the algorithm selects the most constrained site - defined as having the minimal number of permissible valley configurations - from the current backtracking list. For a chosen site, we test each possible bond configuration sequentially. If a configuration is permissible, we commit it and proceed. If it is not, we test the next one. If no configurations are allowed, the algorithm backtracks to the previous decision point.
The newly added valid bond configurations trigger constraint propagation to adjacent sites, where we:

1) Fully determine the valley states and bonds for that site if its bonding configurations can be uniquely specified,

2) Update remaining configuration possibilities for partially constrained sites,

3) Re-sort the backtracking list based on updated permissible configuration counts.

When encountering forbidden configurations, the algorithm immediately prunes that branch of the search tree and retreats to the previous decision point. This constraint-driven approach ensures efficient exploration of the configuration space by focusing computational effort on the most critically constrained regions of the lattice.

The practical implementation of this algorithm requires a finite lattice, which introduces boundary effects. While certain boundary conditions can prevent a perfect tiling of the lattice with the six ground-state patterns, the energy contribution of the boundary is negligible relative to the bulk energy in the thermodynamic limit. 
We therefore
focus on the bulk pattern in this study.

\begin{figure}
    \centering
\includegraphics[width=0.7\linewidth]{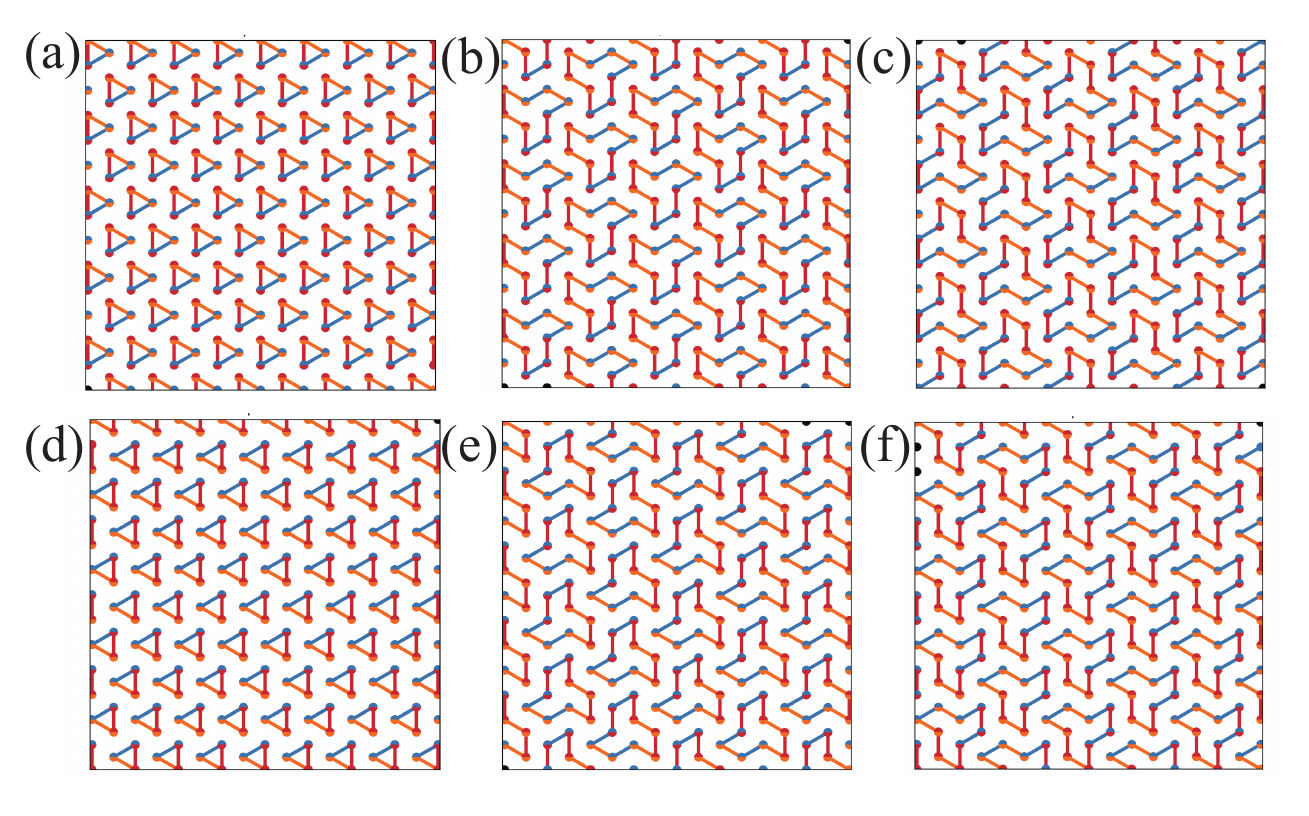}
    \caption{The distinct ground states configurations module translation obtained numerically at $\delta V_{\eta\eta'}(\Delta\vR)=0$ (a-f). Each dot represents a unit cell of the system. Each unit cell is filled with two electrons of two valleys which have been marked by two colors. The dots connected by different colors denote the spin chain formed by spins in different valleys. It is consistent with the ground states we obtained analytically in \cref{fig:spin_config_nu2}.}
    \label{fig:spin_config_nu2_num} 
\end{figure}

To mitigate finite-size effects, we perform our simulations on a large circular lattice that includes all sites satisfying $|\RR| \le 14|\bm{a}_{M,1}|$.
We refer to the central circular region with $|\RR| \le 10|\bm{a}_{M,1}|$ as the bulk, and focus solely on its properties.
This setup effectively isolates the bulk properties from edge interference. 
Within the central circular region ($|\RR|\le 10|\bm{a}_{M,1}|$), we numerically identify exactly six distinct ground state configurations, modulo lattice translations.
As shown in \cref{fig:spin_config_nu2_num}, this result is in perfect agreement with our analytical predictions. 



\subsubsection{Comparison between exact solutions of the spin model and Hartree-Fock simulations}

We now discuss the comparison between the exact solutions of the spin model and the results of the HF simulations at filling $\nu=2$. We begin by considering the strong coupling limit (red region in \cref{fig:phase_diagram_AA}), with a relatively small $12/\epsilon$, such that the Heisenberg coupling $J$ dominates. For example, at twist angle $\theta=3.89^\circ$, this condition corresponds to $0.14>12/\epsilon>0.04$. In this limit, the spin model suggests a VBS state. As discussed in \cref{sec:compare_hf_exact}, HF simulations fail to capture the spin-singlet formation as spin singlets are not Slater determinants. 
However, the formation of spin singlets arises from antiferromagnetic coupling. At the mean-field level, this coupling effectively leads to antiferromagnetic order, as discussed near \cref{eq:h_bond_strong_coupling}. 
We can therefore replace each spin-singlet dimer in the exact ground state (\cref{fig:spin_config_nu2}) obtained from the strong-coupling spin model (at $\delta V_{\eta\eta'}(\Delta\vR)=0$) with an antiferromagnetic state to construct the corresponding mean-field ground state.
Written explicitly, we let 
\begin{align}
    |\uparrow,\downarrow\rangle - |\downarrow,\uparrow\rangle \rightarrow   |\uparrow,\downarrow\rangle\text{ or }|\downarrow,\uparrow\rangle\, ,
\end{align}
We find that the resulting antiferromagnetic ground state is consistent with what we find in HF simulations shown in~\cref{fig:nu=2HF}(a). The antiferromagnetic state in~\cref{fig:nu=2HF}(a) has translational symmetry with respect to a $\sqrt{3}\times \sqrt{3}$ supercell, and a long-range antiferromagnetic order of each valley as shown in~\cref{fig:nu=2HF}.
In addition, one could also act the SU(2) rotational symmetry of each valley, or the $C_{3z}$ rotational symmetry to the state given in \cref{fig:nu=2HF}(a) to generate other degenerate Hartree-Fock ground states. 

In the strong coupling limit with relatively large $12/\epsilon$ (for example, $0.58>12/\epsilon>0.14$ with $\theta=3.89^\circ$), the $ \delta V_{\eta\eta'}(\Delta\vR) $ term dominates and favors a VP state. On top of the VP state, the Heisenberg coupling stabilizes an antiferromagnetic order state in the HF simulations and a quantum paramagnetic state in the spin model. 
The ground state configurations obtained from the HF simulations and the strong-coupling spin model are shown in \cref{fig:nu=2HF}(b) and \cref{fig:spin_config_nu2_vp}, respectively.



\subsection{Ground state in the strong coupling limit at $\nu=3$}

\subsubsection{Effective spin model at $\nu=3$} 
We now derive the effective spin model at $\nu=3$. As we discussed in~\cref{sec:spin_mode_nu_1} and around~\cref{eq:SW_ham_h0_h1}, we separate our Hamiltonian into
\begin{align}
    &H = H_0 +H_1 \nonumber\\ 
    &H_0 = H_{U(6)},\quad H_1=H_t +H_{\delta V} 
\end{align}
We treat $H_1$ as perturbation and derive an effective spin model. 

For $H_0$, the ground state at $\nu=3$ is 
\begin{align}
\label{eq:low_energy_state_nu_3}
    \prod_{\vR}\prod_{i=1,2,3} \cre{d}{\vR, \eta_\RR^i,s^{i}_{\vR}}|0\rangle 
\end{align}
where each site $\vR$ is filled by three electrons. 
We introduce the projection operator that only keeps low-energy states with $3$ electron per unit cell and is defined as
\begin{align}
P_L^{\nu=3} 
= \sum_{ \{ \eta_\RR^i, s_\vR^i\} }
 \prod_{\vR}\prod_{i}
 \cre{d}{\vR, \eta_\RR^i,s^{i}_{\vR}}
 |0\rangle 
 \bigg[ \prod_{\vR'}\prod_j \cre{d}{\vR', \eta_{\RR'}^j,s^{j}_{\vR'}}|0\rangle \bigg]^\dagger
\end{align}
Then the effective Hamiltonian at $\nu=3$ reads
\begin{align}
    H_J^{\nu=3}
    =&  P_L^{\nu=3}\bigg[ H_{\delta V}
    \bigg] P_L^{\nu=3}  
    - P_L^{\nu=3} H_t \sum_{|H\rangle } \frac{|H\rangle \langle H|}{E_H} H_t P_L^{\nu=3}
\end{align}
Note that $H_{\delta V}$ describes density-density interactions, thus does not mix $\ket{L}$ with $\ket{H}$.

The high energy state $|H\rangle$ created by acting $H_t$ on the ground states of $H_0$ can be written as 
\begin{align}
\label{eq:charge_0_nu_3}
   &\cre{d}{\vR,\eta,s}  \des{d}{\vR+\Delta\vR,\eta,s}   \prod_{\vR'}\prod_j \cre{d}{\vR', \eta_{\RR'}^j,s^{j}_{\vR'}}|0\rangle   \nonumber\\ 
   &
  \Delta\vR=  \pm  C_{3z}^\eta \bm{a}_{M,2}
  ,\quad (\eta,s) \in \{(\eta_{\vR+\Delta\vR}^i,s_{\vR+\Delta \vR}^i)\}_{i=1,2,3}
  ,\quad 
  (\eta,s) \notin \{(\eta_{\vR}^i,s_{\vR}^i)\}_{i=1,2,3}
\end{align}
with an excitation energy
\begin{align}
\label{eq:charge_0_nu_3_en}
    E_H = V(0) -V(\Delta\vR)
\end{align}
Similarly as~\cref{eq:ham_nu_1_after_SW} and ~\cref{eq:derive_spin_model_nu_2}, we derive the following spin-spin coupling 
\begin{align}
  J\sum_{\vR,\eta,s,s'}\sum_{\Delta\vR = \pm C_{3z}^{\eta}\bm{a}_{M,2}}
        P_L^{\nu=3}
       \bigg[\bm{S}_{\vR,\eta}\cdot \bm{S}_{\vR+\Delta\vR,\eta} +\frac{1}{4}n_{\vR,\eta}n_{\vR+\Delta\vR,\eta}\bigg] 
    P_L^{\nu=3}
\end{align}
with Heisenberg coupling 
\begin{align}
    J= 2t^2/E_H
\end{align}
The full Hamiltonian now reads
\begin{align}
\label{eq:def_spin_model_nu_3}
    H_{J}^{\nu=3} = &   P_L^{\nu=3}\bigg\{ 
    \frac{1}{2} \sum_{\vR,\Delta\vR, \eta,\eta'}\delta V_{\eta\eta'}(\Delta\vR) n_{\vR,\eta} n_{\vR+\Delta\vR,\eta'}
    \nonumber\\ 
    & +J\sum_{\vR,\eta,s,s'}\sum_{\Delta\vR = \pm C_{3z}^{\eta}\bm{a}_{M,2}}
       \bigg[\bm{S}_{\vR,\eta}\cdot \bm{S}_{\vR+\Delta\vR,\eta} +\frac{1}{4}n_{\vR,\eta}n_{\vR+\Delta\vR,\eta}\bigg] 
       \bigg\} 
    P_L^{\nu=3}
\end{align} 
We note that the effective interactions at $\nu = 3$ are the same to those at $\nu = 1, 2$, except for the projection operator.
In the limit where $\delta V_{\eta\eta'}(\Delta\vR) = 0$, the spin model possesses an additional $C_{2y}$ symmetry defined as
\begin{align}
    &C_{2y} \bm{S}_{\vR,\eta=0} C_{2y}^{-1} = 
    \bm{S}_{C_{2y}(\vR+\rr_0)-\rr_0,\eta=0} \nonumber\\ 
    &C_{2y} \bm{S}_{\vR,\eta=1} C_{2y}^{-1} = 
    \bm{S}_{C_{2y}(\vR+\rr_1)-\rr_2,\eta=2} \nonumber\\ 
    &C_{2y} \bm{S}_{\vR,\eta=2} C_{2y}^{-1} = 
    \bm{S}_{C_{2y}(\vR+\rr_2)-\rr_1,\eta=1}.
\end{align}
There is also a $U(2)$ symmetry for each valley and each chain attached to each valley (\cref{eq:additional_U2_symmetry}).

\subsubsection{Ground state of the effective spin model at $\nu=3$}
We now aim to obtain the ground state of the spin model at $\delta V_{\eta\eta'}(\Delta\vR)=0$ for the following Hamiltonian 
\begin{align}
P_L^{\nu=3}J\sum_{\vR,\eta,s,s'}\sum_{\Delta\vR = \pm C_{3z}^{\eta}\bm{a}_{M,2}}
       \bigg[\bm{S}_{\vR,\eta}\cdot \bm{S}_{\vR+\Delta\vR,\eta} +\frac{1}{4}n_{\vR,\eta}n_{\vR+\Delta\vR,\eta}\bigg] 
       \bigg\} \label{eq:intra_valley_coupling_nu_3}
    P_L^{\nu=3}
\end{align}
Similarly, as $\nu=1,2$, $n_{\vR,\eta}$ is a good quantum number. However, since we have three electrons per moir\'e unit cell, unlike $\nu=1,2$, it is impossible to find configurations that make all the electrons form a length-2 dimer.

\begin{figure}
    \centering
    \includegraphics[width=0.5\linewidth]{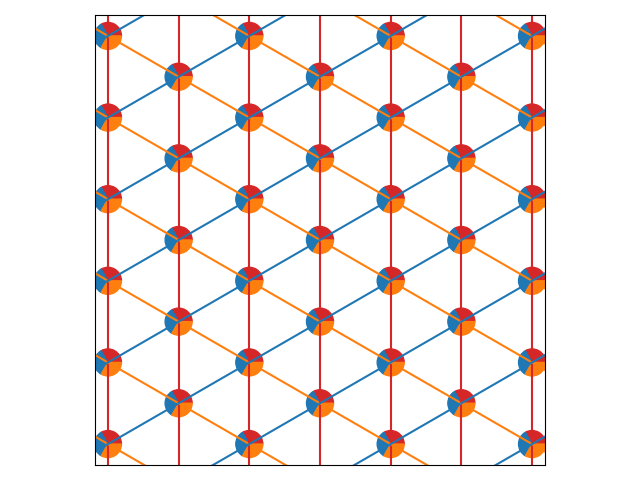}
    \caption{Ground state configurations of the spin model at $\nu = 3$. Each dot represents a unit cell of the system, with each unit cell containing three electrons in three distinct valleys, indicated by three different colors. The dots connected by different colors represent the spin chains formed by spins in different valleys.}
    \label{fig:spin_config_nu3}
\end{figure}

\begin{figure}
    \centering
    \includegraphics[width=0.5\linewidth]{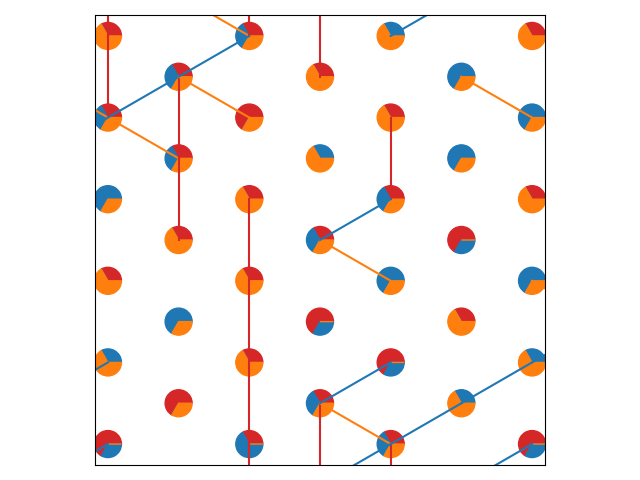}
    \caption{Randomly generated configuration for $\nu = 3$. Each dot represents a unit cell of the system, with each unit cell containing three electrons. The dots connected by different colors represent the spin chains formed by spins in different valleys. For a given dot, if $2/3$ of it is filled with the same color, this indicates that the corresponding unit cell hosts both spin-$\uparrow$ and spin-$\downarrow$ electrons from the same valley. }
    \label{fig:random_spin_config_nu3}
\end{figure}

One candidate ground state corresponds to a configuration in which each site hosts one electron per valley. In this case, the system can be described as a collection of infinite one-dimensional spin chains, with the corresponding configuration illustrated in~\cref{fig:spin_config_nu3}. 
 Besides the configuration shown in \cref{fig:spin_config_nu3}, we can also let two electrons from the same valley, but with opposite spin indices, occupy the same site. 
 This then breaks the infinite one-dimensional spin chain structure, allowing the formation of the finite-length spin chain. One particular randomly generated configuration, where the doubly occupied valley is allowed, has been shown in \cref{fig:random_spin_config_nu3}.

We now analyze the energy contributions of two scenarios, allowing or forbidding double occupancy of a single valley within a single unit cell, in a system with open boundary conditions. Our goal is to demonstrate that the configuration without such double occupancy (see \cref{fig:spin_config_nu3}) corresponds to the ground state.

\begin{figure}
    \centering
    \includegraphics[width=0.5\linewidth]{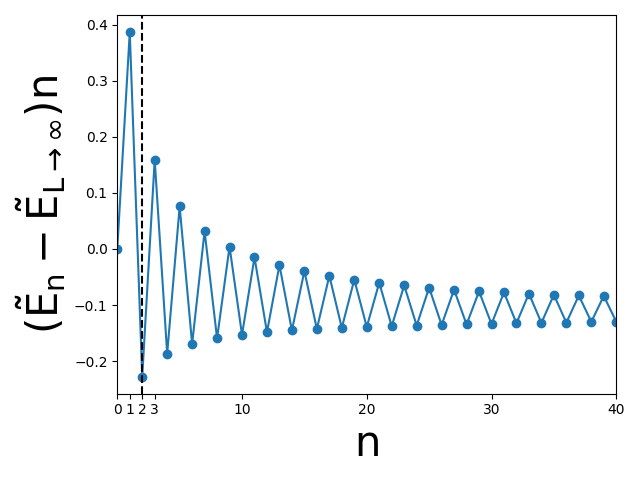}
    \caption{ $(\tilde{E}_{n}- \tilde{E}_{L\rightarrow\infty})n$ as a function of $n$. 
    The minimum is realized at $n=2$ as marked by black dashed line. 
    }
    \label{fig:energy_comp_nu_3}
\end{figure}

We let configuration $I$ (\cref{fig:spin_config_nu3}) denote the configuration where each moir\'e unit cell has $1$ electron per valley. 
We use configuration $II$ to denote a random configuration where each moir\'e unit cell has three electrons with arbitrary valley and spin flavors. 
We aim to show that, for any given configuration \( II \), its energy is higher than or equal to that of configuration \( I \), with equality if and only if configuration \( II \) is identical to configuration \( I \). We use $|I\rangle, |II\rangle$ to denote the corresponding states. 
Since the energy is determined by the intra-valley interaction terms (\cref{eq:intra_valley_coupling_nu_3}), we could calculate the energy contribution from each valley separately. Without loss of generality, we focus on the valley $0$. We note that, for valley $0$, the Hamiltonian coupling is only along $\Delta\vR = \pm \bm{a}_{M,2}$ directions. Therefore, we could calculate the energy for each 1D line characterized by index $n\in[0,...,L-1]$ (where $L$ is the length of the system). The 1D line is defined by  
\begin{align}
\label{eq:line_n}
    l_n^{\eta=0} = \{ 
    \RR = m\bm{a}_{M,2} + n\bm{a}_{M,1} |m\in[0,1,...,L-1] \}
\end{align}
We consider the energy contribution from the valley $0$ electron within the line $l_n^{\eta=0}$. The corresponding Hamiltonian is 
\begin{align}
\label{eq:AA_strong_coupl_nu_3_H_eta_n}
    H_{\eta=0,n} = 
    P_L^{\nu=3}J\sum_{\vR,\vR' \in l_n^{\eta=0}} 
    (\delta_{\vR-\vR', \bm{a}_{M,2}}
    +\delta_{\vR-\vR', -\bm{a}_{M,2}})
       \bigg[\bm{S}_{\vR,\eta=0}\cdot \bm{S}_{\vR',\eta=0} +\frac{1}{4}n_{\vR,\eta=0}n_{\vR',\eta=0}\bigg] 
       \bigg\} 
    P_L^{\nu=3}
\end{align} 
For configuration $I$, the electron of valley $I$ forms a spin chain with length $L$, and the corresponding energy is (see \cref{eq:def_energy_of_spin_chain})
\begin{align}
   E_I = \langle I | H_{\eta=0,n} |I\rangle 
   = L \tilde{E}_L 
\end{align}
where $\tilde{E}_L$ denotes the energy per site for an open-boundary spin chain with length $L$ (see \cref{eq:def_energy_of_spin_chain}). 
We now consider configuration $II$, in which the site $\vR \in l_n^{\eta=0}$ can have either $0$, $1$ or $2$ electrons in valley $0$. Without loss of generality, we assume that there are $M$ unit cells within $l_n^{\eta=0}$ that are occupied by either $0$ or $2$ electrons with valley $0$. We let their positions be 
\begin{align}
    \RR_i = m_i \bm{a}_{M,2} + n\bm{a}_{M,1},\quad i =1,...,M
\end{align}
and $0 \le m_1 < m_2 < ... < m_M < L$. 
All other sites within $l_n^{\eta=0}$ are then occupied by $1$ electron with flavor $0$. 
We now calculate the energy of configuration $II$. Due to empty and doubly occupied states, the chain $l_n^{\eta=0}$ has been decomposed into $M+1$ spin chains. Each spin chain is characterized by sites
\begin{align}
    &\text{Spin chain}_0 = \{  x\bm{a}_{M,2} + n\bm{a}_{M,1}| x\in \mathbb{Z}, 0 \le x < m_1 \} \nonumber\\ 
    &\text{Spin chain}_i = \{  x\bm{a}_{M,2} + n\bm{a}_{M,1}| x\in \mathbb{Z}, m_{i} < x < m_{i+1} \} ,\quad \text{for}\quad 1<i< M \nonumber\\ 
    &\text{Spin chain}_{M} = \{  x\bm{a}_{M,2} + n\bm{a}_{M,1}| x\in \mathbb{Z}, m_{M} < x < L \} 
\end{align} 
The energy of each spin chain is then $\tilde{E}_{L_i} L_i$ with $L_i$ the length of the $i$-th spin chain. Then the total energy contribution from the $M+1$ spin chains is 
\begin{align}\label{eq:E_II_spin}
    E_{\text{Spin chain}} = &\sum_{i=0,...,M+1} \tilde{E}_{L_i}L_i \nonumber\\ 
    =& \tilde{E}_{m_1}m_1 
    + \sum_{i=1,...,M-1}\tilde{E}_{m_{i+1}-m_i-1}(m_{i+1}-m_i-1) 
    + \tilde{E}_{ L-m_{M}-1 }(L-m_M-1)
\end{align}
Then, the energy of configuration $II$ is 
\begin{align}
\label{eq:E_II_sep}
    E_{II} = \langle II | H_{\eta=0,n}|II\rangle = 
    E_{\text{Spin chain}} + E_{\text{doublon}}
\end{align}
where $E_{\text{doublon}}$ denotes the additional energy contribution from the doubly occupied states. We now discuss such a contribution in more detail. 
When we calculate $E_{\text{Spin chain}}$, we only consider the interactions between sites within the same chain $\text{Spin chain}_i$, which are all singly occupied (of valley $0$). However, there is an additional energy contributions that come from the interaction between the doubly-occupied site (of valley $0$) and its nearby site. 
We now include such a contribution to the energy. 
We assume that $\vR$ is the site with either $0$ or $2$ electrons of valley $\eta=0$. 
The coupling between electrons at site $\vR$ of valley $0$ and its nearby site $\RR'$ is described by (\cref{eq:AA_strong_coupl_nu_3_H_eta_n})
\begin{align}
\frac{J}{4} \bigg( \bm{S}_{\vR,\eta=0}\cdot\bm{S}_{\vR',\eta=0} + \frac{1}{4}n_{\vR,\eta=0} n_{\vR',\eta=0} \bigg)    
\end{align}
Since valley $0$ of $\RR$ is either empty or doubly occupied, the spin-spin coupling gives zero energy since $\bm{S}_{\vR,\eta=0} |II\rangle =0$. Therefore, only density-density interactions will contribute, and such a contribution only appears for the doubly-occupied states of valley $0$. 
As we discussed, $n_{\vR,\eta}$ is a good quantum number, therefore, the state we considered - $|II\rangle $ - is also an eigenstate of $ \frac{1}{4}n_{\vR,\eta=0} n_{\vR'=0}$ density-density interaction term. The energy contribution from $ \frac{1}{4}n_{\vR,\eta=0} n_{\vR'=0} $ is non-negative. This suggests $E_{doublon}\ge 0 $, 
which allows us to introduce a lower bound of $E_{II}$ (via \cref{eq:E_II_spin,eq:E_II_sep}), which is 
\begin{align}
\label{eq:inequality_EII_and_E_chain}
    E_{II} \ge E_{\text{Spin chain}} 
\end{align} 
We now aim to show $E_{II}$ is always larger than or equal to $E_I$. We find 
\begin{align}
\label{eq:energy_bound}
    E_{II} -E_I \ge & E_{\text{Spin chain}} -E_I \nonumber\\ 
    =& 
    \tilde{E}_{m_1}m_1 
    + \sum_{i=1,...,M-1}\tilde{E}_{m_{i+1}-m_i-1}(m_{i+1}-m_i-1) 
    + \tilde{E}_{ L-m_{M}-1 }(L-m_M-1) 
    - L \tilde{E}_L \nonumber\\ 
    =& 
     \tilde{E}_{m_1}m_1 
    + \sum_{i=1,...,M-1}\tilde{E}_{m_{i+1}-m_i-1}(m_{i+1}-m_i-1) 
    + \tilde{E}_{ L-m_{M}-1 }(L-m_M-1)  \nonumber\\
    &
    -  \tilde{E}_L [m_1 + \sum_{i=1}^{M-1} (m_{i+1}-m_i -1) 
    + L-m_M -1 
    + M 
    ] \nonumber\\ 
    =& \sum_{i=1}^{M+1}
    (\tilde{E}_{m_i-m_{i-1}-1}-\tilde{E}_L)(m_i-m_{i-1}-1)
    - M\tilde{E}_L
\end{align}
where in order to obtain a compact formula, we have defined $m_{0}=-1$, and $m_{M+1}=L$. As shown in \cref{fig:energy_comp_nu_3}, we find, at the thermodynamic limit with $L\rightarrow \infty$, 
\begin{align}
   \text{argmin}_n\bigg[(\tilde{E}_{n}- \tilde{E}_{L\rightarrow\infty})n\bigg]  = 2 
\end{align} 
Therefore 
\begin{align}
\label{eq:middle_chain_contribution}
    (\tilde{E}_{m_i-m_{i-1}-1}-\tilde{E}_L)(m_i-m_{i-1}-1)\ge
    2(\tilde{E}_{2}- \tilde{E}_{L\rightarrow\infty}) 
\end{align}
On the other hand, as $L\rightarrow \infty$, the $M+1$-th spin chain must have infinite length with
\begin{align}
     &\text{Spin chain}_{M+1} = \{  x\bm{a}_{M,2} + n\bm{a}_{M,1}| x\in \mathbb{Z}, m_{M} < x < L \} 
     \rightarrow \{  x\bm{a}_{M,2} + n\bm{a}_{M,1}| x\in \mathbb{Z}, m_{M} < x < \infty \} 
\end{align}
Then its contribution becomes $0$ with
\begin{align}
\label{eq:last_chain_contribution}
     (\tilde{E}_{L-m_{M}-1}-\tilde{E}_L)\xrightarrow{L\rightarrow \infty} 0 
\end{align}
Therefore, we conclude that (from \cref{eq:last_chain_contribution,eq:middle_chain_contribution,eq:last_chain_contribution}), 
\begin{align}
    E_{II} -E_I \ge&  \sum_{i=1}^{M} 
    (\tilde{E}_2 - \tilde{E}_{\infty})2   - M \tilde{E}_{\infty}  \nonumber\\ 
    =& M ( 2\tilde{E}_2 -3\tilde{E}_{\infty})
\end{align} 
From \cref{eq:energy_length_2_spin_chain,eq:energy_infinit_spin_chain},  we find 
\begin{align}
    2\tilde{E}_2 - 3\tilde{E}_{\infty} \approx 0.158 >0 
\end{align}
Thus 
\begin{align}
    E_{II} -E_I \ge 0
\end{align}
where the equality can only be realized when $M=0$, which indicates there are no doubly occupied or empty states of valley $\eta=0$ in the configuration $II$ of chain $l_n^{\eta=0}$.

We thus that configuration $II$, with doubly occupied or empty states of valley $0$, always has higher energy than configuration $I$ when we consider $H_{\eta=0,n}$ term. Following the same logic, we could then generalize the conclusion to all the chains $l_n^{\eta=0}$ for valley $0$. 
Since three valleys are equivalent, we could also generalize the statement to all valleys. 
Then, using the fact that $[H_{\eta,n}, H_{\eta',n'}] = 0 $, we reach the conclusion that the configuration where each valley of each moir\'e unit cell is occupied by exactly 1 electron (\cref{fig:spin_config_nu3}) has the lowest energy.

Including weak hoppings that deviate from a purely one-dimensional structure induces an intra-valley spin-spin coupling between chains. 
Within the weak spin-spin inter-chain coupling, the low-energy physics of each valley can be described by an anisotropic triangular Heisenberg model. Previous investigations\cite{PhysRevB.103.235132,PhysRevB.64.094425,PhysRevB.59.11398,sun2025singlebandtriangularlatticehubbard} have explored such a model and have identified different phases such as spin-liquid, spiral magnetic order, collinear magnetic order, and so on. However, due to the lack of exact solutions, the precise phase diagram remains uncertain.

\subsubsection{Comparison between exact solutions of the spin model and Hartree-Fock simulations}
In the Hartree-Fock simulation of the strong coupling region with $0.08 < 12/\epsilon < 7.84$ for $\theta = 3.89^\circ$, $0.12 < 12/\epsilon < 3.56$ for $\theta = 4.41^\circ$, $0.2 < 12/\epsilon < 1.84$ for $\theta = 5.09^\circ$, $0.28 < 12/\epsilon < 1.4$ for $\theta = 6.01^\circ$, and $0.6 < 12/\epsilon < 1.48$ for $\theta = 7.34^\circ$ (\cref{fig:phase_diagram_AA_mom}), the ground state pattern is given in Fig.~\ref{fig:nu=3HF}(a) which is consistent with \cref{fig:spin_config_nu3}. In the Hartree-Fock simulations, each unit cell is filled by three electrons from three different valleys. Within each valley, we observe a long-range antiferromagnetic order. We comment that the HF simulations cannot describe the spin-liquid states of the exact ground states, but capture the antiferromagnetic correlations that drive the 1D spin liquid states.

\section{AB stacking SnSe$_2$}
\label{sec:app:AB}
We now discuss the phase diagram of AB-stacked SnSe$_2$. 
The system can be separated into two limits: the strong-coupling limit, where interactions dominate the physics, and the weak-coupling limit, where the interaction strength is smaller than or comparable to the bandwidth, making the Fermi surface more relevant to the low-energy behavior. 
It is difficult to quantitatively distinguish between the strong-coupling and weak-coupling regimes based on the angles of the system, as the interaction strengths also depend on the value of the dielectric constant $\epsilon$. 
However, numerically, we have observed that (\cref{tab:parameter_val_AB}), for small angles $\theta= 3.89^\circ,  4.41^\circ$, the interaction over bandwidth at $\epsilon=12$ is around $8$. However, at large angles, $\theta= 6.01^\circ, 7.34^\circ $, the interaction over bandwidth at $\epsilon=12$ is smaller than $4$. 
Therefore, qualitatively, the strong-coupling limit is more likely to be realized at smaller twist angles (e.g., $\theta = 3.89^\circ$, $4.41^\circ$), while the weak-coupling regime is expected to occur at larger angles (e.g., $6.01^\circ$, $7.34^\circ$).

\subsection{Strong coupling limit}
\label{sec:ab_strong_coupling}
In this section, we discuss AB-stacked SnSe$_2$ in the strong coupling limit, where the interaction strength significantly exceeds the hopping strength.

\subsubsection{Classical charge model}
\label{sec:ab_strong_coupling_charge_model}
In the strong coupling limit, where the hopping is relatively small, the density-density interactions dominate. 
The interaction model with only density-density interactions and chemical potential $\mu$ can be written as 
\begin{align}
\label{eq:strong_coupling_charge_model}
    H_{charge} =& \frac{1}{2}\sum_{\vR,\Delta\vR,\eta\eta'}V_{\eta\eta'}(\Delta\vR) n_{\eta}(\vR) n_{\eta'} (\vR+\Delta\vR) +\sum_{\vR,\eta,s}(-\mu) n_{\eta} (\vR)
\end{align}
where 
\begin{align}
    n_{\eta }(\vR) =\sum_s \cre{d}{\vR,\eta,s}\des{d}{\vR,\eta,s}
\end{align}
For the AB stacked SnSe$_2$, the Wannier centers of three valleys approximately form a kagome lattice (~\cref{fig:wannier_AB}). This suggests that we could map the strong coupling model (\cref{eq:strong_coupling_charge_model}) to an effective kagome Ising model, where the charge operator can be identified as an effective Ising operator. 

We now provide a detailed demonstration of the mapping.
We consider filling $0 < \nu< 3 $. 
From \cref{tab:parameter_val_AB}, we observe that the on-site interaction $U$ is more than twice the nearest-neighbor repulsion $V_1$ at twist angles $3.89^\circ$ and $4.41^\circ$, and approximately $1.6$ to $2.0$ times $V_1$ at angles $6.01^\circ$ and $7.34^\circ$. 
Due to the large on-site interactions(\cref{tab:parameter_val_AB}), the system does not favor double-occupied states, and we could assume 
\begin{align}
\label{eq:strong_coupl_AB_no_double}
 n_{\eta}(\vR)  \in \{0,1\} 
\end{align}
To make the mapping clearer, we also introduce the following Ising variables
\begin{align}
\label{eq:map_ising_to_n}
    \sigma_{\vR,\eta} =  2n_{\eta }(\vR)-1  \in \{-1,1\} 
\end{align} 
Then the charge-charge interaction model can be mapped to an effective Ising model with a magnetic field. 
The Hamiltonian reads
\begin{align}
    H_{Ising} = \sum_{\vR,\Delta\vR,\eta,\eta'} \frac{V_{\eta\eta'}(\Delta\vR) }{8}\sigma_{\vR,\eta} 
    \sigma_{\vR+\Delta\vR,\eta'} +h\sum_{\vR,\eta} \sigma_{\vR,\eta} 
\end{align}
where 
\begin{align}
\label{eq:chem_pot_to_h}
    h = \frac{-\mu}{2} + \frac{1}{12}\sum_{\eta\eta'}V_{\eta\eta'}(\Delta\vR)
\end{align}
The effective magnetic field $h$ tunes the total magnetization of Ising variables which is equivalent to the filling of the system
\begin{align}
\label{eq:density_op_avg_to_Ising_op_avg}
\frac{1}{N_M}\sum_{\vR,\eta} \langle \sigma_{\vR,\eta}\rangle= \frac{1}{N_M}\sum_{\vR,\eta} \langle 2n_{\vR,\eta}-1\rangle = 2 \nu -3 
\end{align}

\begin{figure}
    \centering
    \includegraphics[width=1.0\linewidth]{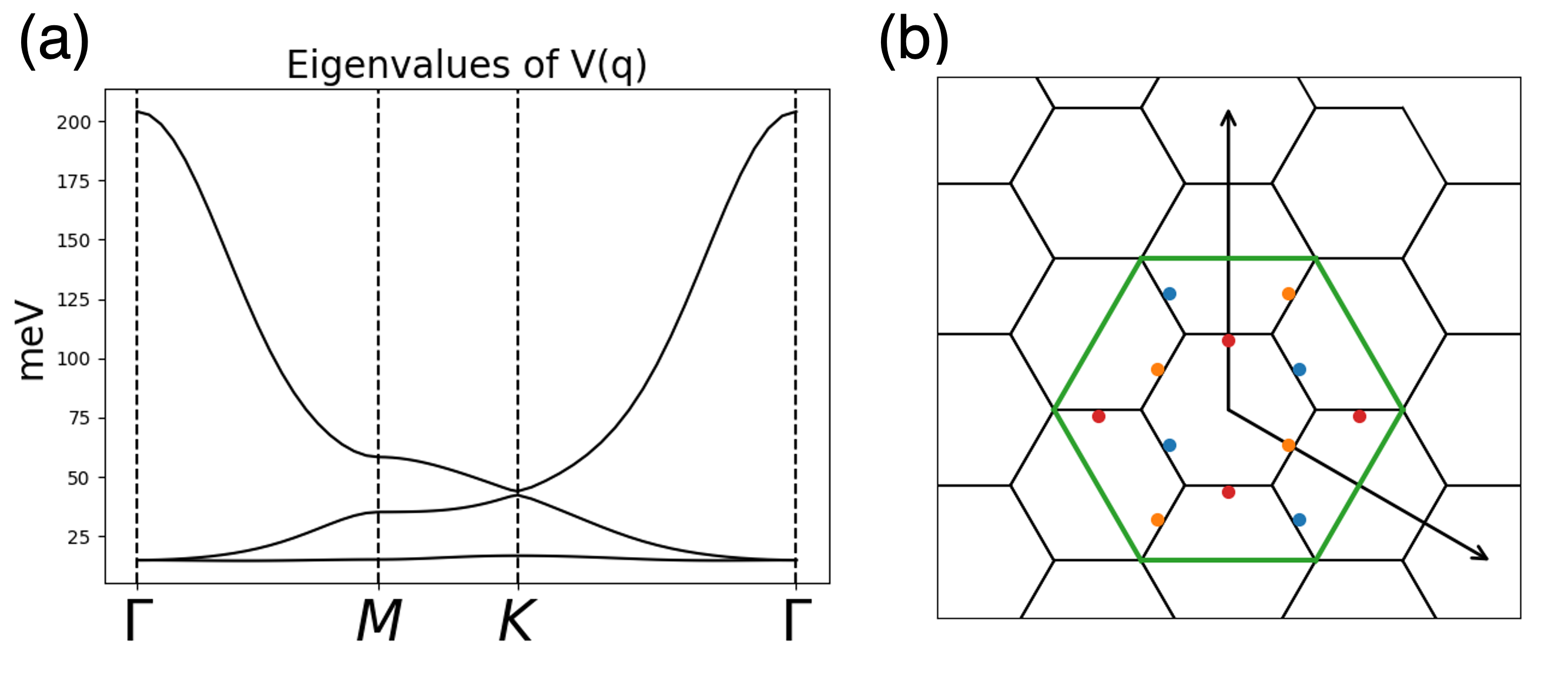}
    \caption{(a) The eigenvalues of the interaction matrix (\cref{eq:AB_Vq_def}) at $\theta=3.89^\circ$. (b) 12 dots in the figure characterize the supercell we picked in the calculations. Three colors label three valleys (red: valley-0; blue: valley-1; orange: valley-2). The black lines mark the boundary of each moir\'e unit cell.  The Green lines mark the boundary of one supercell, with two arrows indicating the lattice vectors of the supercell.
    }
    \label{fig:charge_config_AB}
\end{figure}

To understand the properties of the Ising model, we perform the following Fourier transformation 
\begin{align}
    V_{\eta\eta'}(\vq) = \frac{1}{N}\sum_{\Delta\vR} e^{i\vq\cdot \Delta\vR} V_{\eta\eta'}(\Delta\vR)
    \label{eq:AB_Vq_def}
\end{align}
In the momentum space, the Hamiltonian can also be written as 
\begin{align}
    &H_{Ising} =N\bigg[ \sum_{\qq,\eta,\eta'} 
    \frac{V_{\eta\eta'}(\qq)}{8}\sigma_{-\qq,\eta}\sigma_{\qq,\eta'} + h\sum_\eta \sigma_{\qq,\eta} \bigg] 
    \nonumber\\ 
    &\sigma_{\qq,\eta} = \frac{1}{N}\sum_\qq \sigma_{\RR,\eta}e^{-i\qq\cdot\RR}
\end{align}
We show the eigenvalues of $V(\vq)$ in~\cref{fig:charge_config_AB} (a). Due to the kagome structures, we observe a flat band, which also indicates strong fluctuations in the charge channel. 
Qualitatively, the flat eigenvalues of $V(\vq)$ indicate that various types of long-range order, associated with different wave-vectors $\qq$ and characterized by the condensation of specific $\sigma_{\qq,\eta}$ fields, are energetically degenerate. This suggests strong thermal fluctuations in the charge degrees of freedom at finite temperature and leads to ground state degeneracy at zero temperature.

We also note a key difference between our model and the Ising model (with Ising-Ising interactions and external magnetic fields): in our case, each Ising configuration can exhibit additional degeneracy, whereas in the usual Ising model, each Ising configuration corresponds to a unique state without further degeneracy. 
This can be seen from the following mapping between Ising variables and the density operators: 
\begin{align}
\label{eq:charge_ising_mapping}
    &\sigma_{\vR,\eta} =-1 \Leftrightarrow n_{\eta \uparrow}(\vR) = n_{\eta \downarrow}(\vR) = 0 \nonumber\\ 
    &\sigma_{\vR,\eta} = +1 \Leftrightarrow n_{\eta \uparrow}(\vR)= 1-n_{\eta \downarrow}(\vR)= 1\quad 
    \text{or} \quad n_{\eta \uparrow}(\vR)= 1-n_{\eta \downarrow}(\vR)= 0
\end{align}
We can observe that singly occupied spin $\uparrow$ and spin $\downarrow$ states of the original $d$ electrons correspond to the same Ising state with $\sigma_{\vR,\eta}=+1$. 
Therefore, each configuration of Ising variables, characterized by $\{\sigma_{\vR,\eta}\}_{\vR,\eta}$, corresponds to 
$2^{\sum_{\vR,\eta} (\sigma_{\vR,\eta} + 1)/2}$ configurations of density operators due to the additional spin degeneracy of $d$ electrons. 
Since our strong-coupling model (\cref{eq:strong_coupling_charge_model}) does not introduce interactions in the original physical spin channels (as opposed to the effective Ising spins) and depends only on the charge configuration ${n_\eta(\RR)}_{\eta,\RR}$, all states sharing the same charge distribution but differing in spin configurations are energetically degenerate. As a result, this additional spin degeneracy does not influence the ground-state energy but contributes to the entropy of the system at the strong-coupling limit. 
In addition, we comment that a perturbative term arising from finite hopping leads to a Heisenberg-type interaction between spins, with strength proportional to $t^2/U$. This additional coupling can further stabilize long-range spin order and quench the entropy associated with the spin degrees of freedom.


We also comment that, in what follows, we will use density operators $n_{\eta}(\vR)$ instead of Ising variables $\sigma_{\vR,\eta}$. In this subsection, we introduce Ising variables to more clearly illustrate the mapping between the charge-charge interaction model and the Ising model.

\subsubsection{Ground state of the classical charge models}
\label{sec:gnd_classical_charge}
Since the model (\cref{eq:strong_coupling_charge_model}) is classical, it can be solved exactly under certain assumptions. For any given charge configuration, the energy of the system can be computed directly. However, the total number of possible charge configurations scales as $2^{3N}$, where $N$ is the number of moir\'e unit cells and $3$ comes from three valleys. 
To simplify the problem, we restrict our attention to configurations that respect the translational symmetry of a supercell, as defined in~\cref{fig:charge_config_AB}(b). 
In general, larger supercells allow for a more comprehensive exploration of competing states, leading to more reliable results. 
Notably, within our current choice of supercell, the resulting ground state still preserves the translational symmetry of the original moir\'e unit cell. We thus expect that enlarging the supercell further is unnecessary in this case. 

To be more specific, we use $C_{\mathbb{\mathcal{R}}}$ to denote the supercell characterized by $\mathbb{\mathcal{R}} \in \mathbb{Z}\bm{a}_{SC,1} + \mathbb{Z}\bm{a}_{SC,2}$ with $\bm{a}_{SC,i=1,2}$ the lattice vectors for the supercells ($\bm{a}_{SC,1}=2 \bm{a}_{M,1} , \bm{a}_{SC,2}=2 \bm{a}_{M,2}$). 
For a given super-cell $C_{\mathbb{\mathcal{R}}}$, 
we use $\{n_i = 0, 1\}_{i=1,...,12}$ to characterize the corresponding charge configurations, where $i$ is the index of the site within the supercell with position $\rr_i$ and valley $\eta_i$ correspondingly, and $n_i=0,1$ denote the sites are filled by 0 or $1$ electrons respectively.
Due to the strong on-site repulsion, we do not consider the doubly occupied states, which are high-energy states. Therefore, for a given configuration $\{n_{i}\}_{i=1,...,12}$, we could construct the following state, which has translational symmetry characterized by the supercell in ~\cref{fig:charge_config_AB} (b)
\begin{align}
\label{eq:eigen_state_AB_charge_charge}
\prod_{\mathcal{R}}\prod_i (d_{\mathcal{R}+\rr_i,\eta_i, s_{\rr_i,\eta_i}}^\dag )^{n_i}|0\rangle 
\end{align}
with arbitrary spin indices $\{s_{\rr_i,\eta_i}\}_i$.
Since the Hamiltonian we considered only has density-density interactions, the state given in \cref{eq:eigen_state_AB_charge_charge} is also the eigenstate of the Hamiltonian given in \cref{eq:strong_coupling_charge_model}.
Also, with only density-density interaction, the energy of the state will not depend on the spin indices $\{s_{\rr_i,\eta_i}\}_i$, and only depends on $\{n_i\}_{i=1,...,12}$

We now aim to obtain the ground state at integer filling $\nu=1,2,3$. To do so, we first generate all possible characteristics of $\{n_i \}_{i=1,...,12}$. There will be $2^{12}$ of them ($n_i\in \{0,1\}$). We then pick out the configurations with the filling we considered. In other words, we only consider the configurations with $\sum_i n_i = 4\nu$ (the area of the supercell is 4 times the area of the original unit cell). We then calculate the energy of all configurations that we picked out, and find the configurations with the lowest energy, which corresponds to the ground state. 

By comparing the energy of all possible configurations with translational symmetry characterized by the supercell (\cref{fig:charge_config_AB} b), we identify the corresponding ground state. 
At filling $\nu=1$, the ground states we found are valley-polarized states with 
\begin{align}
\label{eq:ab_charge_gnd_nu1}
    \prod_{\vR} \cre{d}{\vR,\eta,s_{\vR}}|0\rangle 
\end{align}
for given $\eta$ and arbitrary spin configurations $s_{\vR}$. Here, $\vR$ characterizes the position of the original unit cell instead of the supercell. 
Therefore, the valley-polarized state preserves the translational symmetry of the original moir\'e unit cell.

At filling $\nu=2$, the ground states are valley-polarized states with 
\begin{align}
\label{eq:ab_charge_gnd_nu2}
    \prod_{\vR} \cre{d}{\vR,\eta_1,s_{\vR}}\cre{d}{\vR,\eta_2,s'_{\vR}}|0\rangle 
\end{align}
for given $\eta_1, \eta_2$ with $\eta_1\ne \eta_2$, and arbitrary spin configurations $s_{\vR},s'_{\vR}$.
Here, $\vR$ characterizes the position of the original unit cell instead of the supercell. Therefore, the valley-polarized state preserves the translational symmetry of the original moir\'e unit cell.

At filling $\nu=3$, the ground states are valley-polarized states with 
\begin{align}
\label{eq:ab_charge_gnd_nu3}
    \prod_{\vR} \cre{d}{\vR,0,s_{\vR}}\cre{d}{\vR,1,s'_{\vR}}\cre{d}{\vR,2,s_{\vR}''}|0\rangle 
\end{align}
with arbitrary spin configurations $s_{\vR},s'_{\vR},s''_{\vR}$.
Here, $\vR$ characterizes the position of the original unit cell instead of the supercell. Therefore, the valley-polarized state preserves the translational symmetry of the original moir\'e unit cell.


Finally, beyond the mean-field results, we also show that with only nearest-neighbor interactions and at integer fillings, the ground states given in \cref{eq:ab_charge_gnd_nu1,eq:ab_charge_gnd_nu2,eq:ab_charge_gnd_nu3} are the exact ground state of the charge-charge interaction Hamiltonian (\cref{eq:strong_coupling_charge_model}). 
We now aim to prove that states in \cref{eq:ab_charge_gnd_nu1,eq:ab_charge_gnd_nu2,eq:ab_charge_gnd_nu3} minimize  $H_{charge}$ (\cref{eq:strong_coupling_charge_model}).
We first consider the simplest case, which is $\nu=3$
(\cref{eq:ab_charge_gnd_nu3}). The state given in \cref{eq:ab_charge_gnd_nu3} has a charge configuration ${n}_{\eta}(\RR) = 1$ for all $\eta$ and all $\RR$. Since doubly occupied states are excluded due to the strong on-site Hubbard repulsion, this is the only allowed charge configuration and therefore corresponds to the ground state. However, we note that the spin degeneracy discussed below \cref{eq:charge_ising_mapping} also results in a trivial ground-state degeneracy.

We next consider the case of $\nu = 1$. Due to the equivalence of the three valleys, it suffices to consider the representative state in \cref{eq:ab_charge_gnd_nu1}, where valley 0 is occupied. 
This then leads to the following charge distribution
\ba 
n_{\eta=0}(\RR)=1,\quad n_{\eta=1}(\RR)=n_{\eta=2}(\RR)=0,\quad \text{for all }\RR
\ea
We next consider an arbitrary nearest-neighbor bond connecting two sites, labeled $(\RR, \eta)$ and $(\RR + \Delta\RR, \eta')$. For nearest-neighbor bonds on the kagome lattice, we have $\eta \neq \eta'$, indicating that the two sites belong to different valleys (or sublattices). Therefore, at least one of the two sites must be unoccupied, which leads to
\begin{align}
    \frac{1}{2} V_{\eta\eta'}(\Delta\RR) n_{\eta}(\RR) n_{\eta'}(\RR+\Delta\vR) = 0 
\end{align}
thereby minimizing the energy contribution from this bond (given that the nearest-neighbor charge-charge interaction is repulsive). 
Since the same argument applies to any nearest-neighbor bond, the total energy contribution from the charge-charge interaction term 
\ba 
\label{eq:strong_coupl_AB_H_V_int}
H_{V,int} = \frac{1}{2}\sum_{\vR,\Delta\vR,\eta\eta'}V_{\eta\eta'}(\Delta\vR) n_{\eta }(\vR) n_{\eta'} (\vR+\Delta\vR) 
\ea 
is also minimized. In addition, the chemical potential term 
$
-\mu \sum_{\RR,\eta, s}n_{\eta}(\RR) 
$
is a constant for the fixed total filling.
Therefore, \cref{eq:ab_charge_gnd_nu1} is the ground state at $\nu=1$.

We now consider the case of $\nu = 2$. Due to the equivalence of the three valleys, it suffices to consider the representative state in \cref{eq:ab_charge_gnd_nu2}, where valley 0 and 1 are occupied. 
This then leads to the following charge distribution
\ba 
n_{\eta=0}(\RR)=1,\quad n_{\eta=1}(\RR)=1,\quad n_{\eta=2}(\RR)=0,\quad \text{for all }\RR
\label{eq:charge_config_nu_2}
\ea
To show that the charge configuration in \cref{eq:charge_config_nu_2} minimizes $H_{charge}$ (\cref{eq:strong_coupling_charge_model}) at fixed total filling $\nu = 2$, we introduce the following new variables
\begin{align}
\label{eq:strong_coupl_AB_ph_transf_isign}
    \tilde{n}_{\eta}(\vR) = 1- n_{\eta}(\vR)
\end{align}
The Hamiltonian can then be written as 
\begin{align}
\label{eq:H_V_new_var}
    H_V = &\frac{1}{2}
    \sum_{\vR,\Delta\vR,\eta\eta'}
    V_{\eta\eta'}(\Delta\vR)
    \tilde{n}_\eta(\vR)
    \tilde{n}_{\eta'}(\vR+\Delta\vR) \nonumber\\ 
    &
    + 
    \sum_{\vR,\eta}
    \bigg[ \mu - \frac{1}{2} \sum_{\Delta\vR,\eta'}\bigg(V_{\eta\eta'}(\Delta\vR) 
    +V_{\eta'\eta}(-\Delta\vR) \bigg)
    \bigg] \tilde{n}_{\eta}(\vR)
    -\sum_\RR 3\mu + 
     \frac{1}{2}
    \sum_{\vR,\Delta\vR,\eta\eta'}
    V_{\eta\eta'}(\Delta\vR)
\end{align}
Since we fix the total filling to be $2$ and three valleys are equivalent, the second line of \cref{eq:H_V_new_var} becomes
\begin{align}
   &\sum_{\vR,\eta}
    \bigg[ \mu -  \frac{1}{2} \sum_{\Delta\vR,\eta'}\bigg(V_{\eta\eta'}(\Delta\vR) 
    +V_{\eta'\eta}(-\Delta\vR) \bigg)
    \bigg] \tilde{n}_{\eta}(\vR)
     -\sum_\RR 3\mu + 
     \frac{1}{2}
    \sum_{\vR,\Delta\vR,\eta\eta'}
    V_{\eta\eta'}(\Delta\vR) \nonumber\\
    = &
     -2 \mu N - \frac{N}{3}\sum_\eta \sum_{\Delta\vR,\eta'}\bigg(V_{\eta\eta'}(\Delta\vR) 
    +V_{\eta'\eta}(-\Delta\vR) \bigg)
    +
     \frac{1}{2}
    \sum_{\vR,\Delta\vR,\eta\eta'}
    V_{\eta\eta'}(\Delta\vR)
\end{align}
which is a constant at a fixed total filling. Therefore, it is sufficient to consider 
\begin{align}
\label{eq:strong_coupling_charge_model_int}
 \tilde{H}_{V,int} =  \frac{1}{2}
    \sum_{\vR,\Delta\vR,\eta\eta'}
    V_{\eta\eta'}(\Delta\vR)
    \tilde{n}_\eta(\vR)
    \tilde{n}_{\eta'}(\vR+\Delta\vR) 
\end{align}
The charge configuration that minimizes $ \tilde{H}_{V,int}$ is then the ground state.  
We now show that the charge configuration given in \cref{eq:charge_config_nu_2} minimizes $H_{V,int}$. Since we only consider nearest-neighboring coupling, we take an arbitrary nearest-neighbor bond connecting two sites, labeled $(\RR, \eta)$ and $(\RR + \Delta\RR, \eta')$. For nearest-neighbor bonds on the kagome lattice, we have $\eta \neq \eta'$, indicating that the two sites belong to different valleys (or sublattices). Therefore, at least one of the two sites must be occupied \cref{eq:charge_config_nu_2}. 
From \cref{eq:strong_coupl_AB_ph_transf_isign}, we then have $\tilde{n}_{\eta}(\RR) \tilde{n}_{\eta'}(\RR+\Delta\vR) = 0 $
which leads to
\begin{align}
    \frac{1}{2} V_{\eta\eta'}(\Delta\RR) \tilde{n}_{\eta}(\RR) \tilde{n}_{\eta'}(\RR+\Delta\vR) = 0 
\end{align}
This ensures that the energy contribution from the given bond is minimized (given that the nearest-neighbor charge-charge interaction is repulsive). Since the same argument applies to any nearest-neighbor bond, the energy from $H_{V,\text{int}}$ (\cref{eq:strong_coupling_charge_model_int}) is minimized by the charge configuration in \cref{eq:charge_config_nu_2}. Therefore, \cref{eq:ab_charge_gnd_nu2} corresponds to the ground state at filling $\nu = 2$. 

Finally, we comment that the Wannier centers of the three valleys are distinct, so valley polarization induces a charge modulation. However, this modulation still respects the original moir\'e translational symmetry.


\subsubsection{Finite-temperature behavior}
The eigenvalues of the interaction matrix show a relatively flat band~\cref{fig:charge_config_AB} (a). This indicates strong charge fluctuations. 
For the kagome Ising model with nearest-neighbor coupling and zero magnetic field, it has been shown\cite{kano1953antiferromagnetism} that the strong fluctuations lead to 
a non-zero entropy at zero-temperature indicating a classical spin-liquid phase. 
Here, our charge-charge interaction model can also be mapped to a Kagome Ising model. To investigate the fluctuations of our model, we investigate the entropy of the system using a cluster mean-field calculation.


We first discuss the cluster mean-field solution. We consider the following density-density interaction terms 
\begin{align}
    H_V =& \frac{1}{2}\sum_{\vR,\Delta\vR,\eta\eta'}V_{\eta\eta'}(\Delta\vR) n_{\eta }(\vR) n_{\eta'} (\vR+\Delta\vR) +\sum_{\vR,\eta}(-\mu) n_{\eta} (\vR)
\end{align} 
We introduce supercell denoted by $C_{\mathbb{\mathcal{R}}}$ with $\mathbb{\mathcal{R}} \in \mathbb{Z}\bm{a}_{SC,1} + \mathbb{Z}\bm{a}_{SC,2}$ and $\bm{a}_{SC,i=1,2}=2\bm{a}_{i=1,2}$ the lattice vectors for the super-cells. The choice of one supercell has been shown in~\cref{fig:charge_config_AB} (b) with 12 sites. 
We then separate the Hamiltonian into intra-supercell contributions and inter-supercell contributions
\begin{align}
\label{eq:h_v_strong_coup}
    &H_V = H_{intra} + H_{inter} \nonumber\\ 
    &H_{intra} =  \sum_{\mathcal{R}} 
    \bigg[ \frac{1}{2}\sum_{(\vR,\eta) \in C_{\mathcal{R}}, (\vR+\Delta\vR,\eta')\in C_{\mathcal{R}}}V_{\eta\eta'}(\Delta\vR) n_{\eta }(\vR) n_{\eta'} (\vR+\Delta\vR)  
    + \sum_{(\vR,\eta)\in C_{\mathcal{R}}}(-\mu) n_{\eta} (\vR)\bigg]  \nonumber\\ 
    &H_{inter} =  \frac{1}{2}\sum_{\mathcal{R}\ne \mathcal{R'}} \sum_{(\vR,\eta) \in C_{\mathcal{R}}, (\vR+\Delta\vR,\eta')\in C_{\mathcal{R'}}}V_{\eta\eta'}(\Delta\vR) n_{\eta }(\vR) n_{\eta'} (\vR+\Delta\vR)  
\end{align}

We treat the inter-supercell coupling $H_{inter}$ at mean-field levels, which leads to
\begin{align}
    H^{MF}_{inter} = & \frac{1}{2}\sum_{\mathcal{R}\ne \mathcal{R'}} \sum_{(\vR,\eta) \in C_{\mathcal{R}}, (\vR+\Delta\vR,\eta')\in C_{\mathcal{R'}}} V_{\eta\eta'}(\Delta\vR) \bigg[\langle n_{\eta }(\vR) \rangle n_{\eta'} (\vR+\Delta\vR)  \nonumber\\ 
    &+
n_{\eta }(\vR) \langle n_{\eta'} (\vR+\Delta\vR)  \rangle 
    -\langle  n_{\eta }(\vR)\rangle \langle  n_{\eta'} (\vR+\Delta\vR)  \rangle \bigg] 
\end{align}
where the expectation value is taken with respect to the mean-field Hamiltonian 
\begin{align}
    H_{MF} = H_{inra}  +H_{inter}^{MF}
\end{align}
After the mean-field decoupling, each cluster is decoupled from all others (in the sense that the density operators of one cluster only couple to the expectation value of the density operators of another cluster), allowing it to be solved exactly by directly diagonalizing the Hamiltonian. In other words, each cluster is still influenced by the surrounding clusters, but this effect is treated only at the mean-field level.
The mean-field $\langle n_\eta(\vR)\rangle$ will be determined self-consistently. 

We now describe the calculation of entropy within the mean-field approximation. In this framework, the clusters are decoupled at mean-field level, allowing the partition function to be expressed as:
\begin{align}
    Z = \bigg[Z_{SC} \bigg]^{N_{SC}},\quad Z_{SC}= \sum_{\mathcal{N}} d[\mathcal{N}] e^{-\beta \mathcal{N}} 
\end{align}
where $N_{SC}$ is the number of supercells considered. The partition function for each supercell, $Z_{SC}$, is computed using the mean-field approximation. For each given charge configuration $\mathcal{N} = \{n_{\eta}(\vR)\}_{\vR}$ of a single cluster, we can solve the mean-field Hamiltonian ($H_{{MF}}$) of that cluster and obtain the corresponding energy $E_{\mathcal{N}}$. However, for each charge configuration $\mathcal{N}$, there exists an additional degeneracy due to the doubly degenerate singly-occupied states (see the discussion near \cref{eq:charge_ising_mapping}). The degeneracy factor is given by:
\begin{align}
    d[\mathcal{N}]= \prod_{\vR,\eta} 2^{n_{\eta}(\vR)}
\end{align}
We only consider the singly occupied and empty site. States with doubly occupied states are not included in the summation of the charge configurations ($\sum_{\mathcal{N}}$) due to the large on-site Hubbard repulsion. 
Then the partition function $Z_{SC}$ now is
\begin{align}
Z_{SC}= \sum_{\mathcal{N}} d[\mathcal{N}] e^{-\beta \mathcal{N}} 
\end{align} 
The entropy of the system is then
\begin{align}
\label{eq:entropy_charge_model}
    S = -\frac{\partial}{\partial \frac{1}{\beta}}\bigg[ -\frac{1}{\beta}\log(Z)
    \bigg] = N_{SC}\sum_{\mathcal{N}}\frac{d[{\mathcal{N}}] e^{-\beta E_{\mathcal{N}}}}{Z_{SC}}
    \bigg[\log(Z_{SC}) - \beta E_{\mathcal{N}}
    \bigg]
\end{align}
We observe that the entropy calculated above also includes contributions from spin degeneracy.
Since the Hamiltonian we considered at the strong coupling limit (\cref{eq:h_v_strong_coup}) does not have any spin-spin couplings, such a degeneracy is trivial. We are mostly interested in the entropy contribution from the strong fluctuations in the charge degrees of freedom. 
To isolate the entropy arising from the charge degrees of freedom, we introduce the following quantity to characterize the degeneracy associated with the charge configurations
\begin{align}
\label{eq:def_S_charge}
    S_{charge} = N_{SC}\sum_{\mathcal{N}}\frac{e^{-\beta E_{\mathcal{N}}}}{Z_{charge}}
    \bigg[\log(Z_{charge}) + \beta E_{\mathcal{N}}
    \bigg], \quad Z_{charge} =  \sum_{\mathcal{N}} e^{-\beta \mathcal{N}} 
\end{align}
$S_{charge}$ is constructed by setting the spin degenerate factor $d[\mathcal{N}]$ to $1$ in the definition of entropy $S$ (\cref{eq:entropy_charge_model}). 

\subsubsection{Numerical results}

\begin{figure}
    \centering
    \includegraphics[width=0.8\linewidth]{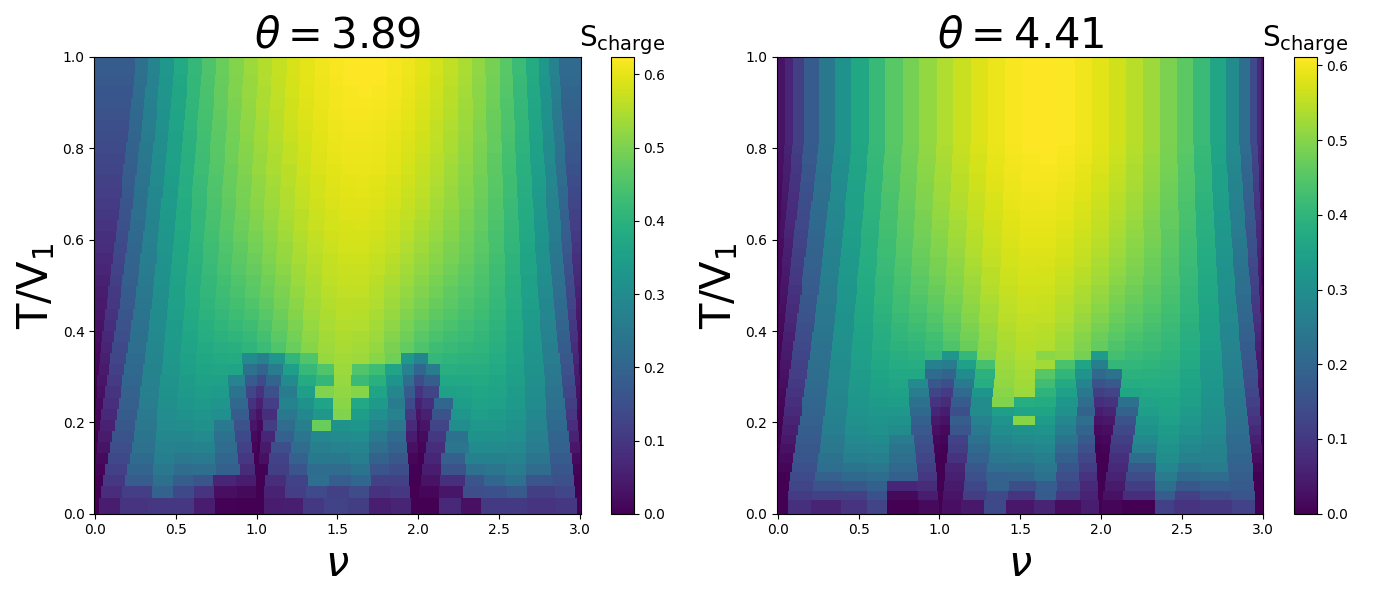}
    \caption{Entropy $S_{charge}$ (\cref{eq:def_S_charge}) at small angles for various fillings and temperatures. The color indicates the value of the entropy. $V_1$ denotes the interaction strength of the nearest-neighbor interactions (see \cref{tab:parameter_val_AB}). }
    \label{fig:entropy_AB_Ising}
\end{figure}

We illustrate the entropy calculated using the cluster mean-field approximation in~\cref{fig:entropy_AB_Ising} for the two smallest angles ($\theta= 3.89^\circ, 4.41^\circ$), where the strong coupling limit is more likely to be realized. 

Enhanced entropy is observed near the filling $\nu = 3/2$. This enhancement occurs because $\nu = 3/2$ corresponds to the kagome Ising model at zero magnetic field, where frustration increases the degeneracy. This mapping is given by \cref{eq:map_ising_to_n}.
At filling $\nu = 3/2$, the average expectation value of the Ising variable per unit cell per valley is given by:
\begin{align}
    \frac{1}{3N}\sum_{\vR,\eta} \langle \sigma_{\vR,\eta} \rangle = 
    \frac{1}{3N}\sum_{\vR,\eta} \langle  2n_{\eta}(\vR)-1 \rangle = \frac{2\nu -3}{3} =0 
\end{align}
This results in zero magnetization in terms of the Ising variables, corresponding to the zero magnetic field in the kagome Ising model. At low temperatures, where $T/V_1 \lesssim 0.1$, the entropy is observed to approach zero. This behavior arises because the realistic model includes longer-range density-density interactions, such as next-nearest-neighbor terms, that cause deviations from the ideal nearest-neighbor kagome Ising model. Within our mean-field solutions, these longer-range interactions stabilize a long-range ordered state and give a vanishing entropy at low temperatures.

Near integer fillings ($\nu = 1, 2$), the system tends to develop a valley-polarized state (see \cref{sec:gnd_classical_charge}). This state suppresses the system's entropy. At $\nu = 0$ and $\nu = 3$, all three valleys are either entirely empty or completely filled with one electron each, respectively. This quenching of charge fluctuations reduces the entropy at these fillings.

\end{document}